\documentclass[prd,aps,floatfix,showpacs,superscriptaddress]{revtex4}

\usepackage{graphicx}
\usepackage{amsmath}
\usepackage[psamsfonts]{amssymb}
\usepackage{longtable}

\newcommand{\be}{\begin{equation}}
\newcommand{\ee}{\end{equation}}
\newcommand{\bea}{\begin{eqnarray}}
\newcommand{\eea}{\end{eqnarray}}
\newcommand{\bi}{\begin{itemize}}
\newcommand{\ei}{\end{itemize}}

\def\simge{
    \mathrel{\rlap{\raise 0.511ex
        \hbox{$>$}}{\lower 0.511ex \hbox{$\sim$}}}}
\def\simle{
    \mathrel{\rlap{\raise 0.511ex
        \hbox{$<$}}{\lower 0.511ex \hbox{$\sim$}}}}

\newcommand{\Pslash}{p \kern -2mm /}
\def\del  {\partial}
\newcommand{\mnuc}{M_N}

\begin{document}



\title{
Nucleon form factors from quenched lattice QCD 
with domain wall fermions
}

\author{Shoichi Sasaki}
\email{ssasaki@phys.s.u-tokyo.ac.jp}
\affiliation{Department of Physics, The University of Tokyo, \\
Hongo 7-3-1, Tokyo 113-0033, Japan}

\author{Takeshi Yamazaki}
\email{yamazaki@phys.uconn.edu}
\altaffiliation[Present address: ]{Yukawa Institute for Theoretical Physics, Kyoto University,
Kitashirakawa-Oiwakecho, Sakyo,
Kyoto 606-8502, Japan}
\affiliation{Physics Department, University of Connecticut, \\
Storrs, Connecticut 06269-3046, USA}

\date{\today}

\begin{abstract}
We present a quenched lattice calculation of the weak nucleon form factors:
vector ($F_V(q^2)$), induced tensor ($F_T(q^2)$), axial-vector ($F_A(q^2)$) and induced pseudo-scalar ($F_P(q^2)$) form factors. 
Our simulations are performed on three different lattice sizes
$L^3 \times T=24^3 \times 32$, $16^3 \times 32$
and $12^3 \times 32$ with a lattice cutoff of  $a^{-1}\approx 1.3$ 
GeV and light quark masses down to about 1/4 the strange quark mass 
($m_\pi \approx 390$ MeV) using a combination of the DBW2 gauge action 
and domain wall fermions. The physical volume of our largest lattice is about 
$(3.6\;{\rm fm})^3$, where the finite volume effects on form factors 
become negligible and the lower momentum transfers 
($q^2 \approx 0.1\;{\rm GeV}^2$) are accessible. 
The $q^2$-dependences of form factors in the low $q^2$ region 
are examined. It is found that the vector, induced tensor, axial-vector form factors 
are well described by the dipole form, while the induced pseudo-scalar form factor
is consistent with pion-pole dominance.
We obtain the ratio of axial to vector coupling 
$g_A/g_V=F_A(0)/F_V(0)=1.219(38)$
and the pseudo-scalar coupling $g_P=m_{\mu}F_P(0.88m_{\mu}^2)=8.15(54)$,
where the errors are statistical errors only.
These values agree with experimental values 
from neutron $\beta$ decay and muon capture on the proton.
However, the root mean-squared radii of the vector, induced tensor and axial-vector 
underestimate the known experimental values by about 20 \%.
We also calculate the pseudo-scalar nucleon matrix element
in order to verify the axial Ward-Takahashi identity in terms of 
the nucleon matrix elements, which may be called as the
generalized Goldberger-Treiman relation.
\end{abstract}


\maketitle


 
\section{Introduction}
\label{sec:intro}
A comprehensive understanding of hadron structure, especially nucleon structure,
based on quantum chromodynamics (QCD) is one of our ultimate goals in lattice QCD
calculations. The latest lattice calculations of nucleon structure have been
greatly developed with increasing accuracy~\cite{Review}.
So far, large efforts by lattice QCD simulations have been mostly devoted 
to studies of electro-magnetic structure of the nucleon and 
either unpolarized or polarized parton distributions in deep 
inelastic scattering~\cite{{Review},{Gockeler:2003ay},{Hagler:2003jd},{Alexandrou:2006ru},
{Hagler:2007xi}}.
However, there are only a few lattice studies to be completed
for the weak nucleon form factors~\cite{{Liu:1994dr},{Alexandrou:2007xj}}, 
which are associated with weak probes of nucleon structure. In this paper, we 
present results from our 
intensive study of the nucleon matrix elements of the weak current in quenched 
lattice QCD calculations with domain wall fermions (DWFs).

Experimentally, weak processes meditated by the weak charged current 
like neutron beta decay $n\rightarrow p+e^{-}+\bar{\nu}_e$, muon 
capture on the proton $\mu^{-} + p \rightarrow \nu_{\mu} + n$ or 
quasi-elastic neutrino scattering $\bar{\nu}_\mu + p \rightarrow \mu^{+} + n$
are mainly exploited for studying the weak nucleon form factors, while
available information obtained from the experiment of the neutral current 
weak process such as semileptonic elastic scattering $\nu + p \rightarrow \nu + p$ 
is still limited. The weak current is known to be described by a linear combination 
of the vector and axial-vector currents. 
In general, four form factors appear in the nucleon matrix elements of
the weak current. Here, for example, we consider the matrix element for neutron 
beta decay. In this case, the vector and axial-vector currents are
given by $V_{\alpha}^+(x)={\bar u}(x)\gamma_{\alpha} d(x)$ and 
$A_{\alpha}^{+}(x)={\bar u}(x)\gamma_{\alpha}\gamma_{5} d(x)$ and 
then the matrix element is expressed by
\bea
\langle p|V_{\alpha}^+(x) + A_{\alpha}^{+}(x)|n\rangle 
&=&\bar{u}_p\left[
\gamma_{\alpha}F_V(q^2) + \sigma_{\alpha \beta} q_{\beta}F_T(q^2)
\right. \nonumber \\
&&\left. +
\gamma_{\alpha}\gamma_5F_A(q^2) +i q_{\alpha}\gamma_5F_P(q^2)
\right]u_n e^{iq\cdot x},
\label{Eq:beta_decay}
\eea
where $q=P_n - P_p$ is the momentum transfer between the proton ($p)$ 
and neutron ($n$). 
The vector ($F_V$) and induced tensor ($F_T$) form factors are introduced 
for the vector matrix element, and also the axial-vector ($F_A$) and induced pseudo-scalar ($F_P$) form factors for the axial-vector matrix element.
The vector part of weak processes
are related to the nucleon's electro-magnetic form factors, which are well measured 
up to large momentum transfer by electron scattering~\cite{Thomas:2001kw},
through an isospin rotation.  
Based on the conserved-vector-current hypothesis, the vector and induced tensor form factors are well understood by knowledge of the electro-magnetic structure of the nucleon.

In the axial-vector part of the weak process, the axial-vector coupling $g_A=F_A(q^2=0)$ 
is most accurately measured by neutron beta decay, where the extremely
small momentum transfer is accessible due to a very small mass difference 
of the neutron and proton. The $q^2$-dependence of $F_A(q^2)$ can be 
determined by other processes such as quasi-elastic neutrino scattering experiments
and charged pion electroproduction experiments. It has been observed 
that the dipole form is a good description for low and moderate momentum transfer,
$q^2< 1\;{\rm GeV}^2$~\cite{Bernard:2001rs}.
On the other hand, the induced pseudo-scalar form factor $F_P(q^2)$ is rather 
less known experimentally. The main source of information on $F_P(q^2)$ 
stems from muon capture. The induced pseudo-scalar coupling, $g_P=m_{\mu}F_P(q^2)$
evaluated at $q^2 =0.88 m_{\mu}^2$, where $m_{\mu}$ is the muon mass,  is
measured by ordinary muon capture (OMC) or radiative muon
capture (RMC). Although there is some discrepancy between the OMC result and 
the RMC result~\cite{{Bernard:2001rs},{Gorringe:2002xx}},
the new precise OMC measurement by the MuCap collaboration, 
which is  nearly independent of $\mu$-molecular effect,  
yields $g_P=7.3\pm 1.1$~\cite{Andreev:2007wg}. 
Only a few of the other $q^2$ data points on $F_P(q^2)$ 
are measured in the low $q^2$ region by a single experiment 
of pion electroproduction~\cite{Choi:1993vt}. 

Theoretically, in the axial part of such weak processes at low energies, one may consider
that spontaneous chiral symmetry breaking, which is induced by the strong interaction, 
plays an essential role. In other words, the axial structure of the nucleon would be 
highly connected with the physics of chiral symmetry and its spontaneous breaking, 
which ensures the presence of pseudo Nambu-Goldstone particles such as the pion.
This is empirically known as the partially conserved axial-vector current (PCAC) 
hypothesis~\cite{Thomas:2001kw}, 
where the divergence of the axial-vector current is proportional to the pion field.
Applying this idea to the axial-vector part of Eq.(\ref{Eq:beta_decay}), 
there appears a specific relation between the residue of the pion-pole structure in $F_P(q^2)$
and the axial-vector coupling $g_A$ known as the Goldberger-Treiman 
relation~\cite{Goldberger:1958tr}.
 
There was the long standing disagreement between experiment and lattice calculations
about the axial-vector coupling $g_A$. However, the RBC Collaboration finally 
resolved this puzzle using quenched DWF simulations~\cite{{Sasaki:2001tha},{Sasaki:2003jh}}.
DWFs are expected to provide an implementation of lattice fermions with 
exact chiral symmetry~\cite{{Kaplan:1992bt},{Shamir:1993zy},{Furman:1995ky}}. 
In the limit where the fifth-dimensional extent $L_s$ is taken to infinity, 
DWFs preserve the axial Ward-Takahashi identity, even at a finite lattice 
spacing~\cite{Furman:1995ky}. Although not sufficiently large $L_s$ loses
the virtues of DWFs, the explicit chiral symmetry breaking with moderate sizes 
of $L_s$ can be attributed to a single universal ``residual mass" parameter
$m_{\rm res}$, acting as an additive quark mass in the axial Ward-Takahashi identity as 
$\del_{\alpha}A_{\alpha}^a\approx 2(m_f+m_{\rm res})P^a$~\cite{{Blum:2000kn},{Aoki:2002vt}}.
A very small value of $m_{\rm res}$, which is typically smaller than
10\% of the quark mass, is always achieved at a given $L_s$ around 10-20 
with the help of improved gauge actions~\cite{Aoki:2002vt}. 
This fact greatly simplifies the nonperturbative determination of the renormalization of quark bilinear currents~\cite{Blum:2001sr}. 
For a calculation of the axial-vector coupling $g_A$, 
the chiral symmetry is very useful because
the renormalization factors of local vector and local axial-vector current operators are equal, $Z_V=Z_A$~\cite{Blum:2001sr}.
This means that the ratio of the nucleon axial-vector and vector couplings,
$g_A/g_V$, calculated on the lattice is not renormalized~\cite{Sasaki:2003jh}. 
Therefore, in DWF simulations, the ambiguity in the renormalization of
quark currents, which is present 
in other fermions such as Wilson-type fermions, is eliminated.
In Ref.~\cite{Sasaki:2003jh}, $g_A=1.212(27)$ in the chiral limit
is obtained from quenched DWF simulations. It underestimates 
the experimental value of 1.2695(29)~\cite{Yao:2006px} by less than 5\%.
It has also shown that there is a significant finite volume effect between 
the axial-vector couplings calculated on lattices with $(1.2\;{\rm fm})^3$ and 
$(2.4\;{\rm fm})^3$ volumes. This observation strongly indicates that 
the axial-vector coupling is particularly sensitive to finite volume effects.
Subsequently, the LHPC Collaboration has evaluated the axial-vector coupling
using domain wall valence fermions with improved staggered sea quark configurations
with physical volume as large as $(3.5\;{\rm fm})^3$ and obtained $g_A=1.226(84)$ 
at the physical pion mass~\cite{Edwards:2005ym}. Its value again agrees with 
experiment within 5\%.

In this paper, we naturally extend the quenched DWF calculation 
for exploring the axial structure of the nucleon, namely the axial-vector 
form factor and the induced pseudo-scalar form factor at low $q^2$ as well 
as the electro-magnetic structure of the nucleon. Especially, to evaluate the induced 
pseudo-scalar coupling $g_P$ is one of our main targets, 
since no intensive study has been done to determine 
this particular quantity in lattice QCD.
Recall that the induced pseudo-scalar form factor is 
assumed to be dominated by a pion pole, which give rises to very rapid 
$q^2$-dependence at low $q^2$. The larger physical volume, where the 
lower momentum transfers are accessible, is required. We therefore
utilize $(3.6\;{\rm fm})^3$ volume where the smallest momentum squared 
($q^2\approx 0.1\;{\rm GeV}^2$) is smaller than measured pion mass squared
($m_{\pi}^2>0.15\;{\rm GeV}^2$).
We also re-examine the finite volume effect on the axial-vector coupling 
using three different volumes, which include $(3.6\;{\rm fm})^3$ together 
with smaller ones $(1.8\;{\rm fm})^3$ and $(2.4\;{\rm fm})^3$.
Furthermore, we calculate the nucleon matrix element of the pseudo-scalar density 
$\langle p|\bar{u}\gamma_5 d|n\rangle$
to check the axial Ward-Takahashi identity in terms of the nucleon matrix elements, which 
may be called as the generalized Goldberger-Treiman relation~\cite{Weisberger:1966ip}. 

Our paper is organized as follows.
In Section II, we first present a brief introduction
of the weak nucleon form factors and
the status of experimental studies. 
In Section III, details of our Monte Carlo simulations and
some basic results are given.
We also describes the lattice method for calculating the nucleon
form factors.
Section IV presents our results of the four weak form factors 
as well as the pseudo-scalar form factor on lattice with $(3.6\;{\rm fm})^3$ volume. Especially, the $q^2$-dependences of all measured form factors at low $q^2$ 
are discussed with great interest. At the end of this section, we discuss the consequence of 
the axial Ward-Takahashi identity among the axial-vector form factor, the induced pseudo-scalar
form factor and the pseudo-scalar form factor.
In Section V, we discuss the finite volume effects on the form factors using results from three different volumes. Meanwhile, we also check whether approximated forms of $q^2$-dependence
of form factors, which are observed at low $q^2$, are still valid even in the relatively
high $q^2$ region, up to at least $q^2\approx 1.0\;{\rm GeV}^2$, apart from consideration
of the finite volume effects. 
In Section VI, we compare our results with previous works.
Finally, in Section VII, we summarize the present work and 
discuss future directions.

\section{Weak nucleon form factors and experimental status}
\label{sec:general}
In general, the nucleon matrix elements of 
the weak current are given by a linear combination of the vector and axial-vector
matrix elements. Here, let us introduce 
the vector and axial-vector currents, which are expressed in terms of the isospin doublet of 
quark fields $\psi=(u, d)^T$
\bea
V_{\alpha}^a(x)&=&\bar{\psi}(x)\gamma_{\alpha}t^a\psi(x), \\
A_{\alpha}^a(x)&=&\bar{\psi}(x)\gamma_{\alpha}\gamma_5t^a\psi(x),
\eea
where $t^a$ are the $SU(2)$ flavor matrices normalized to 
obey ${\rm Tr}(t^a t^b)=\delta_{a b}$.
Then, the nucleon matrix elements are given by
%
%
\bea
\langle N(P') | J^{\rm wk}_{\alpha}(x) | N(P) \rangle 
&=&
\langle N(P') | V^{a}_{\alpha}(x)+A^{a}_{\alpha}(x) | N(P) \rangle \\
&=&{\overline u}_{_N}(P') 
\left(
{\cal O}^{V}_{\alpha}(q)+{\cal O}^{A}_{\alpha}(q) 
\right)t^a
u_{_N}(P) e^{iq\cdot x},
\label{Eq:Mats}
\eea
where $q\equiv P-P'$ is the momentum transfer
between the initial ($P$) state and
the final state ($P^{\prime}$) and $N$ represents the
nucleon isospin doublet as $N=(p, n)^T$. 
Four form factors are needed to describe these matrix elements:
the weak vector and induced tensor (weak magnetism) form factors for 
the vector current,
%
%
\be
{\cal O}^{V}_{\alpha}(q) 
= \gamma_{\alpha} F_V(q^2) + \sigma_{\alpha \beta}q_{\beta} F_T(q^2)
\label{Eq:VcO}
\ee
and the weak axial-vector and induced pseudo-scalar form factors 
for the axial-vector current~\footnote{In this paper,
we restrict ourselves to considering the iso-spin symmetric case as $m_u=m_d$, where
the second-class form factors do not appear in consequence of 
$G$-parity invariance~\cite{Weinberg:1958ut}.}
%
%
\be
{\cal O}^{A}_{\alpha}(q)
= \gamma_{\alpha}\gamma_5 F_A(q^2) 
+iq_{\alpha} \gamma_5 F_P(q^2),
\label{Eq:AvO}
\ee
which are here given in the Euclidean metric convention~\footnote{
The sign of all form factors is chosen to be positive.
Remark that our $\gamma_5$ definition, 
$\gamma_5\equiv \gamma_x \gamma_y\gamma_z \gamma_t=-\gamma_5^M$,
has the opposite sign relative to that in the Minkowski convention 
($\vec{\gamma}^M=i\vec{\gamma}$ and $\gamma^M_0=\gamma_t$)
adopted in the particle data group~\cite{Yao:2006px}.}.
Thus, $q^2$ denoted in this paper, which stands for Euclidean four-momentum squared, 
corresponds to the time-like momentum squared as $q_M^2=-q^2<0$ in Minkowski space.

The weak matrix elements are related to the electro-magnetic matrix elements
if the strange contribution is ignored under the exact iso-spin symmetry. 
A simple exercise in $SU(2)$ Lie algebra leads
to the following relation between the vector part of the weak matrix elements 
of neutron beta decay
and the difference of proton and neutron electro-magnetic 
matrix elements~\cite{{Thomas:2001kw},{Sasaki:2003jh}}:
\be
\langle p| \bar{u}\gamma_{\alpha}d |n \rangle 
= \langle p | \bar{u}\gamma_{\alpha}u -\bar{d}\gamma_{\alpha}d | p \rangle 
= \langle p| j_{\alpha}^{\rm em}|p\rangle - \langle n| j_{\alpha}^{\rm em}|n\rangle,
\ee
where $j_{\alpha}^{\rm em}=\frac{2}{3}\bar{u}\gamma_{\alpha}u
-\frac{1}{3}\bar{d}\gamma_{\alpha}d$.
This relation gives a connection between the weak vector and induced tensor
form factors and the {\it iso-vector} part of electro-magnetic nucleon 
form factors
\bea
F^{v}_1(q^2)&=& F_V(q^2), \\
F^{v}_2(q^2)&=& 2 \mnuc F_T(q^2),
\eea
where $F^v_1$ ($F^v_2$) denotes the iso-vector combination of
the Dirac (Pauli) form factors of the proton and neutron, which are
defined by
\be
\langle N(P') | j^{\rm em}_{\alpha}(x) | N(P) \rangle 
={\overline u}_{_N}(P') 
\left(
\gamma_{\alpha} F_1^{N}(q^2) + \sigma_{\alpha \beta}\frac{q_{\beta}}{2\mnuc} F_2^{N}(q^2)\right ) u_{_N}(P),
\label{Eq:VcEM}
\ee
where $\mnuc$ denotes the nucleon mass, which is defined as the average of neutron
and proton masses, and $N$ represents $p$ (proton) or $n$ (neutron).
Experimental data from elastic electron-nucleon scattering is usually presented in 
terms of the electric $G_E(q^2)$ and magnetic $G_M(q^2)$ Sachs form factors 
which are related to the Dirac and Pauli form factors~\cite{{Thomas:2001kw},{Hyde-Wright:2004gh}}
\bea
G^N_E(q^2)&=& F^N_1(q^2)-\frac{q^2}{4\mnuc^2}F^N_2(q^2),
\label{Eq:EleFF}
\\
G^N_M(q^2)&=&F^N_1(q^2)+ F^N_2(q^2).
\label{Eq:MagFF}
\eea
Their normalization at $q^2=0$ are given by the proton (neutron) charge
and magnetic moment~\cite{Yao:2006px}:
\be
\begin{array}{lll}
\mbox{Proton:} & G_E^{p}(0)=1, & G_M^{p}(0)=\mu_p=   +2.792847351 (28),\\
\mbox{Neutron:} & G_E^{n}(0)=0, & G_M^{n}(0)=\mu_n= -1.91304273 (45).\\
\end{array}
\ee
Therefore, one finds $F_V(0)=G_E^p(0)-G_E^n(0)=1$ and
$2\mnuc F_T(0)=G_M^p(0)-G_M^n(0)-1=3.70589$.
As for the $q^2$-dependence of the form factors, it is experimentally known
that the standard dipole parametrization $G_D(q^2)=\Lambda^2/(\Lambda^2+q^2)$
with $\Lambda=0.84\;{\rm GeV}$ (or $\Lambda^2=0.71\;{\rm GeV}^2$)
describes well the magnetic form factors of 
both the proton and neutron and also the electric form factor of the proton,
at least, in the low $q^2$ region~\cite{Hyde-Wright:2004gh}.
Here, the current interesting issues of the $q^2$-dependence
of the electro-magnetic form factors at higher $q^2$ are beyond the scope of this paper.
Recent reviews on the experimental situation can be found in Ref.~\cite{Hyde-Wright:2004gh}.
The slopes of the form factors at $q^2=0$ determine
mean-squared radii, which can be related to dipole masses 
as $\langle r_i^2\rangle =12/M_i^2$ ($i=E$ or $M$) in the dipole form $G_i(q^2)=G_i(0)/(1+q^2/M_i^2)$.
The experimental values of the electric root mean-squared 
(rms) radius for the proton and the magnetic rms radii of 
the proton and neutron are compiled in Table~\ref{Table:ExpValues}.
These rms radii are all equal within errors and are in agreement with the 
empirical dipole parameter $\Lambda$. On the other hand, the slope
of the neutron electric form factor $G_E^n(q^2)$ is determined with high 
precision from double-polarization measurements of neutron knock-out
from a polarized $^2 {\rm H}$ or $^3 {\rm He}$ target, 
while only a small deviation from zero is observed for $G_E^n(q^2)$ 
at low $q^2$~\cite{Hyde-Wright:2004gh}.
Combined with all four of 
the electric charge and magnetization radii of the proton and neutron,
we finally evaluate the rms radii
for the weak vector form factor and induced tensor form factor as
$\sqrt{\langle (r_V)^2\rangle}= 0.797(4)$ fm and  
$\sqrt{\langle (r_T)^2\rangle}= 0.879(18)$ fm, which
correspond to the dipole masses, $M_V = 0.857 (8)$ GeV and
$M_T =0.778 (23)$ GeV. See Appendix A for details.

The axial-vector form factor at zero momentum transfer, namely the 
axial-vector coupling $g_A=F_A(0)$, is precisely determined 
by measurements of the beta asymmetry in neutron decay. The value of 
$g_A=1.2695(29)$ is quoted in the 2006 PDG~\cite{Yao:2006px}.
Nevertheless, kinematics of neutron beta decay are quite limited
due to a very small mass difference of the proton and neutron. 
Other experimental methods are utilized for determination of the 
$q^2$-dependence of $F_A(q^2)$. For this purpose, there are basically 
two types of experiment, namely quasi-elastic neutrino scattering 
and charged pion electroproduction experiments. The former suffers
from severe experimental uncertainties concerning the 
incident neutrino flux and the background subtraction of elastic events, 
while model-dependent analysis is somewhat inevitable for the latter~\cite{Bernard:2001rs}.
Both methods reported that the dipole form $F_A(q^2)=F_A(0)/(1+q^2/M_A^2)$
is a good description for low and moderate momentum transfer, 
$q^2< 1\;{\rm GeV}^2$.
The resulting world average of the dipole mass parameter $M_A$ is 
quoted as $M_A=1.026(21)$ GeV from neutrino scattering or 
$M_A=1.069(16)$ GeV from pion electroproduction 
in Ref.~\cite{Bernard:2001rs}. As for a small discrepancy between 
two averages, it has been argued that within heavy-baryon chiral perturbation 
theory (HBChPT) the finite pion mass correction of $-0.055$ GeV to the latter value 
may resolve this discrepancy~\cite{Bernard:2001rs}. Therefore, one can translate
the axial dipole mass into the axial rms radius of $\sqrt{\langle (r_A)^2\rangle}=
0.67(1)\;{\rm fm}$, which is consistently obtained from quasi-elastic neutrino 
scattering experiments and charged pion electroproduction experiments~\cite{Bernard:2001rs}.

On the other hand, the induced pseudo-scalar form factor $F_P(q^2)$ is 
less well-known experimentally~\cite{Gorringe:2002xx}. 
The main source of information on $F_P(q^2)$ stems from ordinary muon 
capture (OMC) on the proton, $\mu^- + p \rightarrow \nu_{\mu}+n$. 
One measures the induced pseudo-scalar 
coupling $g_P=m_{\mu}F_P(q_0^2)$ at the specific momentum transfer for 
the muon capture by the proton at rest as $q_0^2 =0.88 m_{\mu}^2$. 
The induced pseudo-scalar coupling
$g_P$ is also measured in radiative muon
capture (RMC), $\mu^- + p\rightarrow \gamma+\nu_{\mu}+n$.
Before 2006, the Saclay OMC experiment, which was the 
most recent OMC experiment at that time, reported $(g_{P}^{\rm OMC})_{\rm Saclay, original}=8.7 \pm 1.9$~\cite{Bardin:1981cq}. 
Combining with the older OMC experiments including bubble chamber measurements, 
the world average for OMC is obtained as $(g_{P}^{\rm OMC})_{\rm old\;Ave}=8.79 \pm 1.92$, which is given in Refs.~\cite{Bernard:2001rs} and \cite{Bardin:1981cq}. Surprisingly, this value is close to the theoretically 
predicted value by HBChPT, $g_{P}^{\rm ChPT}=8.26 \pm 0.16$~\cite{Bernard:2001rs}.
However, the novel RMC experiment at TRIUMF ~\cite{{Jonkmans:1996my}, {Wright:1998gi}} is
puzzling: their measured value of $g_P^{\rm RMC}= 12.4 \pm1.0$ is quite higher than the theoretical value as is the OMC value as $g_P^{\rm RMC} \approx 1.4 g_P^{\rm OMC}$. This disagreement is reduced by reanalysis with the updated $\mu^+$ 
lifetime~\cite{Gorringe:2002xx}. Then, the updated result of the Saclay OMC experiment 
yields $(g_{P}^{\rm OMC})_{\rm Saclay, updated}=10.6  \pm 2.7$.
Accordingly, the weighted world average for OMC, $(g_{P}^{\rm OMC})_{\rm updated\;Ave.}=10.5 \pm 1.8$ given in Ref.~\cite{Gorringe:2002xx}, is shifted away from the 
theoretical expected value, while the updated average value is in agreement with 
the RMC result within its error. Indeed, there is a caveat that 
the ortho-para transition rate in $\mu$-molecular Hydrogen,
to which either OMC and RMC results are very sensitive, is poorly 
known due to mutually inconsistent results among 
two experiments~\cite{{Bardin:1981cq},{Clark:2005as}} 
and theory~\cite{Bakalov:1980fm}. 
Comprehensive reviews of a history of $g_P$ have 
been given in Refs.\cite{Bernard:2001rs} and \cite{Gorringe:2002xx}.

Recently, a new OMC experiment has been done by the MuCap 
Collaboration~\cite{Andreev:2007wg}. The MuCap result is nearly 
independent of $\mu$-molecular effects in contrast
with the previous OMC experiments and the RMC experiment.
After the electro-weak radiative corrections, which were underestimated in the old literature, 
are correctly taken into account~\cite{Czarnecki:2007th}, the new precise OMC measurement
yields 
\be
g_P^{\rm MuCap}=7.3\pm 1.1.
\ee
Including the new MuCap result and taking into account the electro-weak radiative corrections, 
the new world average of the OMC results becomes 
$(g_{P}^{\rm OMC})_{\rm new Ave}=8.7 \pm 1.0$~\cite{Czarnecki:2007th}. 
As for other $q^2$ data of $F_P(q^2)$, only a few data points are measured in the low $q^2$ region by a single experiment of pion electroproduction at threshold~\cite{Choi:1993vt}. These data are summarized in Table~\ref{Table:ExpPsF}.
Three data points from pion electroproduction at threshold are well fitted 
by the pion-pole dominance form, 
$F_P(q^2)=2\mnuc F_A(q^2)/(q^2+m_\pi^2)$~\cite{Nambu:1960xd}, 
which is also consistent with the value determined by the new OMC result 
at $q^2=0.88 m_{\mu}^2$. Therefore, the pion-pole dominance is confirmed, more or less, through pion electroproduction~\cite{Choi:1993vt}.

\section{Simulation details}
\label{sec:simulation}
We work in the quenched approximation and use domain wall
fermions (DWFs) to compute the nucleon matrix elements of the weak current.
We generate ensembles of the quenched QCD 
configuration with the renormalization group improved, 
DBW2 (doubly blocked Wilson in two-dimensional parameter space) 
gauge action~\cite{{Takaishi:1996xj},{de Forcrand:1999bi}}
at $\beta=6/g^2=0.87$ 
($a^{-1}\approx 1.3$ GeV), where the residual chiral symmetry 
breaking of domain wall fermions is significantly improved with a moderate size
of the fifth-dimension $L_s$ such as $L_s=16$~\cite{Aoki:2002vt}.
Indeed, the residual quark mass for $L_s=16$ is measured as small as 
$m_{\rm res}\sim 5 \times 10^{-4}$ in lattice units~\cite{Aoki:2002vt}, 
which is safely negligible compared with the input quark masses 
in our simulations, $0.02 \le m_f \le 0.08$.
We work with relatively coarse lattice spacing, $a\approx 0.15$ fm~\footnote{
One might worry about the large scaling violation, which is observed for 
the axial-vector coupling $g_A$ in a quenched calculation with overlap fermions 
at the same lattice spacing~\cite{Galletly:2005db}.
However, the previous quenched DWF studies reported that 
there is no appreciate scaling violation in the kaon B-parameter $B_K$~\cite{Aoki:2005ga}
and proton decay matrix elements~\cite{Aoki:2006ib} at $\beta=0.87$ ($a\approx 0.15$ fm) and 1.04 ($a\approx 0.10$ fm). We may deduce that no large scaling violation is ensured for other matrix elements as well in our DWF calculations.}, 
which is determined from the $\rho$ meson mass~\cite{Aoki:2002vt}.

To study finite volume effects, numerical simulations
are performed on three different lattice sizes $L^3\times T$ = $24^3 \times 32$,
$16^3 \times 32$ and $12^3 \times 32$. The spatial extents in our study correspond
to $La\simeq 3.6$, 2.4 and 1.8 fm.
Quark propagators are generated for four bare masses $m_f=0.02$, 0.04, 0.06
and 0.08 for $L=24$ and three bare masses $m_f=0.04$, 0.06 and 0.08 for 
$L=16$ and 12, using DWFs with $L_s=16$ and $M_5=1.8$. 
Details of our simulations are summarized in Table~\ref{tab:simulation_param}.
In Table~\ref{tab:old_simulation_results}, some basic physics results are compiled from Ref.~\cite{Aoki:2002vt}.

The pseudo-scalar meson (pion) masses computed in these calculations are summarized in 
Table~\ref{tab:mass_spect}. 
All fitted values are obtained from the covariant single cosh fit.
It is clear that there is no visible finite-volume effect on the pion mass. Measured values for $L=16$ are in good agreement with the values found in Ref.~\cite{Sasaki:2003jh}, 
where point-to-box quark propagators are used with the mostly same gauge ensembles, 
while the point-to-gauss-smeared quark propagators are utilized in the present study.
Our simulated values of the pion mass range from 0.39 GeV to 0.76 GeV.

\subsection{Nucleon spectra and Dispersion relation}

In order to compute nucleon masses or matrix elements, 
we define the nucleon (proton) operator as
\be
{\chi}^{S}(t, {\bf p}) = \sum_{\bf x} e^{-i{\bf p}\cdot {\bf x}}\varepsilon_{abc}
[u_a^T({\bf y}_1, t)C\gamma_5 d_b({\bf y}_2,t)]u_c({\bf y}_3,t) \times \phi({\bf y}_1 -{\bf x}) \phi({\bf y}_2 -{\bf x}) \phi({\bf y}_3 -{\bf x}),
\ee
where $abc$ and $ud$ have usual meanings as color and flavor indices.
$C$ is the charge conjugation matrix defined as $C=\gamma_t \gamma_y$
and the superscript $T$ denotes transpose. The superscript $S$ of the nucleon 
operator $\chi$ specifies the smearing for the quark propagators.
In this study, we use two types of source: local source as $\phi({\bf y}_i-{\bf x})=\delta({\bf y}_i-{\bf x})$ and Gaussian smeared source.
Here we take ${\bf y}_1={\bf y}_2={\bf y}_3={\bf 0}$ in our calculation. 
As for the Gaussian smeared source, 
we apply the gauge-invariant Gaussian smearing~\cite{{Gusken:1989qx},{Alexandrou:1992ti}}
with $N = 30$, $\omega = 4.35$. Details of our choice of smearing 
parameters are described in Ref.~\cite{Berruto:2005hg}.

We construct two types of the two-point function for the proton. One interpolating
operator at the source location is constructed from Gaussian smeared quark fields, 
while the other interpolating operator at the sink location is either constructed from 
local quark fields (denoted LG) or Gaussian smeared ones (denoted GG):
\be
C_{SG}(t-t_{\rm src}, q)=\frac{1}{4}{\rm Tr}\left\{
{\cal P_+}
\langle 
{\chi}^{S}(t, {\bf q})
 \overline{{\chi}^{G}}(t_{\rm src},-{\bf q})
\right\},
\ee
with $S=L$ or $G$. The projection operator ${\cal P}_+=\frac{1+\gamma_t}{2}$
can eliminate contributions from the opposite-parity state 
for $q^2=0$~\cite{Sasaki:2001nf, Sasaki:2005ug}. 
It is rather expensive to make the Gaussian smeared interpolating operator 
projected onto a specific finite momentum at the source location ($t_{\rm src}$). 
However, it is sufficient to project only the sink operator onto the desired
momentum by virtue of momentum conservation. Thus, the quark fields at the
source location are not projected onto any specific momentum in this
calculation.
For the momentum at the sink location ($t_{\rm sink}$), 
we take all possible permutations of the three-momentum ${\bf q}$ including both 
positive and negative directions.

Nucleon masses and energies are computed by using the LG correlators
with the five lowest momenta: $(0,0,0)$, $(1,0,0)$,
$(1,1,0)$, $(1,1,1)$ and $(2,0,0)$ in units of $2\pi/L$.
All fitted values, which are obtained from the conventional single exponential fit, for
each volume are summarized in Table~\ref{tab:mass_spect}. 
Next, we examine the dispersion relation of the nucleon state
in our simulations. The purpose of this examination is two fold:
1) Our analysis should be restricted to the lower momenta that do
not suffer from large ${\cal O}(a^2)$ errors. 
2) The evaluation of the squared four-momentum transfer $q^2$
requires precise knowledge of the nucleon energies with finite momentum.
The later point can be achieved by an estimation of the energy $E({\bf p})$ 
with the help of the dispersion relation and the measured nucleon rest mass $\mnuc$ that
can be most precisely measured. 

\begin{figure}[htbp]
\vspace{2mm}
\begin{center}
\includegraphics[width=0.52\linewidth,clip]{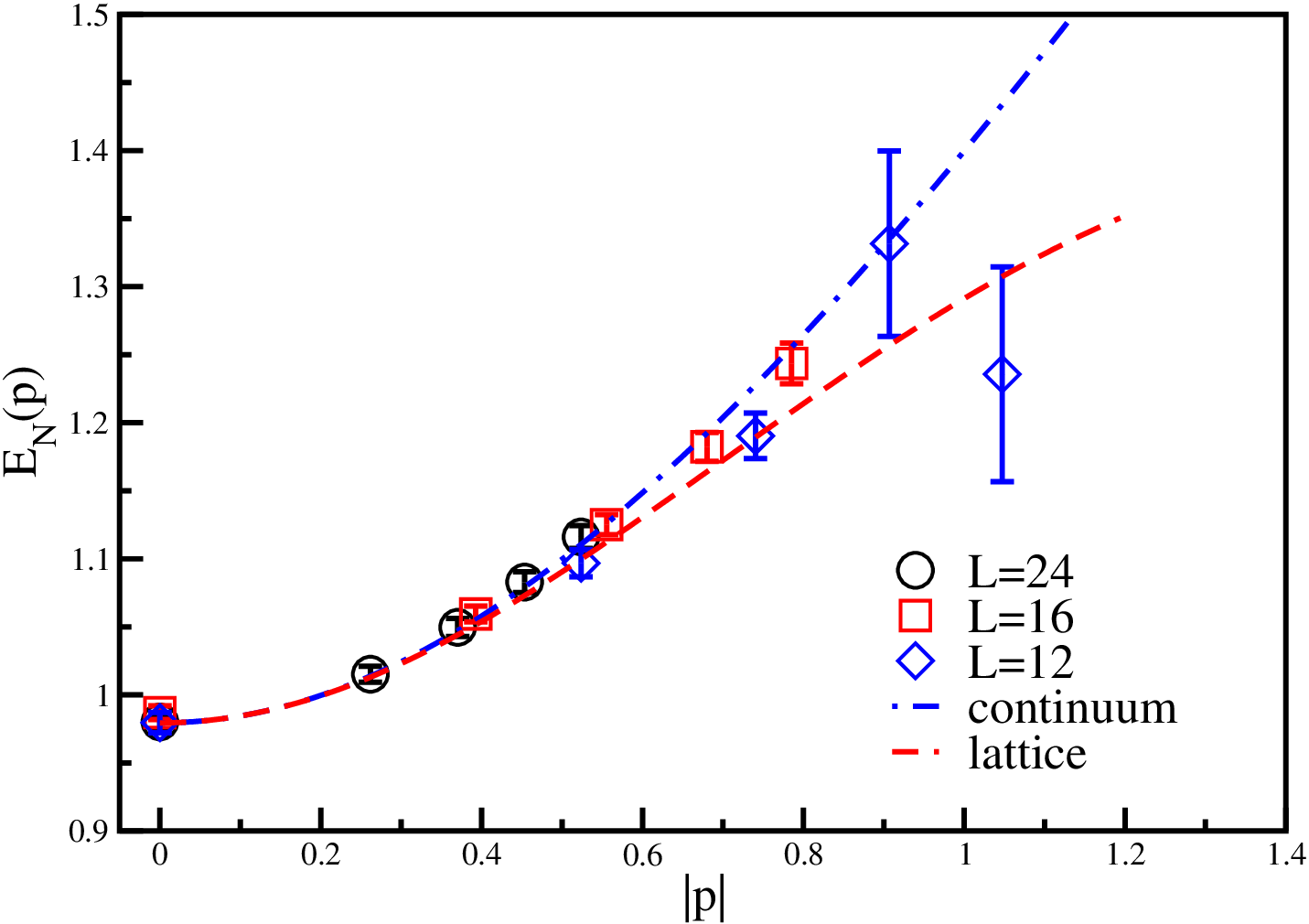}
\end{center}
\caption{Comparison of measured and estimated energies of the nucleon
for $m_f=0.04$ as a function of absolute value of three-momentum $|{\bf p}|$. 
Open circles, squares and diamonds, which corresponds 
to the measured values in lattice units, are obtained from $L=24$, $L=16$ and $L=12$. 
The estimated energies are given by the relativistic dispersion 
formula $E({\bf p})=\sqrt{{\bf p}^2 + M_N^2}$ for continuum-like momenta $p_i=\frac{2\pi}{L}n_i$ (dashed-dotted curve) and lattice momenta $p_i=\sin[\frac{2\pi}{L}n_i]$ (dashed curve) with the rest mass $M_N$ measured 
at $L=24$.}
\label{fig:Disp_m04}
\end{figure}

For $m_f=0.04$ the measured values of the nucleon energy, which are 
obtained from $L=24$ (open circles), $L=16$ (open squares) and $L=12$ 
(open diamonds), are compared with the relativistic dispersion relation
\be
E({\bf p})=\sqrt{{\bf p}^2 + \mnuc^2}
\label{Eq:RelDisp}
\ee
in Fig.~\ref{fig:Disp_m04} with either the naive discrete (continuum-like) 
momentum $p_i=\frac{2\pi }{L}n_i$ or the lattice discrete momentum $p_i=\sin[\frac{2\pi}{L}n_i]$ ($n_i=0, 1, 2, \cdot\cdot\cdot, (L-1)$)
for ${\bf p}=(p_x, p_y, p_z)$. 
We observe that the measured energies $E({\bf p})$ are consistent
with the estimated values from the relativistic dispersion formula for 
continuum-like momenta (dashed-dotted curve) and lattice momenta (dashed curve)
in the range of our admitted momentum except for the largest momentum 
on the lattice with $L=12$ as shown in Fig.~\ref{fig:Disp_m04} .
The difference between either choice of the discrete momentum is mostly
comparable to the statistical errors, while differences increase at the larger
momentum. To restrict ourselves to low $q^2$ region ($q^2 < 1$ ${\rm GeV}^2$), we do not use 
the two largest momenta on the lattice with $L=12$ for the proceeding analysis.
Therefore, it is not a concern to choose what type of the discrete momentum 
in the dispersion relation in our current calculation.
We simply choose the continuum-like momentum 
throughout this paper and then evaluate the values of the squared four-momentum 
transfer $q^2$ with the measured 
rest mass $\mnuc$ and the continuum dispersion relation~(\ref{Eq:RelDisp}).

\subsection{Three point correlation functions}

As discussed in the previous section, under the exact iso-spin symmetry ($m_u=m_d$), 
the $SU(2)$ current algebra leads to the following relations~\cite{{Thomas:2001kw},{Sasaki:2003jh}}
\bea
\langle p| V_{\alpha}^+ |n\rangle &=& 2\langle p| V_{\alpha}^{3} | p\rangle, \\
\langle p| A_{\alpha}^+ |n\rangle &=& 2\langle p| A_{\alpha}^{3} | p\rangle,
\eea
where $V_{\alpha}^3=\frac{1}{2}({\bar u}\gamma_{\alpha} u- {\bar d}\gamma_{\alpha}d)$ and $A_{\alpha}^3=\frac{1}{2}({\bar u}\gamma_{\alpha}\gamma_5 u- {\bar d}\gamma_{\alpha}\gamma_5 d)$.
Thus, we may calculate the weak transition matrix elements by
the iso-vector proton matrix elements. 

First of all, we define the finite-momentum three-point functions for the relevant 
components of either the local vector current (${\cal J}^V_{\alpha}(x)={\bar u}(x)\gamma_{\alpha}u(x)-{\bar d}(x)\gamma_{\alpha}d(x)$) or 
the local axial-vector current (${\cal J}^A_{\alpha}(x)={\bar u}(x)\gamma_{\alpha}\gamma_5 u(x)-{\bar d}(x)\gamma_{\alpha}\gamma_5 d(x)$) with the proton
interpolating operator $\chi$:
%
%
\be
\langle {\chi}(t', {\bf p'}){\cal J}^{\Gamma}_{\alpha}(t, {\bf q}) \overline{\chi}(0, -{\bf p})\rangle
={\cal G}^{\Gamma}_{\alpha}(p,p')\times f(t, t', E({\bf p}), E({\bf p}')) + \cdot\cdot\cdot,
\label{Eq:three-pt}
\ee
where the initial and final proton states carry fixed momenta $\bf p$ and $\bf p'$ respectively
and then the current operator has a three-dimensional momentum transfer ${\bf q}={\bf p}-{\bf p}'$. Here, Dirac indices have been suppressed. The ellipsis denotes excited state contributions which can be ignored in the case of $t'-t\gg 1$ and $t\gg 1$. We separate the correlation 
function into two parts: ${\cal G}_{\alpha}^{\Gamma}(p,p')$ which is defined as
%
%
\be
{\cal G}^{\Gamma}_{\alpha}(p,p')=(-i\gamma\cdot p' + \mnuc) {\cal O}^{\Gamma}_{\alpha}(q)
(-i\gamma\cdot p + \mnuc),
\label{Eq:spinor-part-of-three-pt}
\ee
where ${\cal O}^{\Gamma}_{\alpha}(q)$ corresponds to either Eq.(\ref{Eq:VcO})
or Eq.(\ref{Eq:AvO}), and the factor $f(t,t',E({\bf p}), E({\bf p}'))$
which collects all the kinematical factors, normalization of states, and time 
dependence of the correlation function. The trace of ${\cal G}^{\Gamma}_{\alpha}(p, p')$ 
with some appropriate projection operator ${\cal P}$ for specific combinations
of $\Gamma$ and $\alpha$ yields some linear 
combination of form factors in each $\Gamma$ channel. On the other hand, 
all time dependences of the factor $f(t,t',E({\bf p}), E({\bf p}'))$ 
can be eliminated by the appropriate ratio of three- 
and two-point functions~\cite{Gockeler:2003ay,Hagler:2003jd}
\be
{\cal R}(t)=\frac{C^{{\cal P}}_{\Gamma, \alpha}(t, {\bf p}', {\bf p})}
{C^{GG}(t_{\rm sink}-t_{\rm src}, {\bf p}')}
\left[
\frac{C^{LG}(t_{\rm sink}-t, {\bf p})C^{GG}(t-t_{\rm src}, {\bf p}')C^{LG}(t_{\rm sink}-t_{\rm src}, {\bf p}')}
{C^{LG}(t_{\rm sink}-t, {\bf p}')C^{GG}(t-t_{\rm src}, {\bf p})C^{LG}(t_{\rm sink}-t_{\rm src}, {\bf p})} 
\right]^{\frac{1}{2}},
\label{Eq:LHPCratio}
\ee
where 
\be
C_{\Gamma, \alpha}^{{\cal P}}(t, q)=\frac{1}{4}{\rm Tr}
\left\{
{\cal P}
\langle {\chi}^{G}(t_{\rm sink}, {\bf p}')
{\cal J}^{\Gamma}_{\alpha}(t, {\bf q})
 \overline{{\chi}^{G}}(t_{\rm src},-{\bf p})
\rangle
\right\},
\ee
which are calculated by the sequential source method described 
in Ref.~\cite{Sasaki:2003jh}:

In this study, we consider only the case at the rest frame of the final 
state (${\bf p}'={\bf 0}$), which leads to ${\bf q}={\bf p}$. Therefore, the squared 
four-momentum transfer 
is given by $q^2=2\mnuc(E({\bf q})-\mnuc)$. 
Nucleon energy $E({\bf q})$ is simply abbreviated as $E$, hereafter. In this kinematics,
${\cal G}^{\Gamma}_{\alpha}(p,p')$ is represented by a simple notation as ${\cal G}^{\Gamma}_{\alpha}(q)$. 
Then, the ratio~(\ref{Eq:LHPCratio}) gives the asymptotic form as a function 
of the current-operator insertion time $t$,
\be
{\cal R}(t)\rightarrow \frac{1}{4}{\rm Tr}\{
{\cal P}{\cal G}_{\alpha}^{\Gamma}(q) \}
\times\frac{1}{\sqrt{2\mnuc^2E(E+\mnuc)}}
\label{Eq:RatioAsym}
\ee
in the limit when
the Euclidean time separation between all operators is large, $t_{\rm sink}\gg
t\gg t_{\rm src}$ with fixed $t_{\rm src}$ and $t_{\rm sink}$.

We choose particular combinations of the projection operator ${\cal P}$ and 
the current operator ${\cal J}^{\Gamma}_{\alpha}$ ($\Gamma=V$ or $A$).
We consider two types of the projection operator, 
${\cal P}^t={\cal P}_{+}\gamma_{t}$
and ${\cal P}_5^{z}={\cal P}_{+}\gamma_5\gamma_{z}$ in this study.
The latter projection operator implies that the $z$-direction is chosen as the 
polarized direction.
We then obtain some linear combination of desired form factors from
the projected correlation functions,
%
%
\bea
\frac{1}{4}{\rm Tr}\left\{ {\cal P}^{t}{\cal G}^{V}_{t}(q)\right\} &=& \mnuc(E + \mnuc) \left[
F_V(q^2)-(E - \mnuc) F_T(q^2)
\right],
\label{Eq:3pt_vec_time}
\\
\frac{1}{4}{\rm Tr} \left\{{\cal P}_5^{z}{\cal G}^{V}_i(q) \right\}&=& - i\varepsilon_{i j z}q_j \mnuc
\left[
F_V(q^2)+2\mnuc F_T(q^2)
\right],
\label{Eq:3pt_vec_trans}
\eea
for the vector currents ${\cal J}^{V}_{t}$ and ${\cal J}^{V}_{i}$ $(i=x,y,z)$.
Similarly, we get 
%
%
\bea
\frac{1}{4}{\rm Tr}\left\{ {\cal P}_5^{z}{\cal G}^{A}_{i}(q)\right\} &=& \mnuc(E + \mnuc) \left[
F_A(q^2)\delta_{iz} 
- \frac{q_iq_z}{E+\mnuc} F_P(q^2)
\right],
\label{Eq:3pt_axial}
\eea
for the axial-vector current ${\cal J}^{A}_{i}$ $(i=x, y, z)$. 
In this calculation, we use at most the four non-zero three-momentum
transfer ${\bf q}=\frac{2\pi}{L}{\bf n}$ (${\bf n}^2=1$, 2, 3, 4).
All possible permutations of the lattice momentum including both positive and 
negative directions are properly taken into account. 
All three-point correlation functions are calculated with a source-sink separation of 10
in lattice units, which is the same in the previous DWF calculation of the axial-vector coupling $g_A$~\cite{Sasaki:2003jh}. For $L=24$, we calculate three-point correlation 
functions with three different sequential sources generated with source-sink locations,
$[t_{\rm src}, t_{\rm sink}]=[0,10]$, $[10, 20]$, and  $[20,30]$
on a given gauge configuration~\footnote{
We treat our data sets as 70 independent measurements after taking average of 
the multiple source results on each configuration.}, 
while only a single sequential source with $[t_{\rm src}, t_{\rm sink}]=[0,10]$ is utilized for $L=12$ and $L=16$ calculations.

\begin{figure}[htbp]
\vspace{2mm}
\begin{center}
\includegraphics[width=0.52\linewidth,clip]{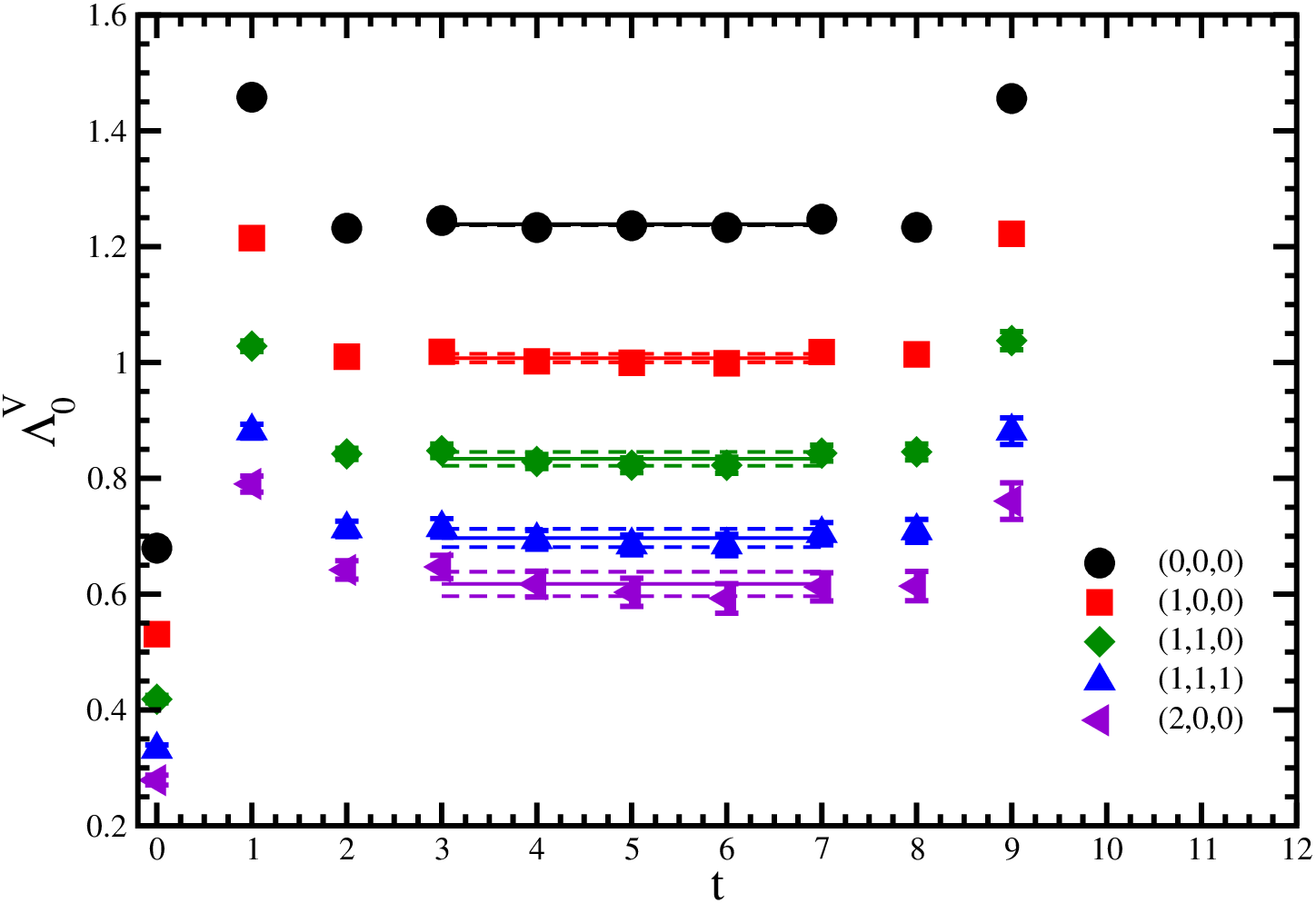}
\includegraphics[width=0.52\linewidth,clip]{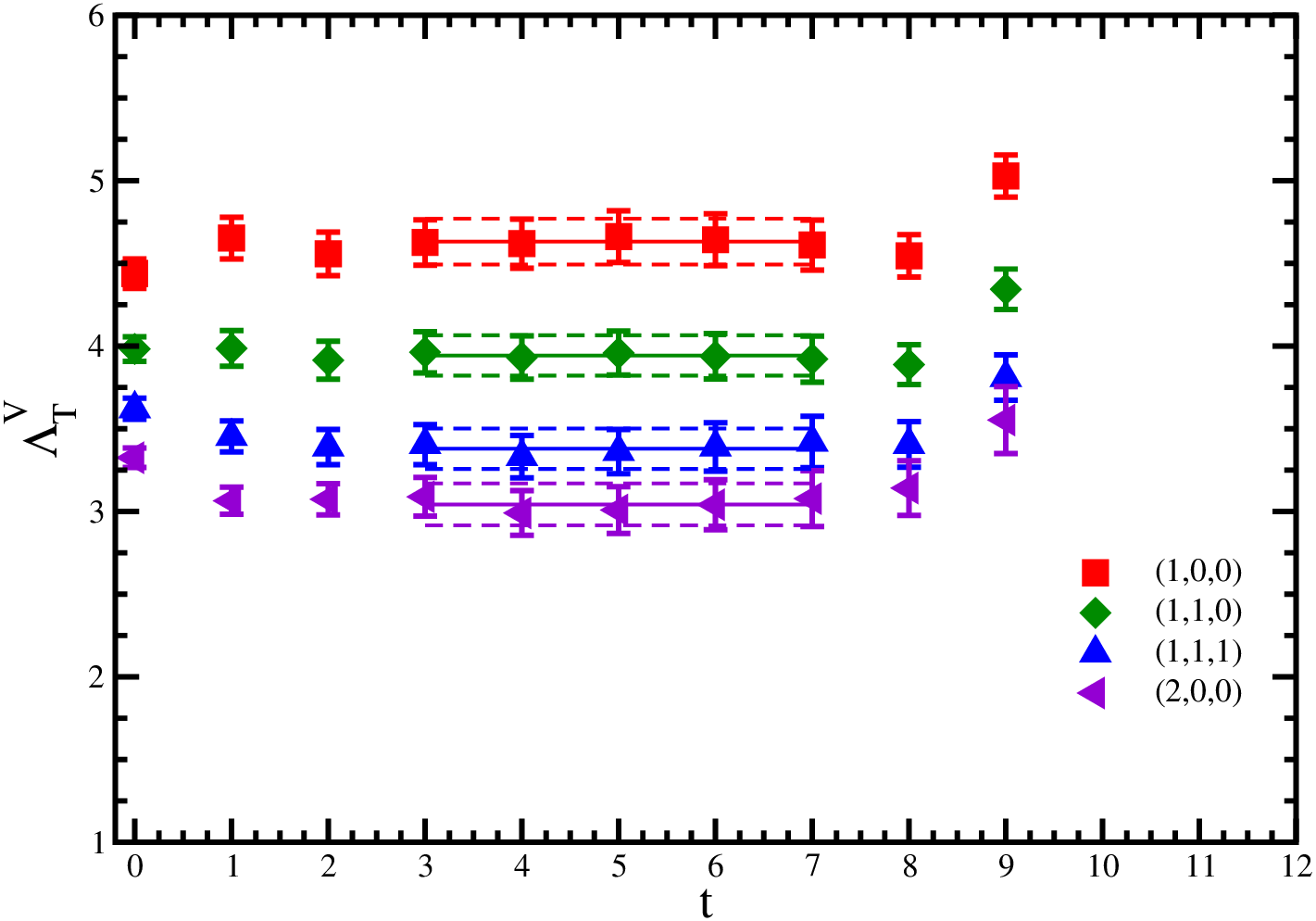}
\end{center}
\caption{
Relevant ratios of three- and two-point functions, ${\Lambda}^{V}_{0}$ (top) and ${\Lambda}^{V}_{T}$ (bottom), for all possible three-momentum
transfer ${\bf q}$
as a function of the current insertion time slice at $m_f=0.04$.
}
\label{fig:3pt_V}
\end{figure}

In Fig.~\ref{fig:3pt_V}, we plot the dimensionless projected correlators
%
%
\bea
{\Lambda}^{V}_{0}&=&\frac{\frac{1}{4}{\rm Tr}
\{{\cal P}^t {\cal G}^{V}_{t}(q)\}}
{\mnuc(E+\mnuc)}, \\
{\Lambda}^{V}_{T}&=&-\frac{1}{2}\left(
\frac{\frac{1}{4}{\rm Tr} 
\{{\cal P}^z_{5} {\cal G}^{V}_{x}(q)\}}{iq_y\mnuc} 
-\frac{\frac{1}{4}{\rm Tr}
\{ {\cal P}^z_{5} {\cal G}^{V}_{y}(q)\}}{iq_x\mnuc} 
\right),
\eea
as a function of the current insertion time slice $t$ for $m_f=0.04$ on the largest volume ($L=24$) as typical examples. 
Good plateaus for all squared three-momentum transfer 
are observed in the middle region between 
the source and sink points. The quoted errors are estimated by a single
elimination jack-knife method.
The lines plotted in each figure represent the average value (solid lines) 
and their one standard deviations (dashed lines) in the time-slice range $3\le t \le 7$. 

\begin{figure}[htbp]
\vspace{2mm}
\begin{center}
\includegraphics[width=0.52\linewidth,clip]{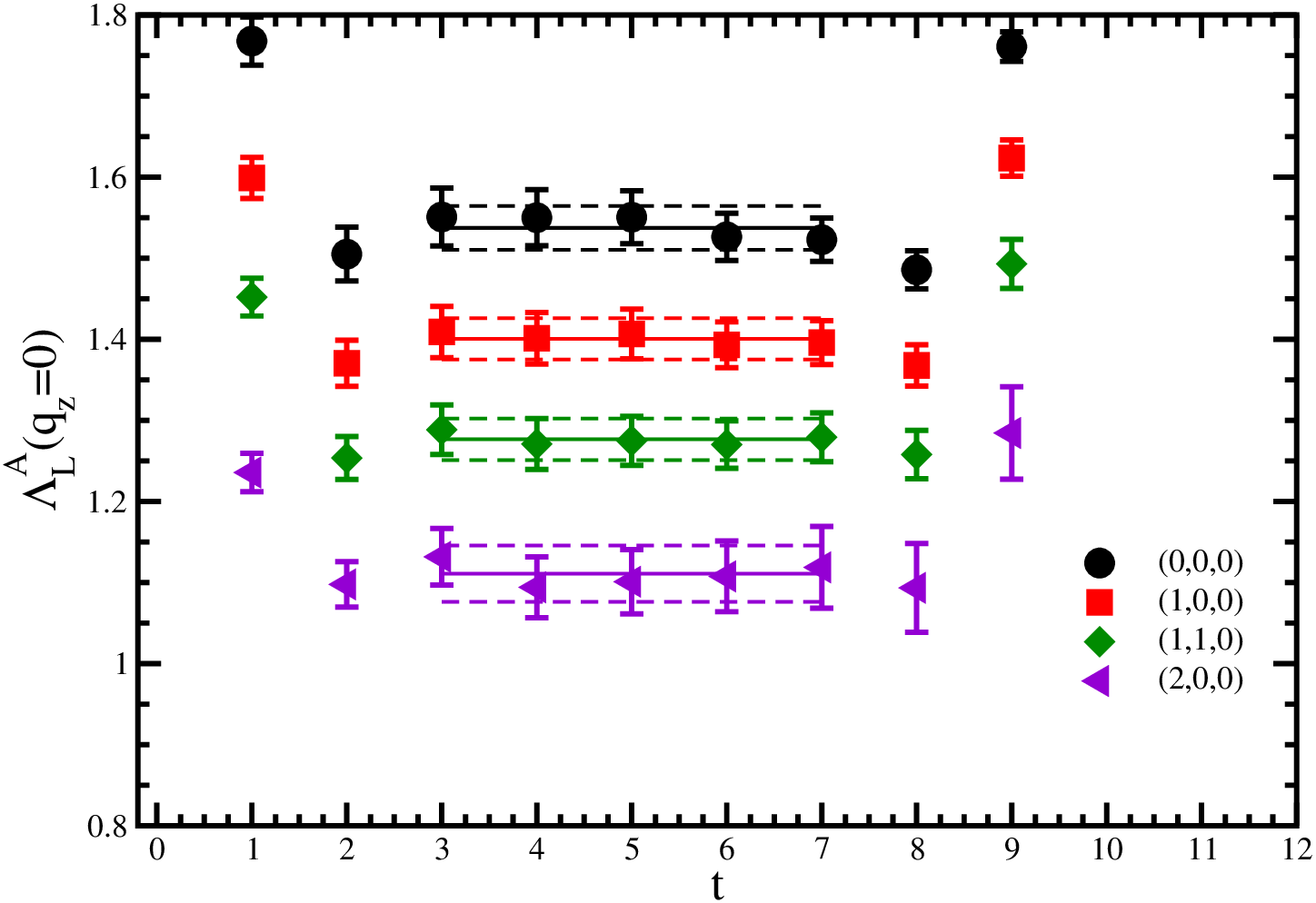}
\includegraphics[width=0.52\linewidth,clip]{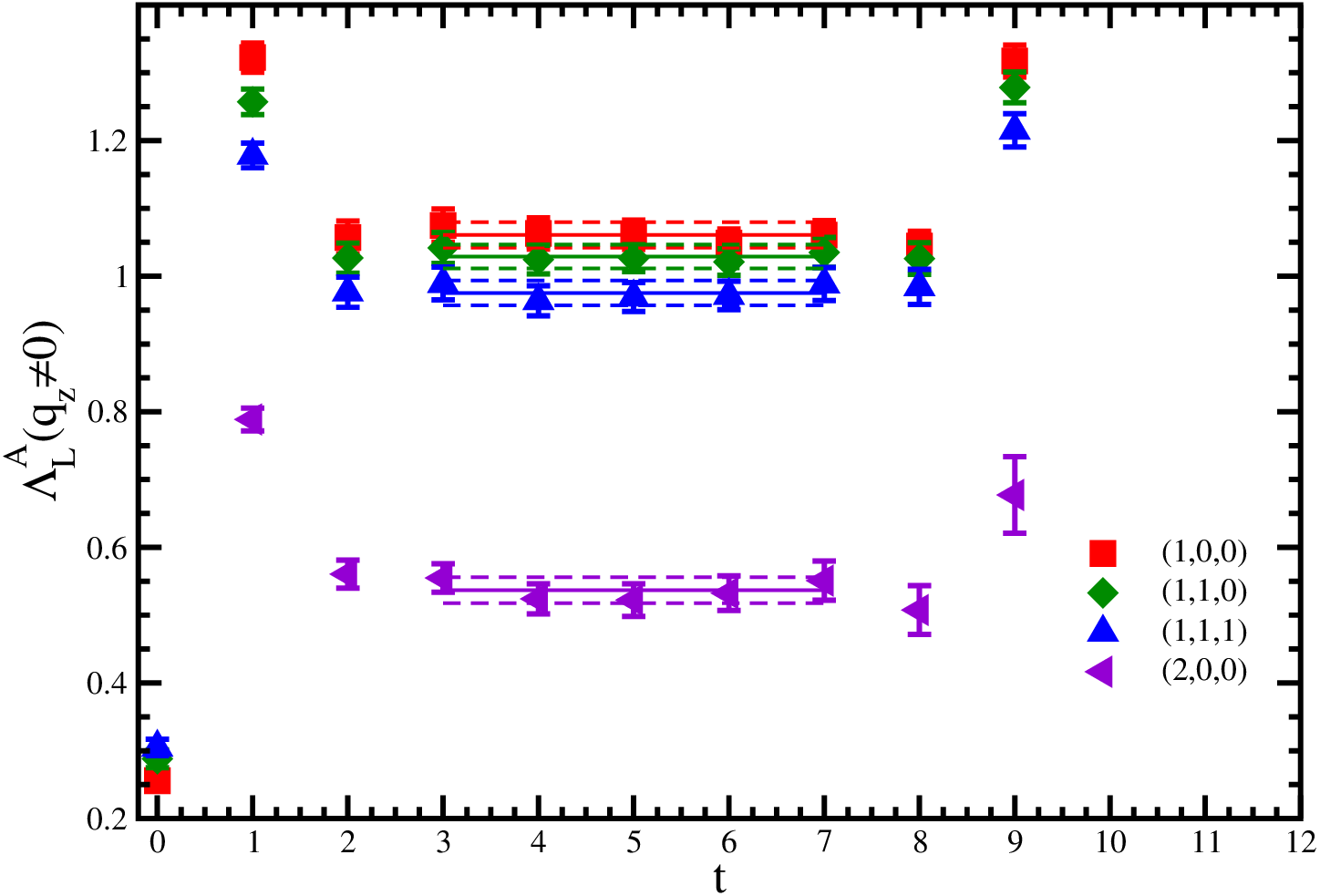}
\includegraphics[width=0.52\linewidth,clip]{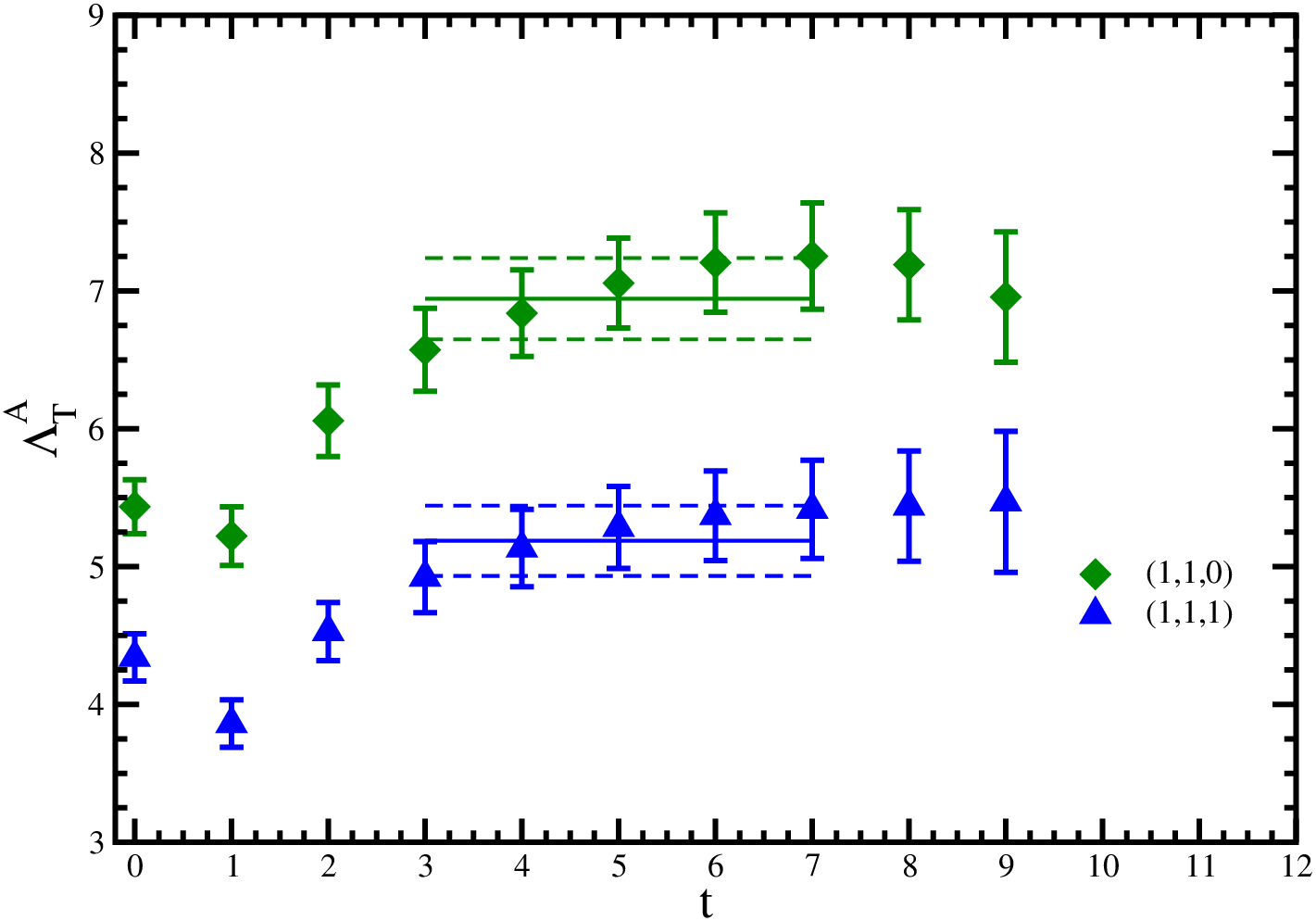}
\end{center}
\caption{
Relevant ratios of three- and two-point functions, $\Lambda_{L}^{A}(q_z=0)$ (top),
$\Lambda_{L}^{A}(q_z\neq0)$ (middle) and $\Lambda_{T}^{A}$ (bottom),
for all possible three-momentum transfer ${\bf q}$ as a function of the current
insertion time slice at $m_f=0.04$.
}
\label{fig:3pt_A}
\end{figure}

Similarly, Fig.~\ref{fig:3pt_A} shows $\Lambda^{A}_{L}$ and $\Lambda^{A}_{T}$
for the axial vector current, which are defined by
%
%
\bea
\Lambda^{A}_{L}&=&\frac{\frac{1}{4}{\rm Tr} 
\{{\cal P}^z_{5} {\cal G}^{A}_{z}(q)\}}
{\mnuc(E+\mnuc)}, \\
\Lambda^{A}_{T}&=&-\frac{1}{2}\left(
\frac{\frac{1}{4}{\rm Tr} 
\{{\cal P}^{z}_{5} {\cal G}^{A}_{x}(q)\}}{q_zq_x} 
+\frac{\frac{1}{4}{\rm Tr} 
\{{\cal P}^{z}_{5} {\cal G}^{A}_{y}(q)\}}{q_zq_y} \right).
\eea
It is worth noting that  in the axial-vector channel 
the $z$-direction is chosen as the polarized direction in this study. 
Therefore, the longitudinal momentum ($q_z$) dependence explicitly appears 
in Eq.~(\ref{Eq:3pt_axial}). This fact provides two kinematical constraints on 
determination of the three-point functions in our calculation.
First, there are two types of kinematics, $q_z\neq 0$ and $q_z=0$ 
in the longitudinal component ($i=z$) of Eq.~(\ref{Eq:3pt_axial}), except 
for the case of ${\bf n}^2=3$ where $q_z$ is always non-zero. 
Secondly, the transverse components ($i=x$ or $y$)
of Eq.~(\ref{Eq:3pt_axial}) are prevented from vanishing by the kinematics 
only if ${\bf n}^2=2$ and 3, where two components of the momentum 
including the polarized  direction ($z$-direction) are non-zero. 
These are the reasons why $\Lambda^A_L(q_z=0)$ and $\Lambda^A_L(q_z\neq0)$ 
are separately plotted in Fig.~(\ref{fig:3pt_A}) and results of $\Lambda^A_L(q_z=0)$ 
at ${\bf n}^2=3$ and $\Lambda^A_T$ at ${\bf n}^2=1$, 4 are missing there.

Finally, we recall that the lattice local operators ${\cal J}_{\alpha}^{\Gamma}(x)$ 
($\Gamma=V$ or $A$), which are represented as the quark bilinear currents, receive finite renormalizations relative to their continuum 
counterparts. Thus the renormalized form factors require some independent 
estimation of $Z_\Gamma$, the renormalization of the quark bilinear currents,
$[{\cal J}_{\alpha}^{\Gamma}]^{\rm ren}=
Z_{\Gamma} [{\cal J}_{\alpha}^{\Gamma}]^{\rm lattice}$. As mentioned previously,
good chiral properties of DWFs  ensure that the lattice renormalizations of the local 
currents are equal, $Z_V=Z_A$, up to terms of order ${\cal O}(a^2)$ in the chiral limit 
and neglecting explicit chiral symmetry breaking due to the moderate size of 
the fifth-dimensional extent $L_s$~\cite{Blum:2001sr}. In this paper, we evaluate
$Z_V$ at each quark mass from the inverse of $F_V(0)$ that should be unity in the 
continuum under
the exact $SU(2)$ iso-spin symmetry and multiply four weak form factors by this renormalization factor $Z_V$ to estimate the renormalized form factors in the chiral limit.

\subsection{Vector coupling $g_V$ and axial-vector coupling $g_A$}
\label{Sec4:coupling}

\begin{figure}[htbp]
\vspace{2mm}
\begin{center}
\includegraphics[width=0.52\linewidth,clip]{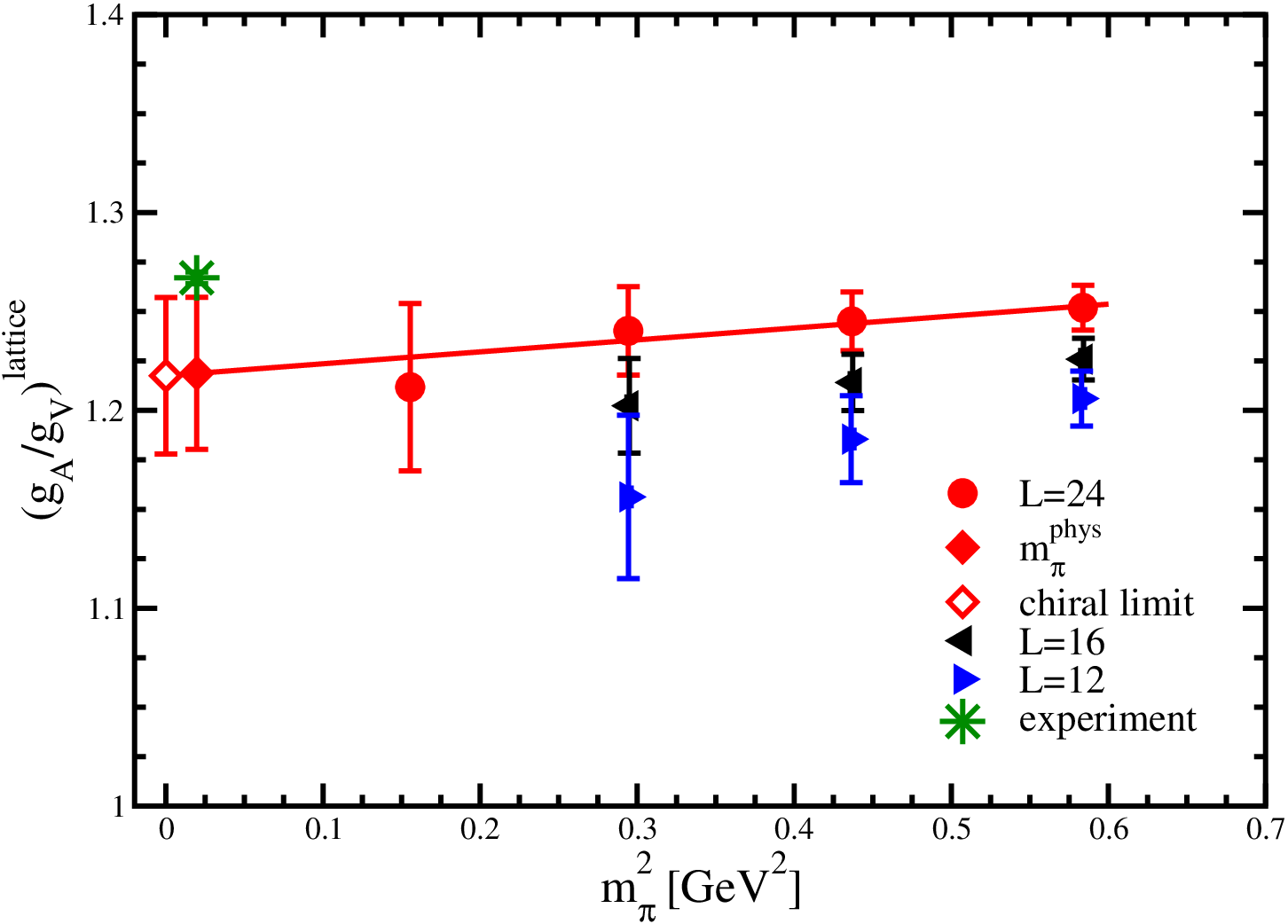}
\end{center}
\caption{The physical ratio of couplings $g_A/g_V$ as
a function of the pion mass squared.
Results on the largest volume $(3.6\;{\rm fm})^3$ (circles)
exhibit milder quark mass dependence, while the smaller
volume results (right-oriented triangles) show a slow downward tendency toward
the chiral limit away from the experimental point (asterisk).
}
\label{fig:FV_gagv}
\end{figure}

At zero three-momentum transfer $|{\bf q}|=0$, only $\Lambda^V_0$ and $\Lambda^A_L$ 
are calculable.  Then, these directly yield the values of $(g_V)^{\rm lattice}$ and 
$(g_A)^{\rm lattice}$
respectively. Our results of $(g_V)^{\rm lattice}$, $(g_A)^{\rm lattice}$
and their ratio $(g_A)^{\rm ren}=(g_A/g_V)^{\rm lattice}$ obtained in this calculation
are summarized in Table~\ref{tab:ga_gv_all}, where old results for $L=8$ and $L=16$ calculated in Ref.~\cite{Sasaki:2003jh} are also tabulated. 
In Fig.~\ref{fig:FV_gagv}, we show
the ratios of the axial to the vector coupling $(g_A/g_V)^{\rm lattice}$ calculated for
three different volumes as functions of pion mass squared. Clearly, the finite
volume effect on $(g_A/g_V)^{\rm lattice}$ can be observed. 
The larger volume results exhibit milder quark mass dependence, while the smallest
volume results show a slow downward tendency toward the chiral limit away from the 
experimental point. Therefore, for the largest volume results, we simply adopt a linear extrapolation with respect to the pion mass squared to take the chiral limit.
We obtain the axial-vector coupling $g_A^{\rm ren}=1.219(38)$ at the physical point ($m_{\pi}=0.14$ GeV).

\begin{figure}[htbp]
\vspace{2mm}
\begin{center}
\includegraphics[width=0.7\linewidth,clip]{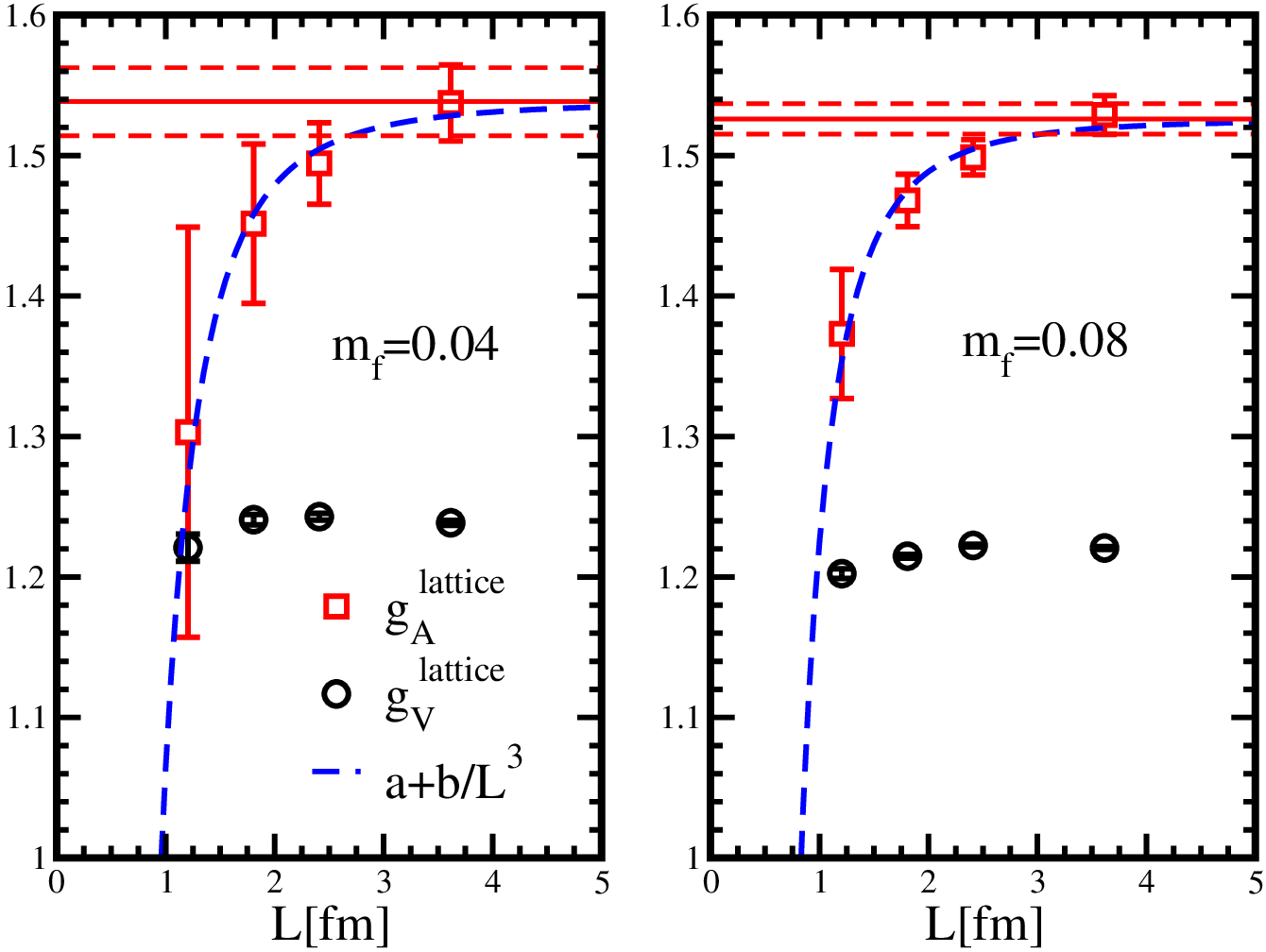}
\end{center}
\caption{
The vector coupling $(g_V)^{\rm lattice}$ and axial-vector coupling 
$(g_A)^{\rm lattice}$ as functions of spatial lattice size for $m_f=0.04$ 
(left figure) and $m_f=0.08$ (right figure). Dashed
curves are fits of the form $g_A^{\rm lattice}(L)=g_A^{\rm lattice}(\infty)+bL^{-3}$.
}
\label{fig:FV_ga_and_gv}
\end{figure}

We next examine more details of the finite volume effect on the vector and 
axial-vector couplings separately. 
Combined results from three different spatial sizes, $La\approx1.8\;{\rm fm}$, 
$2.4\;{\rm fm}$ and $3.6\;{\rm fm}$ together with old results 
from $La\approx1.2\;{\rm fm}$, we plot
$(g_V)^{\rm lattice}$ and $(g_A)^{\rm lattice}$ against the spatial lattice
size in the physical unit in Fig.~\ref{fig:FV_ga_and_gv}.
The quoted errors in the figure represent only the statistical
errors, which are obtained by a single elimination jack-knife method.
The left (right) figure is for lighter (heavier) pion mass.
The axial-vector coupling $(g_A)^{\rm lattice}$ shows the significant spatial-size 
dependence, while we do not see any serious finite volume 
effect on the vector coupling $(g_V)^{\rm lattice}$. This indicates
that the observed finite volume effect in Fig.~\ref{fig:FV_gagv} stems from
that of $(g_A)^{\rm lattice}$. Clearly, it is observed that 
$(g_A)^{\rm lattice}$ decreases monotonically with decreasing spatial size $L$. 
Therefore, we simply utilize the power-law formula to estimate the infinite volume limit of 
the axial-vector coupling as
\be
g_A^{\rm lattice}(L)=g_A^{\rm lattice}(\infty) + b L^{-n}
\ee
with the power three ($n=3$). Horizontal lines in figures represent the values
in the infinite volume limit and their one standard deviations. 
The values obtained from the largest volume, $(3.6\;{\rm fm})^3$,
are quite close to the values in the infinite volume limit. At a glance,
the spatial size over 2.5 fm is large enough to accurately calculate 
the axial-vector coupling, at least within the range of our admitted quark mass.

We finally quote 
\be
g_A^{\rm ren}=1.219\pm 0.038({\rm stat})\pm 0.024({\rm norm})\pm 0.002({\rm volume}),
\ee
where the second error is evaluated from a 2 \% error stemming
from $Z_V\neq Z_A$, which was observed previously~\cite{Sasaki:2003jh}
and the third error is estimated from a difference between the extrapolated value in
the infinite volume limit and the largest volume result at $m_f=0.04$.

\section{Results of nucleon form factors on a $(3.6\; {\rm fm})^3$ volume}
\label{sec:numeric}

In this section, we focus on the results obtained from lattice size 
$24^3 \times 32$, which corresponds to physical volume, $V\approx (3.6 {\rm fm})^3$.
The lower momentum is admitted by the larger spatial extent $L$.
Therefore, we can make the shorter extrapolation with respect to $q^2$
toward the forward limit, $q^2=0$, for nucleon form factors, $F_T(q^2)$ and $F_P(q^2)$,
of which values at $q^2=0$ cannot be accessible directly due to the kinematical 
constraint as described before.
We also discuss the finite size effect on the nucleon form factors, which may 
be sensitive to the nucleon ``wave function" or the nucleon ``size" squeezed  
due to the finite spatial extent of lattice volume. 
In the previous section, the spatial lattice-size dependence of the axial-vector coupling
shows that spatial lattice size $La\approx 3.6$ fm is large enough  
to avoid significant finite volume effect on $g_A$.

\subsection{Vector channel}
\label{Sec5:vector}

In the case if spatial momentum transfer $\bf q$ is non-zero, all three-point 
correlation functions defined in Eqs.(\ref{Eq:3pt_vec_time}) and
(\ref{Eq:3pt_vec_trans}) are calculable. 
Two independent form factors $F_V(q^2)$ and $F_T(q^2)$ are obtained by
\bea
F_V (q^2)&=&\frac{2\mnuc}{E+\mnuc}{\Lambda}^{V}_0
+\frac{E-\mnuc}{E+\mnuc}{\Lambda}^{V}_T, \\
F_T (q^2)&=&\frac{1}{E+\mnuc}\left(-{\Lambda}^{V}_0+{\Lambda}^{V}_T\right)
\eea
at finite $q^2$.

\subsubsection{Dirac form factor $F_V(q^2)$}
\label{Sec5:Fv}

\begin{figure}[htbp]
\vspace{2mm}
\begin{center}
\includegraphics[width=0.52\linewidth,clip]{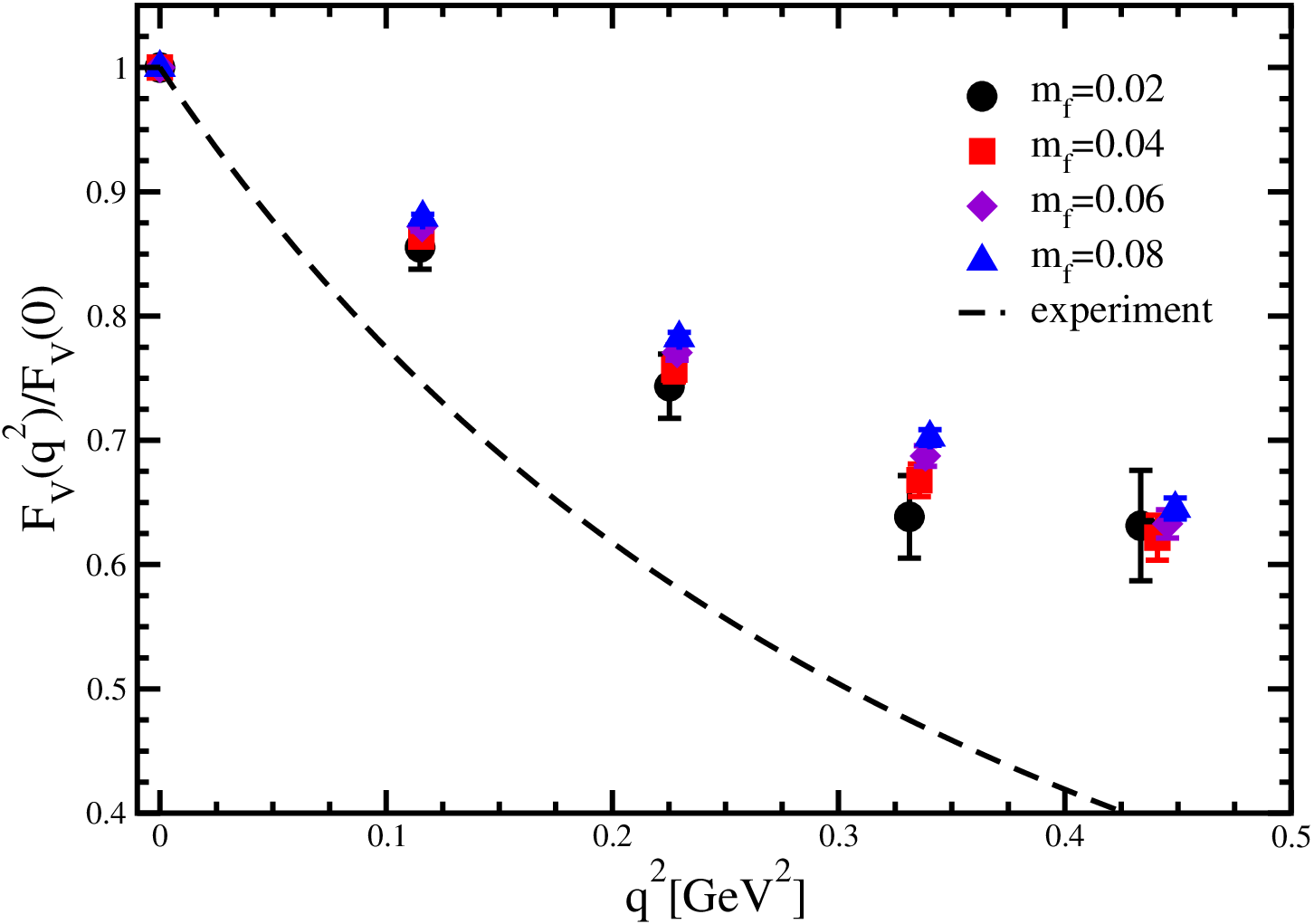}
\end{center}
\caption{The renormalized Dirac form factor, $F^{\rm ren}_V(q^2)=F_V(q^2)/F_V(0)$ as a function 
of four-momentum squared $q^2$.}
\label{fig:Fv_qsqr}
\end{figure}

First, we show quark mass dependence of the Dirac form factor $F_V(q^2)$. 
In Fig.~\ref{fig:Fv_qsqr}, we plot the normalized $F_V(q^2)$ by $F_V(0)$ as
a function of four-momentum squared $q^2$. Different symbols represent
the values obtained from different quark masses $m_f$.
There is no large $m_f$-dependence, while it seems that
the smaller quark mass makes the $q^2$-dependence steeper. 
The Dirac form factor is supposed to be the dipole form at low $q^2$:
\be
F_V(q^2)=\frac{F_V(0)}{(1+q^2/M_V^2)^2},
\label{fig:dipole_fit}
\ee
where $M_V$ denotes the dipole mass. 
A dashed curve in Fig.~\ref{fig:Fv_qsqr} corresponds to the dipole form with
the empirical value of the dipole mass $M_V=0.857(8)$ GeV, which is evaluated
from the electric charge and magnetization radii 
of the proton and neutron as described in Appendix A.

\begin{figure}[htbp]
\vspace{2mm}
\begin{center}
\includegraphics[width=0.52\linewidth,clip]{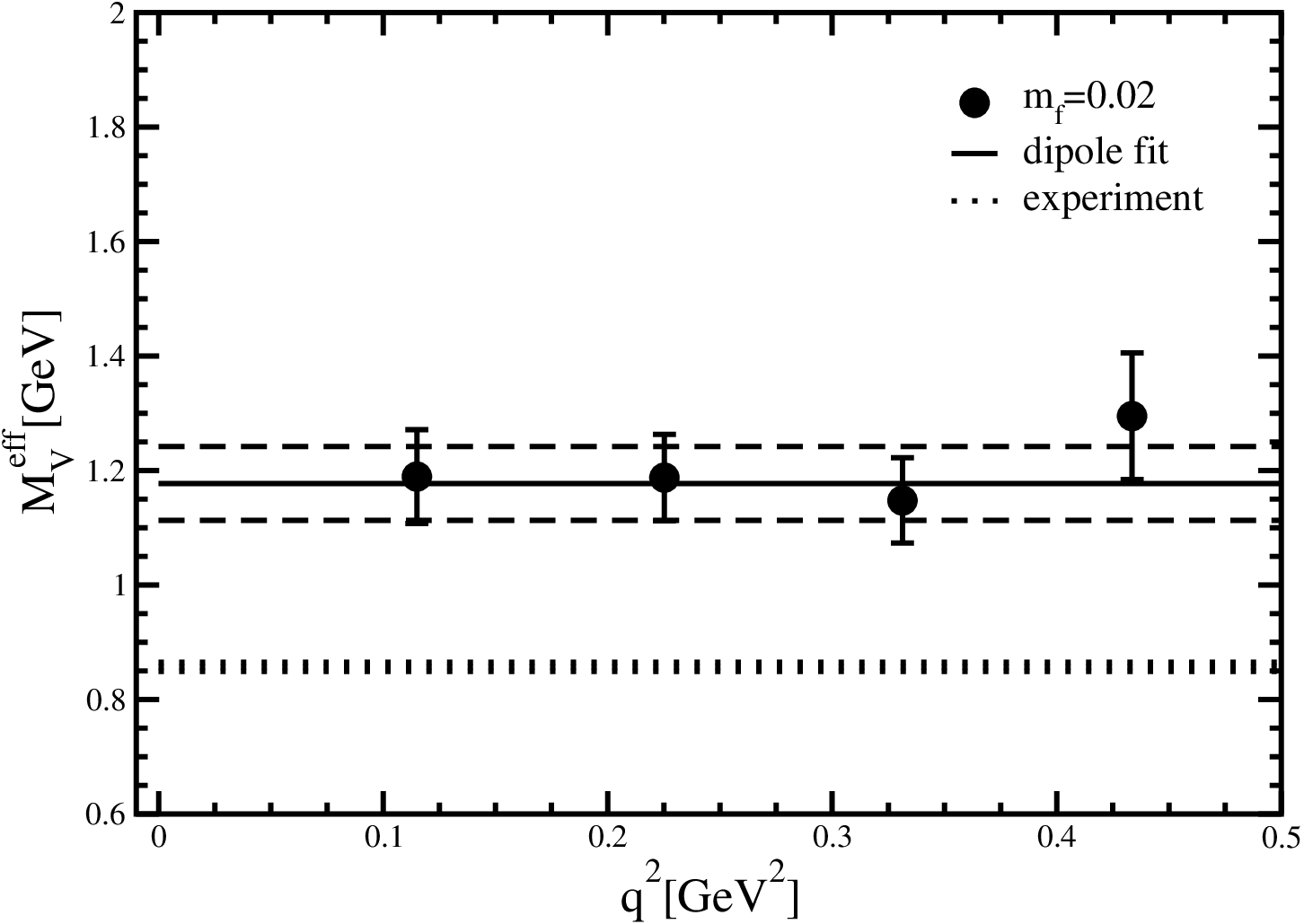}
\end{center}
\caption{Effective dipole-mass plot for the Dirac form factor $F_V(q^2)$ as a function of four-momentum squared  $q^2$ at $m_f=0.02$.}
\label{fig:Effdipole_V}
\end{figure}

In order to see how our measured $F_V(q^2)$ has an expected dipole form, 
we define the following quantity only at nonzero momentum:
%
%
\be
M_V^{\rm eff}(q^2)=\sqrt{
\frac{q^2\sqrt{F_V(q^2)}}
{\sqrt{F_V(0)}-\sqrt{F_V(q^2)}}
},
\label{Eq:EffmassPlot}
\ee
which should provide $q^2$ independent plateau if the $q^2$-dependence of 
$F_V(q^2)$ ensures the dipole form. We call this quantity 
the effective dipole mass hereafter. 
In Fig.~\ref{fig:Effdipole_V}, we show the 
effective dipole-mass plot for the Dirac form factor at $m_f=0.02$ as a typical
example. Horizontal solid and dashed lines represent the fitted dipole mass 
obtained from a correlated fit to $F_V(q^2)$ using the dipole form (\ref{fig:dipole_fit})
and its one standard deviation. The dotted lines shows the empirical dipole mass, $M_V=0.857(8)$ GeV.
Clearly, there is no appreciable $q^2$-dependence of the effective dipole mass within statistical errors. Even at the highest $q^2 \approx 0.44$ ${\rm GeV}^2$, the fitted dipole mass agrees with the value of the effective dipole mass. Therefore, we conclude that
the dipole form describes well the $q^2$-dependence of our measured 
Dirac form factor $F_V(q^2)$. 
This observation is consistent with previous 
studies~\cite{Gockeler:2003ay,Alexandrou:2006ru}.

\begin{figure}[htbp]
\vspace{2mm}
\begin{center}
\includegraphics[width=0.52\linewidth,clip]{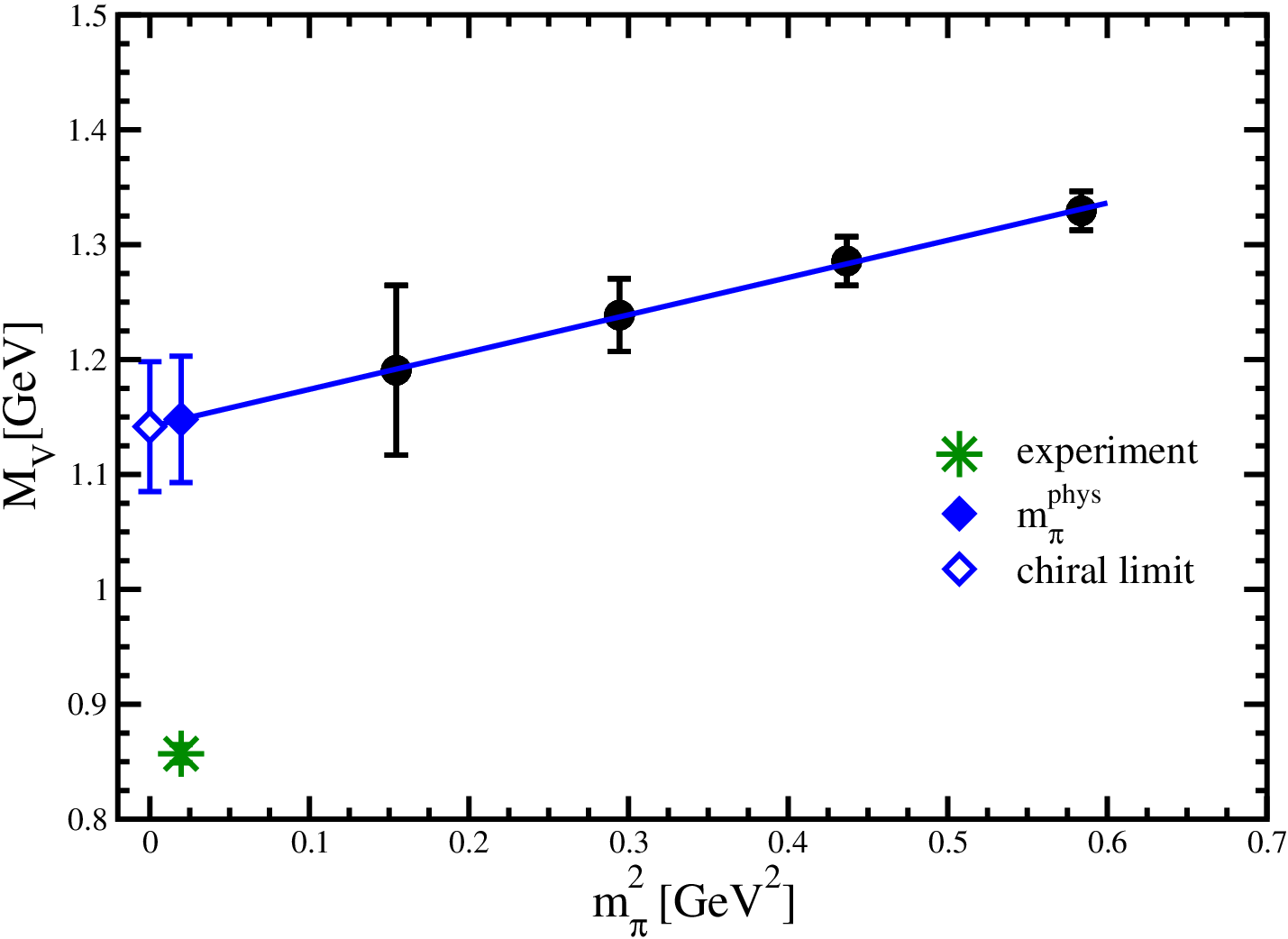}
\end{center}
\caption{Chiral extrapolation of the Dirac dipole mass $M_V$.
The extrapolated points in the chiral limit and at the physical point are represented
by an open diamond and a filled diamond. The experimental value is marked
with an asterisk. }
\label{fig:Mv_chi}
\end{figure}

Fig.~\ref{fig:Mv_chi} shows the quark-mass dependence of the fitted dipole mass.
As seen from this figure, the quark-mass dependence is rather mild and then there is
no appreciable curvature as a function of the pion mass squared.
Therefore, we simply adopt a linear extrapolation with respect to the pion mass
squared to evaluate the value of the dipole mass of 
$F_V(q^2)$ in the chiral limit. Diamond symbols in Fig.~\ref{fig:Mv_chi} 
are extrapolated values for the chiral limit ($m_{\pi}=0$) and the physical point 
($m_{\pi}=0.14$ GeV) and the solid line represents 
the fitted line. Our measured dipole masses of the Dirac form factor are much
larger than the experimental value. Here we recall that the root mean-squared (rms)
radii can be determined with the corresponding dipole mass as $\sqrt{\langle r_V^2\rangle}=\sqrt{12}/{M_V}$. The larger dipole mass $M_V$ means
that the spatial size of the nucleon in our simulations is smaller than the physical one.
This may be attributed to the missing large ``pion-cloud" contribution
since it is well known that the mean-squared radius $\langle r_V^2\rangle$ 
receives a large pion loop correction, which leads to a logarithmic divergence 
in the chiral limit in heavy baryon chiral perturbation theory~\cite{Beg:1973sc}.
Indeed, 
the present calculation is still far from the chiral regime:
our smallest pion mass is around 0.39 GeV which is comparable to the lightest 
pion mass in the most recent lattice study of nucleon electro-magnetic form factors~\cite{Alexandrou:2006ru}.
Although the expected chiral behavior is not guaranteed 
in the quenched approximation, in the present study the estimation
of systematic errors stemming from quenching and a long chiral extrapolation is
beyond the scope of this paper. 
Rather we would like to see how large volume can be fitted for
studying the structure of the nucleon, namely, the nucleon form factors
without significant finite volume effect.
Studies of the finite size effect on the nucleon form factors 
using results obtained from three different lattice sizes will be 
presented in the next section.

\subsubsection{Pauli form factor $F_T(q^2)$}
\label{Sec5:Ft}

\begin{figure}[htbp]
\vspace{2mm}
\begin{center}
\includegraphics[width=0.52\linewidth,clip]{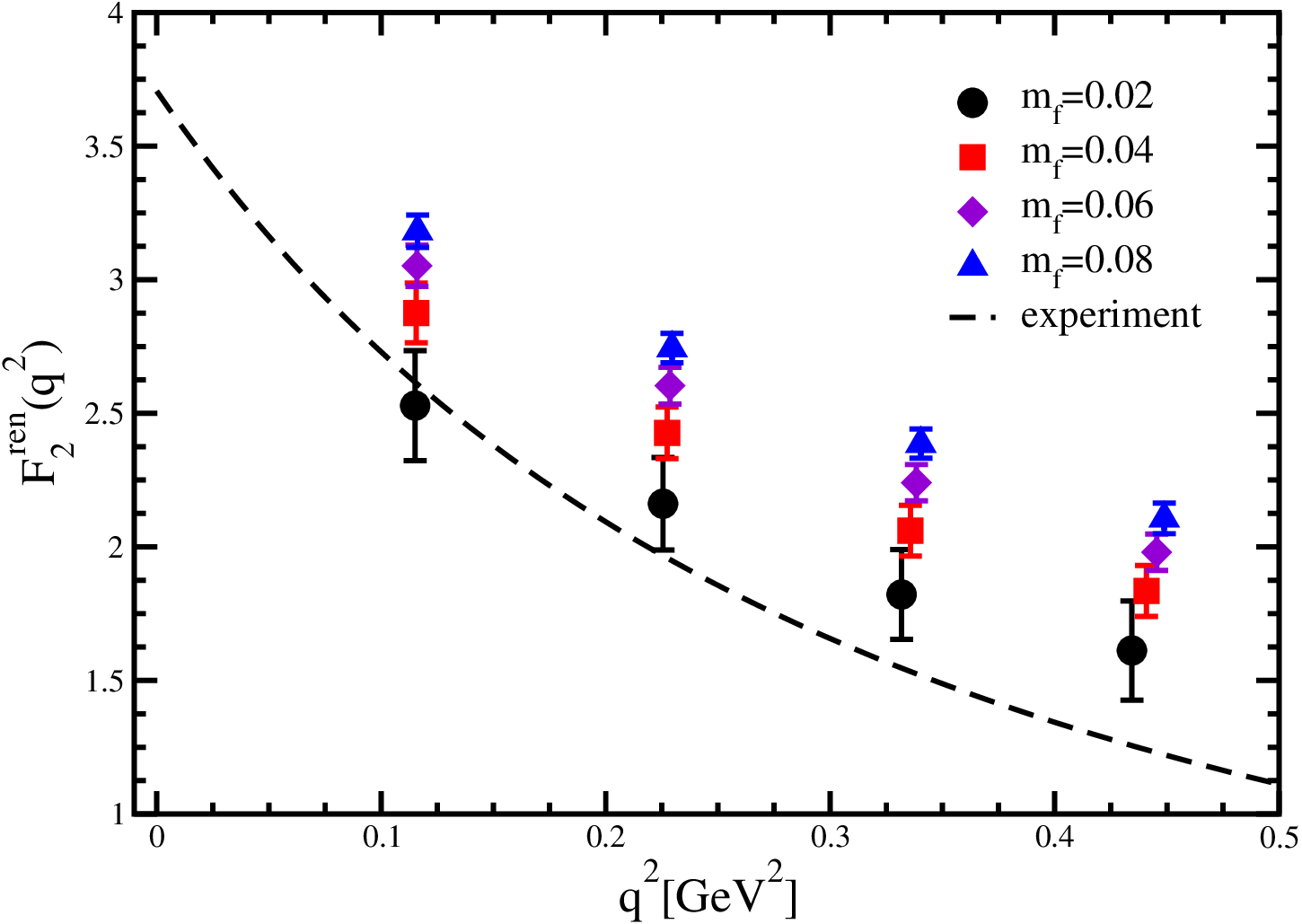}
\end{center}
\caption{
The renormalized Pauli form factor, $F^{\rm ren}_2(q^2)=2MF_T(q^2)/F_V(0)$ as a function 
of four-momentum squared $q^2$.}
\label{fig:Ft_qsqr}
\end{figure}

\begin{figure}[htbp]
\vspace{2mm}
\begin{center}
\includegraphics[width=0.52\linewidth,clip]{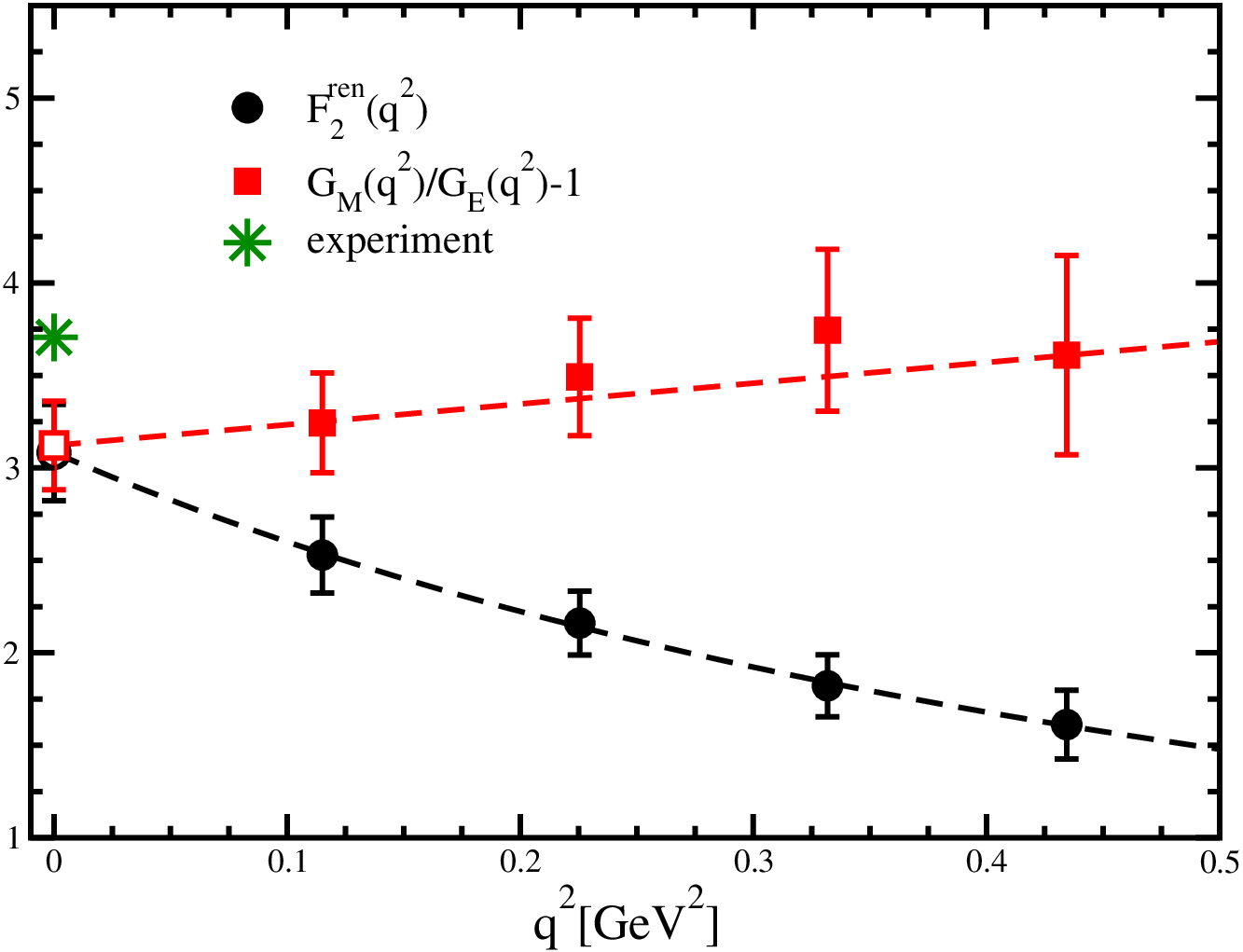}
\end{center}
\caption{
The $q^2$ extrapolation toward $q^2=0$ for either $F_2^{\rm ren}(q^2)$ or 
$G_M(q^2)/G_E(q^2)-1$ for $m_f=0.02$. Those quantities should intersect each other at
$q^2=0$. For $F_2^{\rm ren}(q^2)$, the dipole form is applied, while 
a simple linear extrapolation with respect to $q^2$ is used for $G_M(q^2)/G_E(q^2)-1$ 
thanks to its mild $q^2$-dependence. The extrapolated values from both determinations
agree well with each other, although those underestimate the experimental value.}
\label{fig:Ft_qsqr_extrap}
\end{figure}

In Fig.~\ref{fig:Ft_qsqr}, we show the Pauli form factor $F_2(q^2)=2\mnuc F_T(q^2)$
as a function of four-momentum squared $q^2$. 
The form factor plotted here is scaled by the renormalization factor 
$Z_V=1/F_V(0)$ to get the renormalized one, $F^{\rm ren}_2(q^2)=Z_VF_2(q^2)$.
In contrast with Fig.~\ref{fig:Fv_qsqr}, large $m_f$-dependence is observed.
As well as the Dirac form factor, the Pauli form factor is phenomenologically
supposed to be the dipole form at low $q^2$:
\be
F^{\rm ren}_2(q^2)=\frac{F^{\rm ren}_2(0)}{(1+q^2/M_T^2)^2},
\label{fig:dipole_fit_t}
\ee
where the value of $F^{\rm ren}_2(q^2)$ at $q^2=0$ is associated with the difference of the proton and neutron magnetic moments, $\mu_p - \mu_n =1 + F^{\rm ren}_2(0)$.
This dipole form is commonly adopted as a fitting form of the $q^2$ extrapolation 
to evaluate $F^{\rm ren}_2(0)$ in
published works~\cite{Gockeler:2003ay,Alexandrou:2006ru}.
We also plot the dipole form with the empirical value of the Pauli dipole mass $M_T=0.778(23)$ GeV and the experimental values of $\mu_p$ and $\mu_n$ in the same figure.
Our results of $F_2^{\rm ren}(q^2)$ gradually approach this dipole form as $m_f$ decreases. Indeed, data points 
for $m_f=0.02$ in the range of our calculated $q^2$ follow the experimental curve  
within the statistical error. However, if we apply the dipole form to the data, our obtained
values of $F_2^{\rm ren}(0)$ from dipole fits are somewhat underestimated in comparison with the experimental value $\mu_p-\mu_n-1=3.70589$ 
as shown in Fig.~\ref{fig:Ft_qsqr_extrap}.

In contrast to the Dirac form factor, 
we cannot try the effective dipole mass plot for a justification of the applied dipole form, since we do not have data of $F^{\rm ren}_2(0)$ without the $q^2$ extrapolation. Instead, we consider an independent observation
for the difference of $\mu_p$ and $\mu_n$, which can be derived from the forward limit of 
the ratio of the magnetic form factor $G_M(q^2)$ and the electric form factor $G_E(q^2)$.
The ratio is calculated by a different combination of $\Lambda^{V}_{0}$ and $\Lambda^{V}_{T}$~\cite{Berruto:2005hg}
as
\be
\frac{G_M(q^2)}{G_E(q^2)}=\frac{G^{\rm ren}_M(q^2)}{G^{\rm ren}_E(q^2)}=\frac{\Lambda^{V}_{T}}{\Lambda^{V}_{0}} .
\label{Eq:EMratio}
\ee
Experimentally, it is known that this ratio shows no $q^2$-dependence at low $q^2$
since both form factors are well fitted by the dipole form with the comparable 
dipole masses~\cite{Thomas:2001kw, Hyde-Wright:2004gh}.
Therefore, this ratio may yield the constant value identified to 
$\mu_p -\mu_n=G^{\rm ren}_M(0)$.
Indeed, in our calculation, the proposed ratio (\ref{Eq:EMratio}) exhibits
no appreciable $q^2$-dependence in the range of our calculated $q^2$. We 
may use a simple linear fitting form with respect to four-momentum squared $q^2$ for an alternative evaluation of the value $\mu_p-\mu_n$. Fig.~\ref{fig:Ft_qsqr_extrap} shows
that two determinations to evaluate $\mu_p-\mu_n-1$ are consistent with each other.

\begin{figure}[htbp]
\vspace{2mm}
\begin{center}
\includegraphics[width=0.52\linewidth,clip]{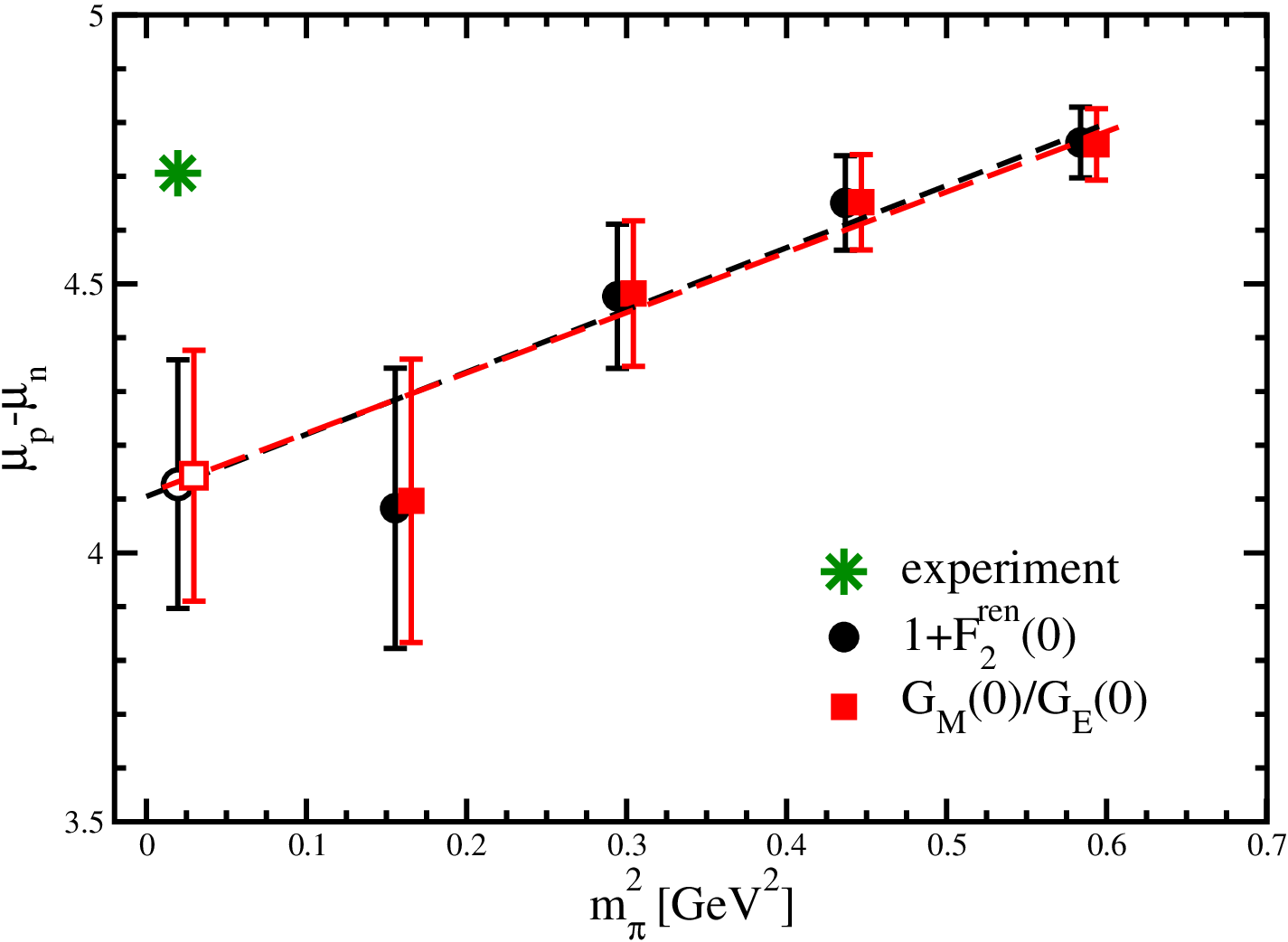}
\end{center}
\caption{
Chiral extrapolation of $\mu_p -\mu_n $.
The square symbols have been moved slightly in the plus x-direction.
}
\label{fig:Ft0_chi}
\end{figure}

In Fig.~\ref{fig:Ft0_chi}, we plot values of $\mu_p-\mu_n$, which are
evaluated by two determinations, $1+F_2^{\rm ren}(0)$ and $G_M(0)/G_E(0)$, 
as a function of pion mass squared $m_{\pi}^2$.
As described above, both determinations fairly agree with each other.
Although the values measured at two heaviest points 
are consistent with the experimental one, a strong $m_{\pi}^2$ dependence appears
near the chiral limit and then the extrapolated value 
tends to somewhat underestimate the experimental data.
Here, we simply adopt a linear fit with respect to $m_{\pi}^2$
regardless of the fact that a slight downward curvature is observed  in Fig.~\ref{fig:Ft0_chi}.

\begin{figure}[htbp]
\vspace{2mm}
\begin{center}
\includegraphics[width=0.52\linewidth,clip]{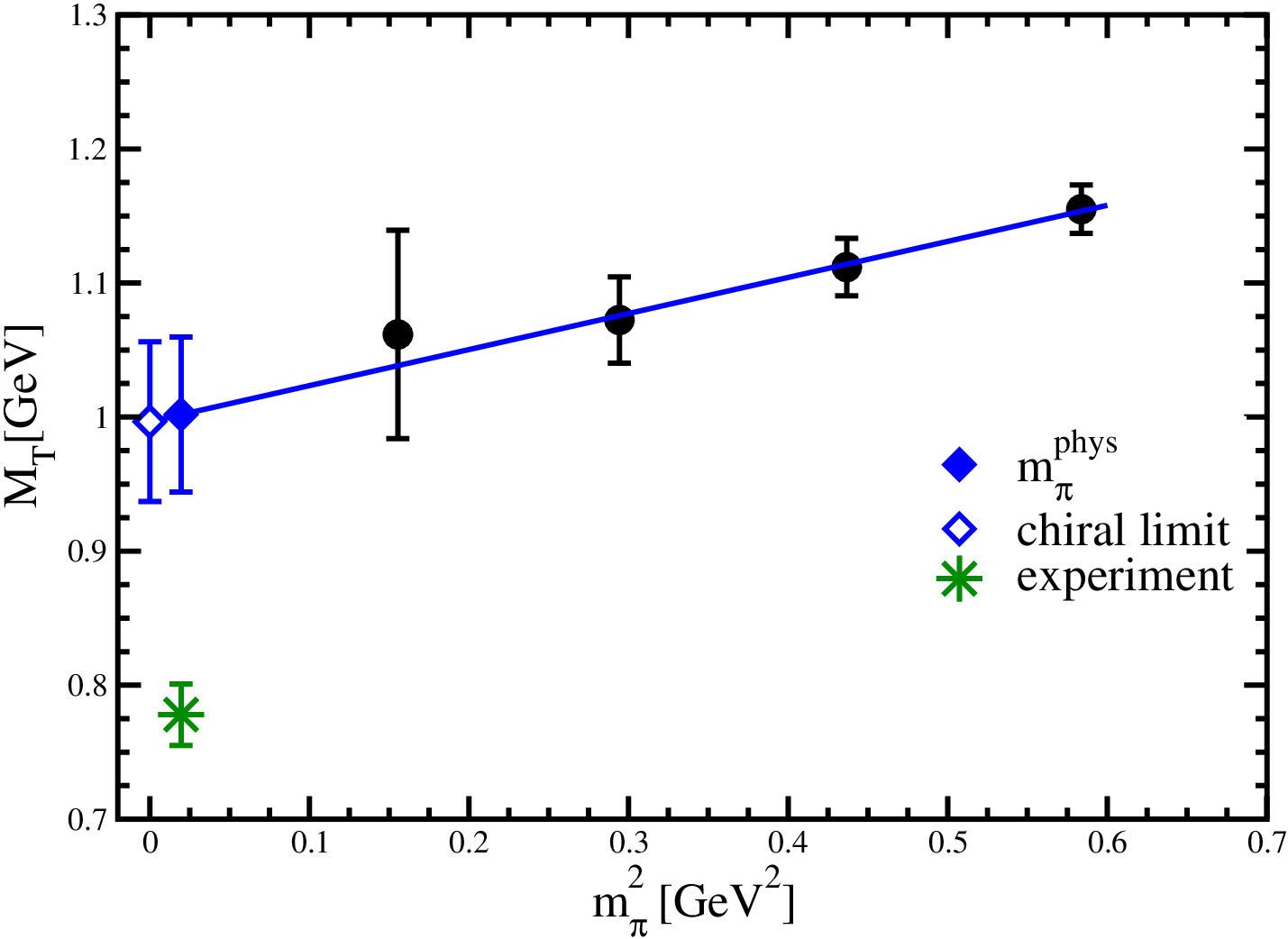}
\end{center}
\caption{
Chiral extrapolation of the Pauli dipole mass $M_T$.
Symbols are defined as in Fig.~\ref{fig:Mv_chi}.
}
\label{fig:Mt_chi}
\end{figure}

We also extrapolate the Pauli dipole mass $M_T$ to the chiral limit in Fig.~\ref{fig:Mt_chi}.
Again, we use a simple linear fitting form for the chiral extrapolation.
The value obtained at the physical point is about a 20\% overestimation in comparison 
with the experimental one, the same as in the case of the Dirac dipole mass. 
This indicates that corresponding rms radii are somewhat smaller than the 
actual nucleon size.
Finally, all fitted results with the dipole form for both form factors 
and their extrapolated values to the chiral limit are summarized 
in Table~\ref{tab:summary_vector}.

\subsection{Axial-vector channel}
\label{Sec5:Axial}

In the axial-vector channel, two independent form factors 
$F_A(q^2)$ and $F_P(q^2)$ can be evaluated separately by
\bea
F_A (q^2)&=&\Lambda_L^A(q_z=0),\\
F_P (q^2)&=&\Lambda_T^A/M_N
\eea
at finite $q^2$. It should be reminded that $\Lambda_L^A(q_z=0)$ at ${\bf n}^2=3$ and $\Lambda_T^A$ at ${\bf n}^2=1$, 4 are not obtained directly from corresponding
three-point functions due to the kinematics as described in the previous section.
However, instead, we can evaluate them by using a relation
\be
\Lambda_L^A(q_z\neq 0)=\Lambda_L^A(q_z=0)-\frac{q_z^2}{M_N(E+M_N)}\Lambda_T^A,
\ee
where $\Lambda_L^A(q_z\neq 0)$ are always calculable at finite $q^2$.

\subsubsection{Axial-vector form factor $F_A(q^2)$}
\label{Sec5:Fa}

\begin{figure}[htbp]
\vspace{2mm}
\begin{center}
\includegraphics[width=0.52\linewidth,clip]{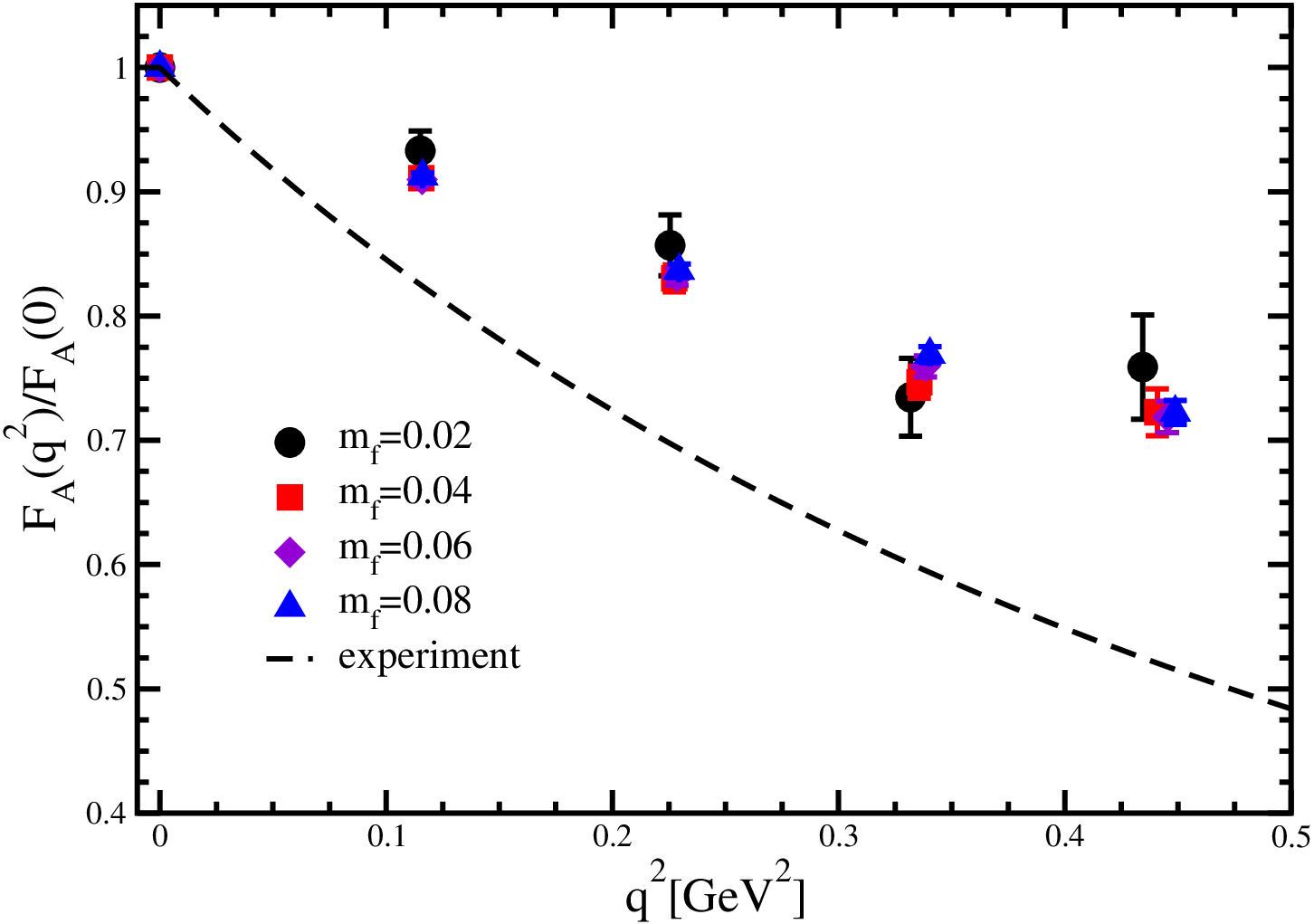}
\end{center}
\caption{
The axial-vector form factor normalized by $F_A(0)$ as a
function of four-momentum squared $q^2$.
}
\label{fig:Fa_qsqr}
\end{figure}

Fig.~\ref{fig:Fa_qsqr} shows quark mass dependence of the axial-vector form factor
$F_A(q^2)$. The vertical axis is normalized by $F_A(0)$ and the horizontal axis denotes
the four-momentum squared $q^2$ in physical units. Different symbols represent
the values obtained from different quark mass $m_f$. 
The axial-vector form factor is phenomenologically fitted with the dipole form,
at least at low $q^2$, as well as the Dirac and Pauli form factors~\cite{Bernard:2001rs}:
\be
F_A(q^2)=\frac{F_A(0)}{(1+q^2/M_A^2)^2},
\label{fig:dipole_fit_a}
\ee
where $M_A$ denotes the axial dipole mass. A dashed curve in Fig.~\ref{fig:Fa_qsqr}
shows the dipole form with an experimental value of the axial dipole mass $M_A=1.026 (21)$
GeV~\cite{Bernard:2001rs}. There is a similarity here in comparison with Fig.~\ref{fig:Fv_qsqr}.
No large $m_f$-dependence is observed. Even at the smallest quark mass $m_f=0.02$, where
the corresponding pion mass is less than 400 MeV, our observed $F_A(q^2)$ is far from 
the experimental curve. Indeed, the axial-vector form factor $F_A(q^2)$ is flatter than the experimental one, similar to what we observe in $F_V(q^2)$ and $F_2(q^2)=2\mnuc F_T(q^2)$.
This again indicates that the nucleon size in coordinate space shrinks away.
The similar observation is reported in Ref.~\cite{Alexandrou:2007xj}.

\begin{figure}[htbp]
\vspace{2mm}
\begin{center}
\includegraphics[width=0.52\linewidth,clip]{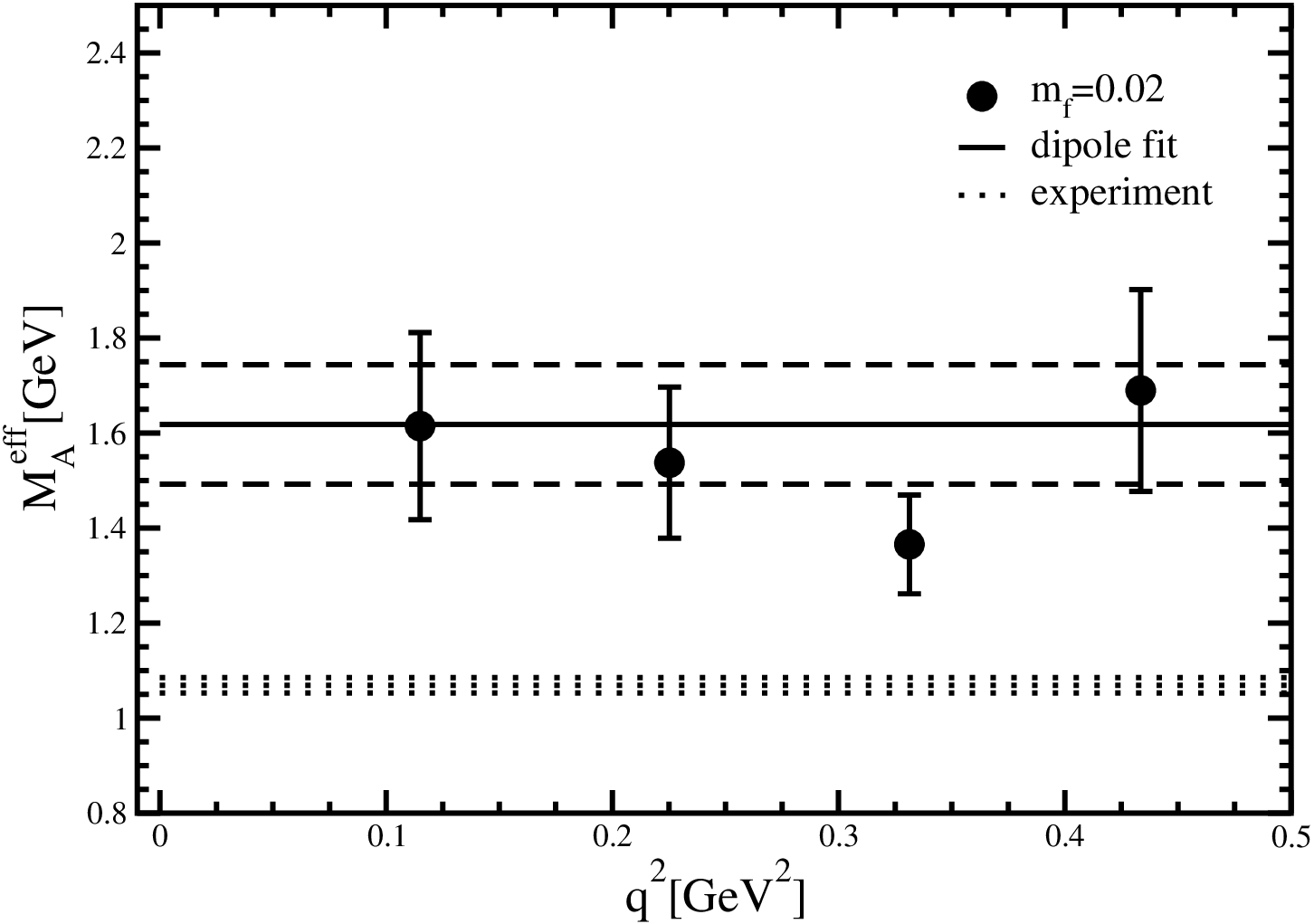}
\end{center}
\caption{
Effective dipole-mass plot for the axial-vector form factor $F_A(q^2)$ as a function of four-momentum squared  $q^2$ at $m_f=0.02$.
}
\label{fig:Effdipole_A}
\end{figure}

Next, to see how the dipole form is fitted to our measured $F_A(q^2)$, we show
the effective dipole mass plot, which was defined similarly to Eq.(\ref{Eq:EffmassPlot}). 
Fig.~\ref{fig:Effdipole_A} is plotted for $m_f=0.02$ as a typical example. 
We also include the fitted $M_A$, which is obtained from a correlated fit to $F_A(q^2)$
using the dipole form (\ref{fig:dipole_fit_a}), with its 
one standard deviation as solid and dashed horizontal lines. All momentum points except
the third one, which deviates from the fitted value by about $2\sigma$, are located within 
the horizontal lines. Here, we remark that the third momentum point of $F_A(q^2)$ 
at $m_f=0.02$ in  Fig.~\ref{fig:Fa_qsqr} is slightly dropped from the values measured
at other quark masses. We then stress that the case of $m_f=0.02$ is the worst example.
Indeed, it is found that the effective dipole mass plot for heavier $m_f$ are quite consistent 
with the fitted values in all $q^2$ range that we measured. From this observation, we 
conclude that the $q^2$-dependence of our measured $F_A(q^2)$ can be well 
described by the dipole form (\ref{fig:dipole_fit_a}) in the range of our utilized $q^2$, 
($q^2\le0.44$ ${\rm GeV}^2$). This is quite consistent with the phenomenological 
knowledge on the $q^2$-dependence of $F_A(q^2)$. 

\begin{figure}[htbp]
\vspace{2mm}
\begin{center}
\includegraphics[width=0.52\linewidth,clip]{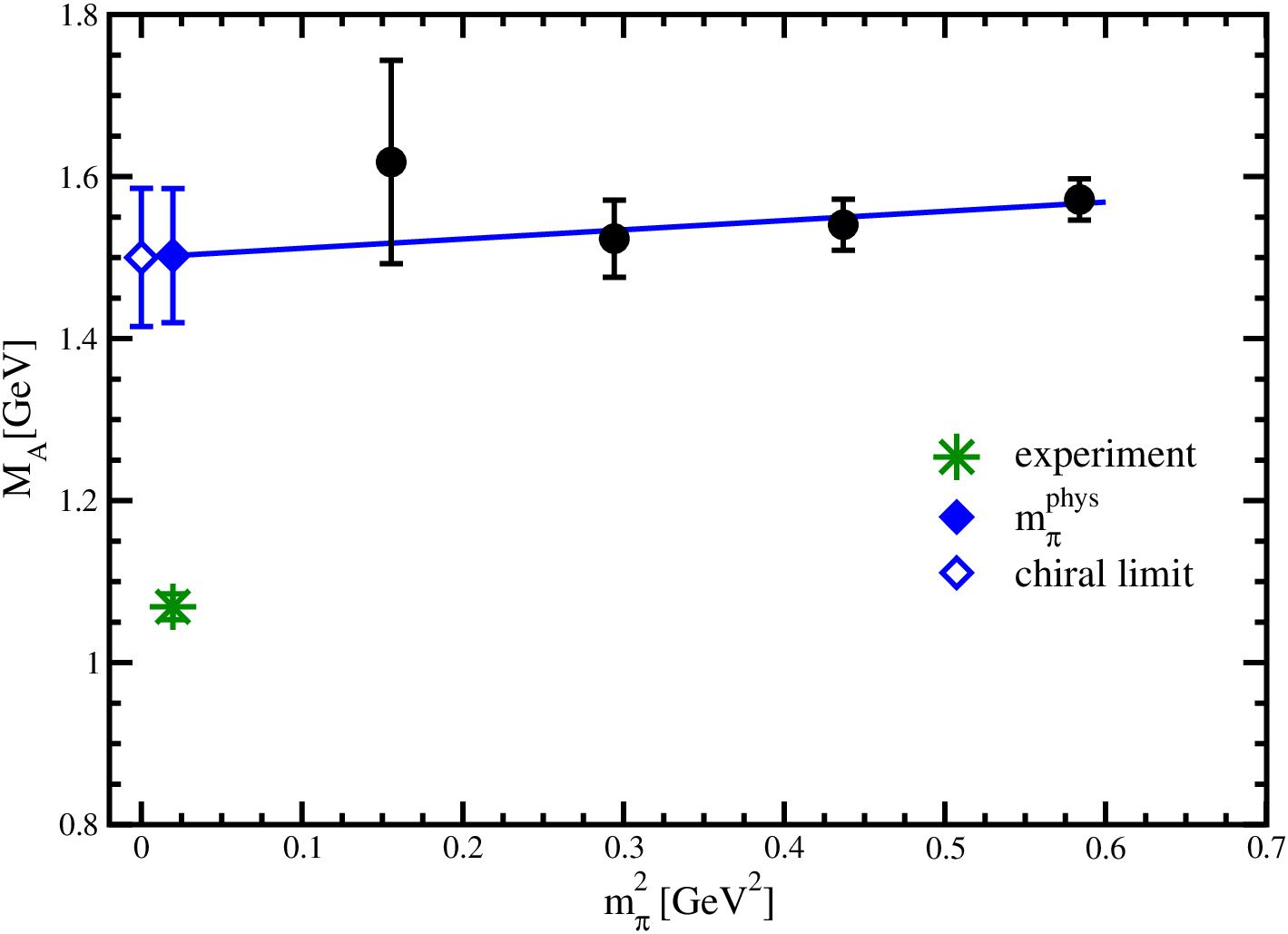}
\end{center}
\caption{
Chiral extrapolation of the axial dipole mass $M_A$.
Symbols are defined as in Fig.~\ref{fig:Mv_chi}.
}
\label{fig:Ma_chi}
\end{figure}

We show the quark-mass dependence of the fitted axial dipole mass as a function
of the pion mass squared in Fig.~\ref{fig:Ma_chi}. 
All measured values are listed in Table~\ref{Tab:FaFp_summary}.
We find that the quark-mass dependence is somewhat milder than the dipole masses for
the Dirac and Pauli form factors. Clearly, there is no appreciable curvature as
a function of the pion mass squared. As before, we simply
adopt a linear extrapolation for the axial dipole mass $M_A$ toward the chiral limit.
The extrapolated values (diamond symbols) overestimate the experimental one marked
by the asterisk in Fig.~\ref{fig:Ma_chi}.

\begin{figure}[htbp]
\vspace{2mm}
\begin{center}
\includegraphics[width=0.52\linewidth,clip]{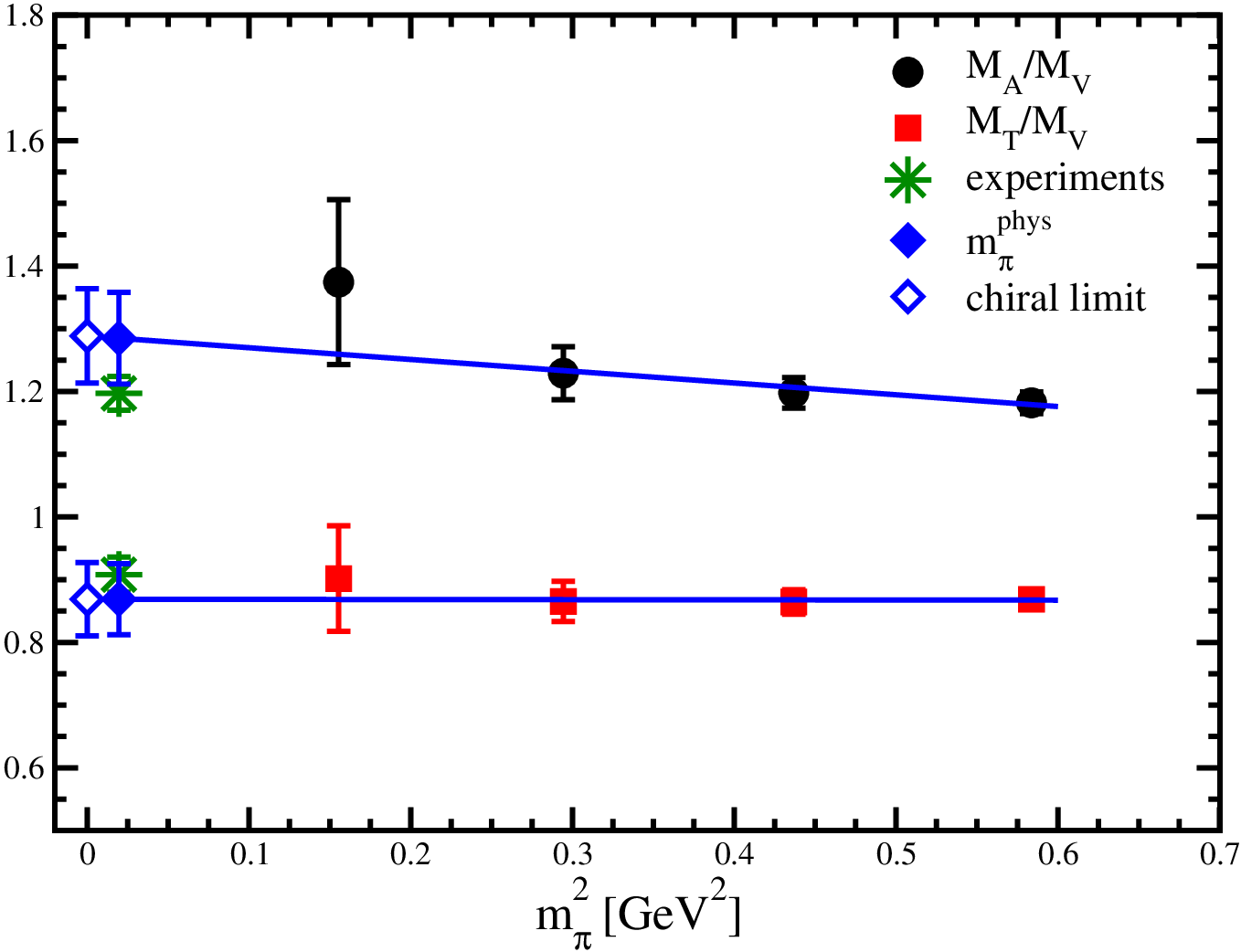}
\end{center}
\caption{
Ratios, $M_A/M_V$ and $M_T/M_V$ as functions of squared pion mass.
The extrapolated points in the chiral limit and at the physical point are represented
by an open diamond and a filled diamond. The experimental values are marked
with asterisk symbols. 
}
\label{fig:DM_ratio}
\end{figure}

As pointed out before, our observed ``size" of 
the nucleon in coordinate space is much smaller than the experimental one. 
We will see that there is no significant finite volume effect, which may cause
the ``size" of the nucleon to be squeezed on the lattice with  $(3.6\;{\rm fm})^3$
volume. 
Thus, this observed ``squeezing", which is evident from the broadening of 
the form factors, may be attributed to the missing contribution of the 
``pion-cloud" surrounding the nucleon 
outside of the chiral regime.
Interestingly, however, the ratio of the axial dipole mass to the 
Dirac dipole mass is in very good agreement with the experiment.
In Fig~\ref{fig:DM_ratio}, we show the ratio $M_A/M_V$ together 
with the ratio $M_T/M_V$ as a function of the pion mass squared.
The quark mass dependence of both ratios is found to be very mild 
in our observed range of $m_{\pi}^2$. All measured values of $M_A/M_V$ 
and $M_T/M_V$, which are listed in Table~\ref{tab:summary_ratio_mdp}, 
are fairly comparable to their respective experimental values. 
We obtain $M_A/M_V=1.285(73)$ and $M_T/M_V=0.869(57)$ 
at the physical point by using a simple linear extrapolation.

\subsubsection{Induced pseudo-scalar form factor $F_P(q^2)$}
\label{Sec5:Fp}

\begin{figure}[htbp]
\vspace{2mm}
\begin{center}
\includegraphics[width=0.52\linewidth,clip]{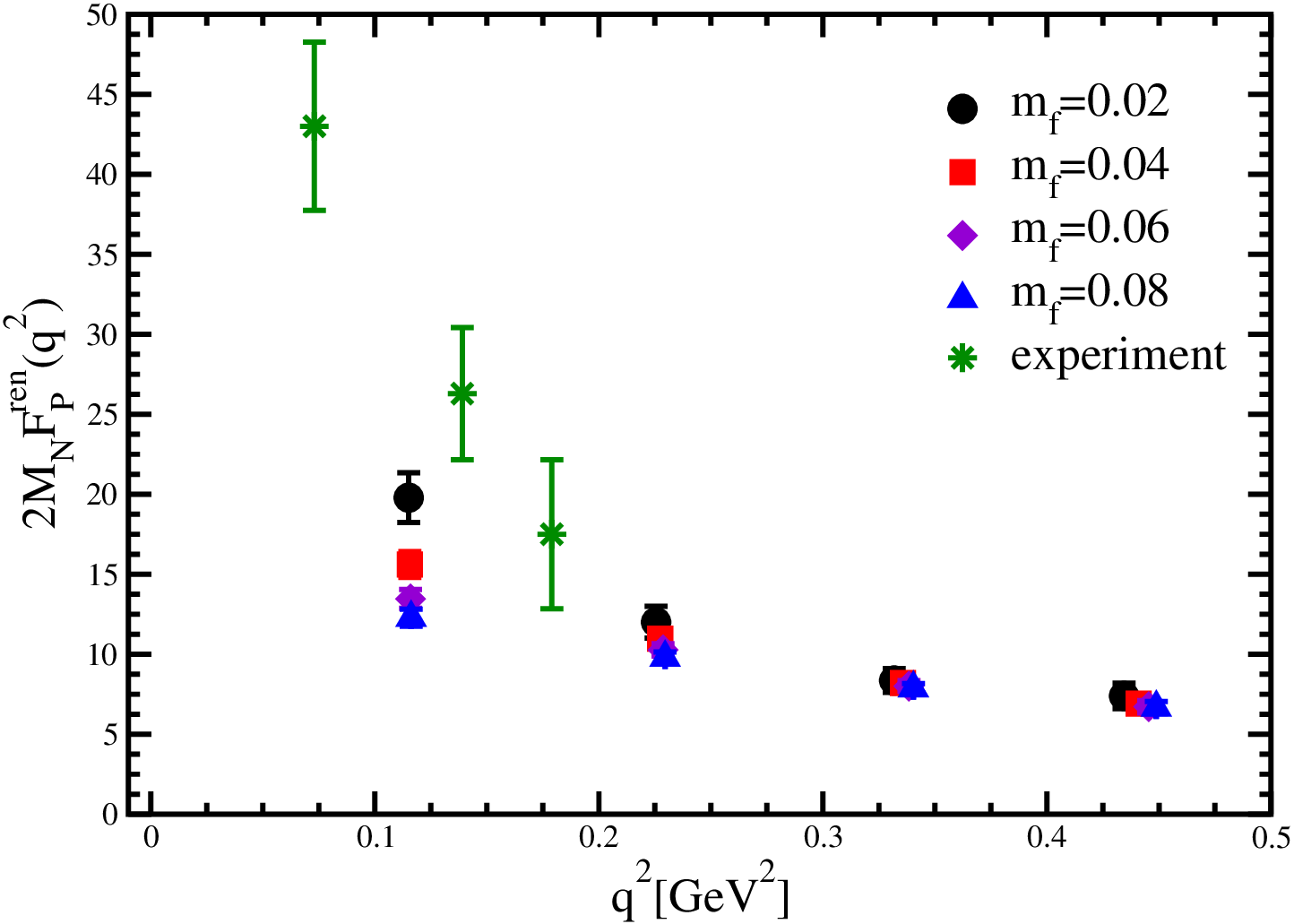}
\end{center}
\caption{The renormalized and dimensionless 
induced pseudo-scalar form factor $2\mnuc F^{\rm ren}_P(q^2)$ as a function of four-momentum squared $q^2$.}
\label{fig:Fp_qsqr}
\end{figure}

First, we show the quark-mass dependence of the induced pseudo-scalar form factor 
$F_P(q^2)$ in Fig.~\ref{fig:Fp_qsqr}. 
In contrast to the axial-vector form factor $F_A(q^2)$, significant $m_f$-dependence
is observed especially in the lower $q^2$ region ($q^2 < 0.3$ ${\rm GeV}^2$). 
This might be associated with the pion-pole contribution to $F_P(q^2)$, which 
is expected theoretically. Indeed, the partially conserved axial-vector current (PCAC)
hypothesis and pion-pole dominance (PPD) predict that the induced 
pseudo-scalar form factor approximately behaves like
\be
F^{\rm PPD}_P(q^2)=\frac{2\mnuc F^{\rm ren}_A(q^2)}{q^2+m_\pi^2},
\label{Eq:Fp_PPD_model}
\ee
which becomes exact
in the chiral limit where the pion is massless ($m_{\pi}=0$)~\cite{Nambu:1960xd,Thomas:2001kw}. 
The single pion electroproduction experiment also supports the PPD form~\cite{Choi:1993vt}. 
Here, to see how the pion-pole behavior is preserved in 
$F_P(q^2)$ measured in the quenched calculation, we consider the following ratio
\be
\alpha_{_{\rm PPD}}=\frac{F^{\rm ren}_P(q^2)}{F^{\rm PPD}_P(q^2)} ,
\label{Eq:ratio_PPD}
\ee
which is inspired by the above PCAC prediction.
If the measured $F_P(q^2)$ has exactly the same form described in Eq.(\ref{Eq:Fp_PPD_model}),
the ratio $\alpha_{_{\rm PPD}}$ yields the value of unity in the entire $q^2$ region.
 
\begin{figure}[htbp]
\vspace{2mm}
\begin{center}
\includegraphics[width=0.52\linewidth,clip]{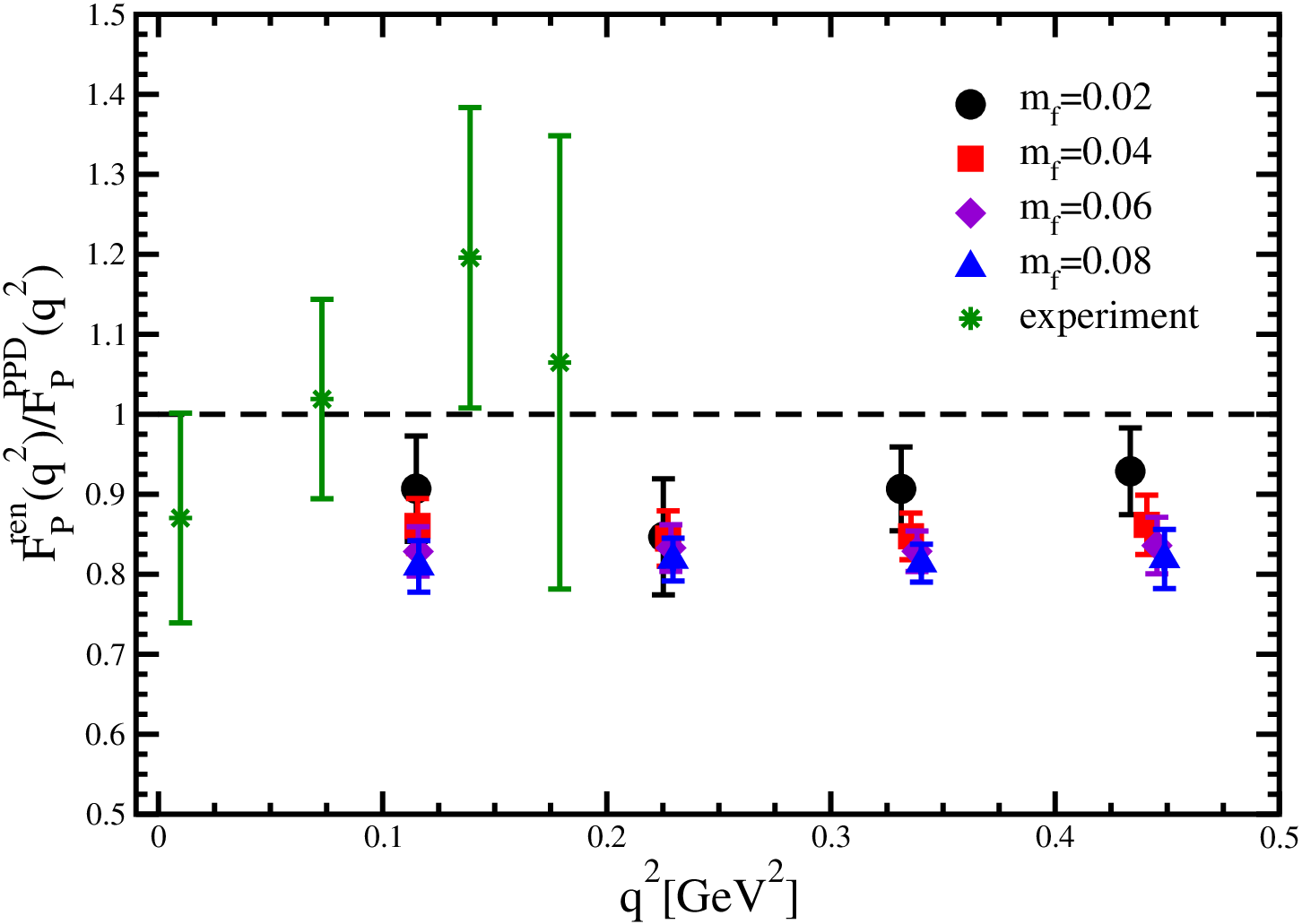}
\end{center}
\caption{The ratio of $F^{\rm ren}_P(q^2)$ and $F_P^{\rm PPD}(q^2)$ as a function of four-momentum squared $q^2$.}
\label{fig:R_PPD}
\end{figure}

\begin{figure}[htbp]
\vspace{2mm}
\begin{center}
\includegraphics[width=0.52\linewidth,clip]{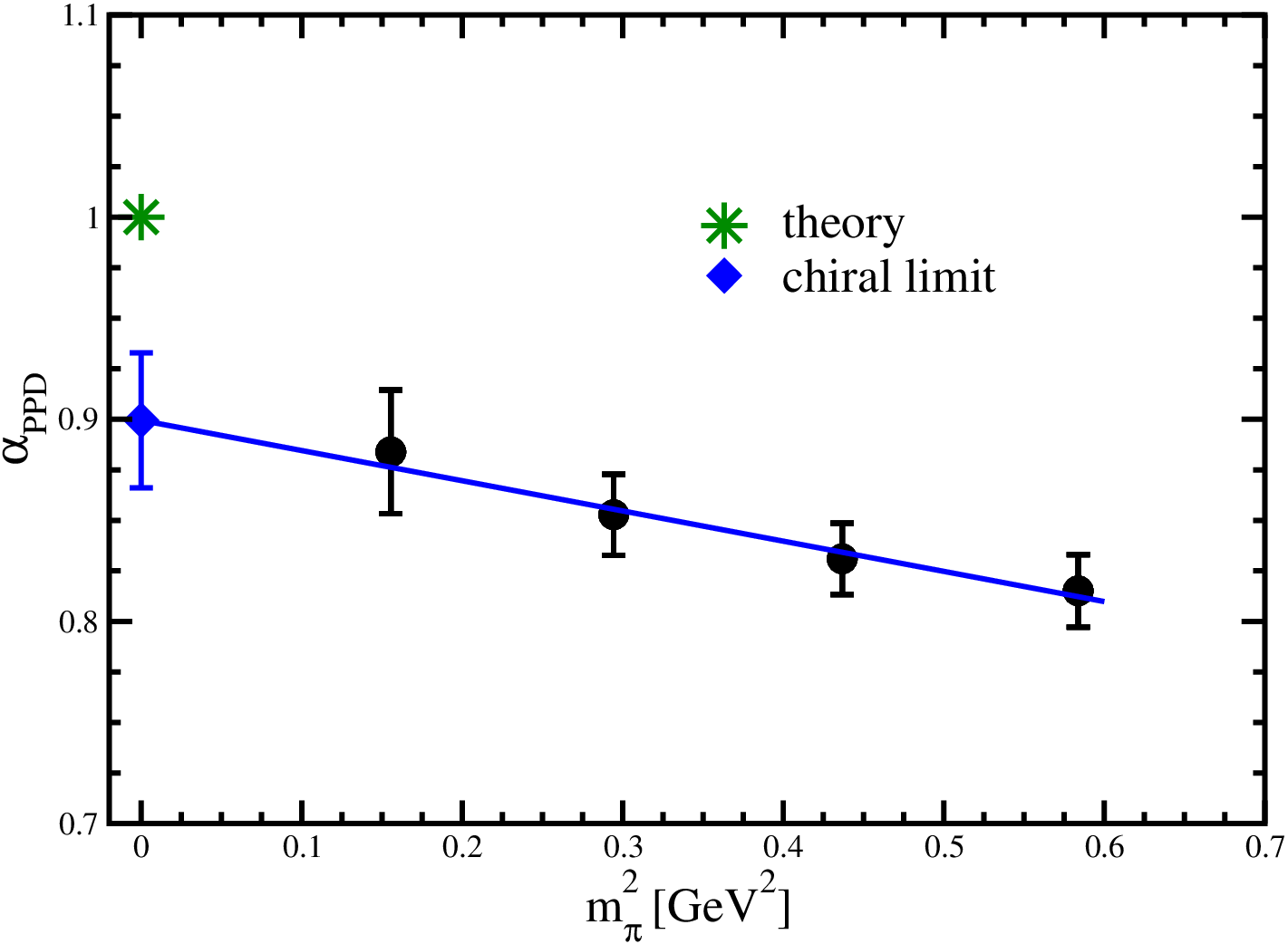}
\end{center}
\caption{The quenching factor $\alpha_{_{\rm PPD}}$ is plotted as a function of $m_\pi^2$.}
\label{fig:R_PPD_chi}
\end{figure}

In Fig.~\ref{fig:R_PPD}, we plot the above defined ratio $\alpha_{_{\rm PPD}}$ as a function of four-momentum squared $q^2$. This figure shows two important features. The significant 
quark-mass dependence observed in Fig.~\ref{fig:Fp_qsqr} almost disappears as expected. 
Furthermore, there is no appreciable $q^2$-dependence in $\alpha_{_{\rm PPD}}$.
Clearly, four different $q^2$ points of $\alpha_{_{\rm PPD}}$ reveal $q^2$ independent plateau within the statistical errors. We simply take the weighted average of 
$\alpha_{_{\rm PPD}}$ within
four measured $q^2$ points, then plot them against the pion mass squared. 
As shown in Fig.~\ref{fig:R_PPD_chi}, the average values of $\alpha_{_{\rm PPD}}$ 
gradually approach unity as the pion mass decreases. 
However, a simple linear extrapolation of $\alpha_{_{\rm PPD}}$
yields a value slightly smaller than 1 even in the chiral limit.
As a result, measured $F^{\rm ren}_P(q^2)$ is quit well described by the PPD form
with a multiplicative (quenching) factor $\alpha_{_{\rm PPD}}< 1$.
\be
F^{\rm ren}_P(q^2)\approx \alpha_{_{\rm PPD}}\times F_P^{\rm PPD}(q^2).
\label{Eq:Fp_PPD_form}
\ee
The validity of the PPD form is also tested by the other analysis.
Following the analysis done in the previous study of the $q^2$-dependence
of $F_P(q^2)$~\cite{Alexandrou:2007xj}, we apply the monopole fit 
to the ratio $F_P(q^2)/F_A(q^2)$. The satisfactory consistency between
the fitted monopole mass and the measured pion mass is observed 
in our DWF calculation~\footnote{
It is worth mentioning that such consistency is
not observed at $L=16$, since the lowest $q^2$ point at $L=16$
may suffer from the finite volume effect as we will describe
in Sec.~\ref{Sec:FVE}.}, while both quenched and unquenched Wilson simulations
fail to exhibit the correct pion-pole structure of $F_P(q^2)$~\cite{Alexandrou:2007xj}.

\begin{figure}[htbp]
\vspace{2mm}
\begin{center}
\includegraphics[width=0.52\linewidth,clip]{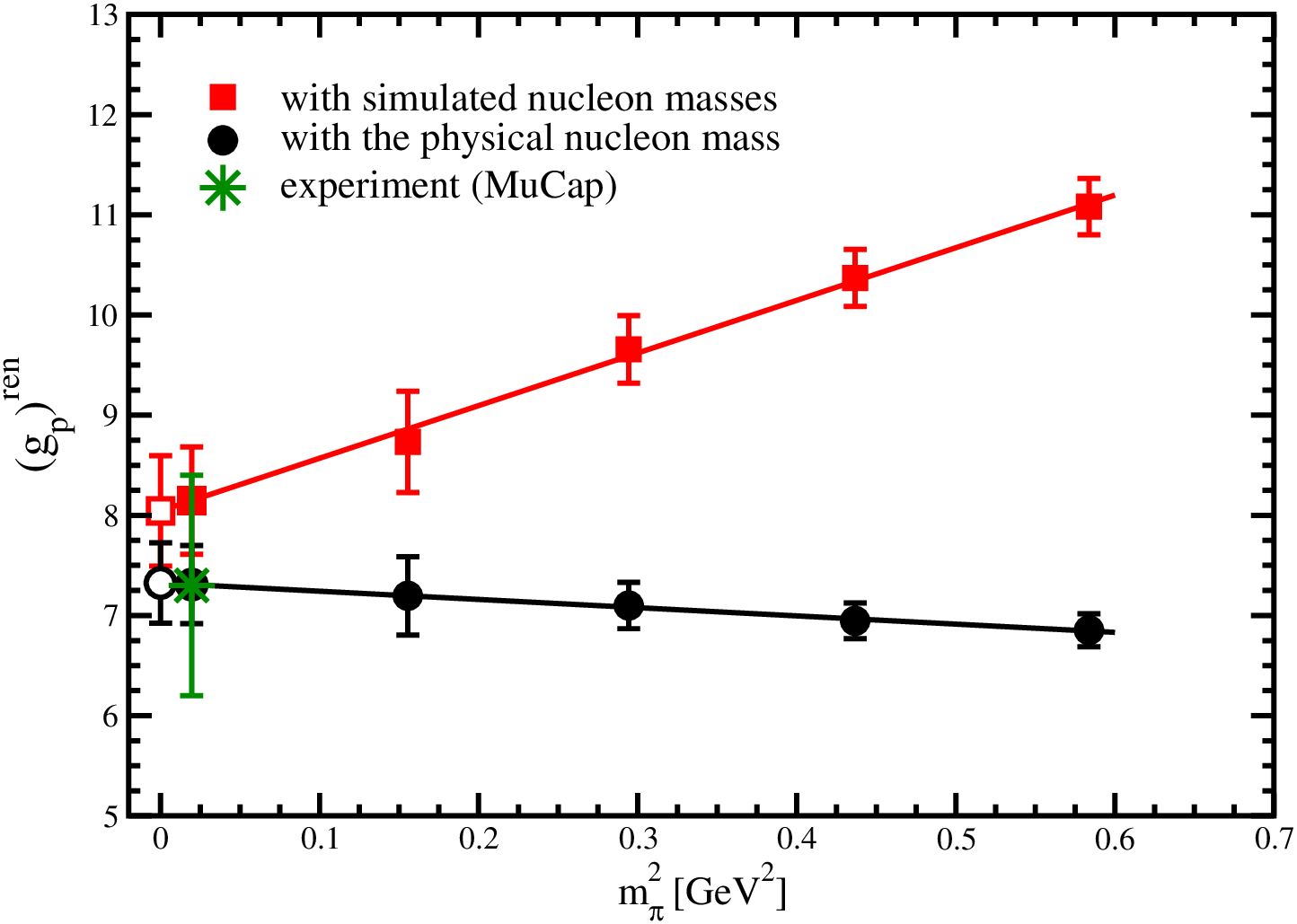}
\end{center}
\caption{The induced pseudo-scalar coupling $(g_P)^{\rm ren}$ evaluated through 
two determinations, where measured values (squares) and the physical value (circles)
are used for the nucleon mass in Eq.(\ref{Eq:ps-coupling}).
}
\label{fig:g_p_chi}
\end{figure}

Next, we evaluate the induced pseudo-scalar coupling, which is defined by $(g_P)^{\rm ren}=m_\mu F_P^{\rm ren}(0.88 m_\mu^2)$
where $m_{\mu}$ is the muon rest mass and $F_P^{\rm ren}(q^2)=F_P(q^2)/F_V(0)$.
The specific momentum transfer for muon capture ($q^2=0.88 m_\mu^2$) is still
far from our lowest momentum transfer ($q^2\approx 0.1\;{\rm GeV}^2$) so that
the determination of $(g_P)^{\rm ren}$ requires the $q^2$ extrapolation of 
$F_P^{\rm ren}(q^2)$. We have already learned that the $q^2$-dependence 
of measured $F_P^{\rm ren}(q^2)$ is well described by the PPD-like 
form (\ref{Eq:Fp_PPD_form}) in the low $q^2$ region. 
Therefore, the induced pseudo-scalar coupling can be evaluated by 
\be
(g_P)^{\rm ren}
=\frac{2m_{\mu}\mnuc }{m_{\pi}^2+0.88 m_{\mu}^2} \times \alpha_{_{\rm PPD}} F^{\rm ren}_A(0.88 m_{\mu}^2),
\label{Eq:ps-coupling}
\ee
where $F_A^{\rm ren}(0.88m_{\mu}^2)$ is precisely determined through 
the $q^2$ interpolation with the help of the dipole form and very accurate 
data $F_A^{\rm ren}(0)$. 
The pion mass in Eq.(\ref{Eq:ps-coupling}) is simply replaced by its physical value
in order to subtract the dominant source of the large $m_f$-dependence.
In Fig.~\ref{fig:g_p_chi}, we plot the resulting value of $(g_P)^{\rm ren}$ 
(square symbols) as a function of the pion mass squared. 
All measured values are listed in Table~\ref{Tab:FaFp_summary}.
Although there still remains the explicit dependence of the quark mass, 
the simple linear extrapolation yields $(g_P)^{\rm ren}=8.15\pm0.54$ 
at the physical point ($m_\pi=0.14$ GeV).

Here, the observed $m_f$-dependence stems from that of measured $M_N$, since both 
$\alpha_{_{\rm PPD}}$ and $F_A^{\rm ren}(q^2)$ have a very mild quark mass 
dependence. To diminish the explicit $m_f$-dependence, we may evaluate
the dimensionless prefactor in Eq.(\ref{Eq:ps-coupling}) with the experimental values of $m_{\mu}=105.7$ MeV, $m_{\pi}=139.6$ MeV and $\mnuc=938.9$ MeV instead 
of using measured values.
We then obtain a more simple form as $(g_P)^{\rm ren}=6.77 \times \alpha_{_{\rm PPD}} F^{\rm ren}_A(0.88 m_{\mu}^2)$, which is similar to the known phenomenological form 
for $g_P$ beside the quenching factor $\alpha_{_{\rm PPD}}$~\cite{Gorringe:2002xx}.
Indeed, evaluated values using this simple formula have no appreciable $m_f$ 
dependence as shown in Fig.~\ref{fig:g_p_chi}.
After the linear extrapolation, we obtained $(g_P)^{\rm ren}=7.31 \pm 0.39$ at the 
physical point. Two determinations provide consistent results within their statistical errors.
Of course, the latter determination is rather phenomenological, then we prefer
to quote the former value for our final result.
We finally quote 
\be
(g_P)^{\rm ren}=8.15\pm0.54({\rm stat})\pm 0.16({\rm norm}),
\ee
where the second error is an estimate of a 2\% error stemming 
from $Z_V\neq Z_A$, the same as in the axial-vector coupling $g_A$.

This value is to be compared with the most recent experimental 
value $g_P^{\rm exp} = 7.3 \pm 1.1$ from the MuCap experiment~\cite{Andreev:2007wg},
where the obtained value of $g_P^{\rm exp}$ is nearly independent of $\mu$-molecular effects.
We also quote the prediction of chiral perturbation theory, $g_P^{\rm ChPT}=8.26 \pm 0.23$ and the new world average of experimental values, $g_P^{\rm exp} = 8.7 \pm 1.0$~\cite{Czarnecki:2007th} obtained from ordinary muon capture including the new MuCap result.

Phenomenologically, the residue of the pion pole in the induced pseudo-scalar form factor
is related to the pion-nucleon coupling $g_{\pi N N}$~\cite{Nambu:1960xd}. 
The induced pseudo-scalar form factor should be expressed as
\be
F^{\rm ren}_P(q^2)\simeq 
\frac{2F_{\pi}g_{\pi N N}}{q^2+m_{\pi}^2}
\ee
near the pion pole ($q^2\approx -m_{\pi}^2$)~~\cite{Nambu:1960xd,Thomas:2001kw} with the renormalized 
pion decay constant $F_{\pi}$, which is defined as
$Z_A\langle 0|\del_{\alpha}A_{\alpha}^a(x)|\pi_{b}(q)\rangle =m_{\pi}^2 F_{\pi}\delta_{ab}e^{iq\cdot x}$~\footnote{Here we use a traditional
convention as $F_{\pi}=f_{\pi}/\sqrt{2}\sim 93 {\rm MeV}$, while
$f_{\pi}$ is quoted in Ref.\cite{Aoki:2002vt}.
}.
This parameterization provides a way to evaluate the pion-nucleon 
coupling $g_{\pi N N}$ from the measured induced pseudo-scalar form factor
as follows:
\bea
g_{\pi N N}&=&\lim_{q^2 \rightarrow -m_{\pi}^2}  (q^2+m_\pi^2) \frac{F^{\rm ren}_P(q^2)}{2F_{\pi}} \\
&=& \frac{\alpha_{_{\rm PPD}}}{F_{\pi}} \times \mnuc F^{\rm ren}_A(-m_{\pi}^2),  
\label{Eq:g_pi_nuc}
\eea
where the second equality follows from our observed
form (\ref{Eq:Fp_PPD_form}) on $F^{\rm ren}_P(q^2)$.
The value of $F_A^{\rm ren}(-m_{\pi}^2)$ is evaluated by the dipole form with
measured $F_A^{\rm ren}(0)$, $M_A$ and $m_\pi$. 
We then obtain 
\be
g_{\pi N N}=10.4 \pm 1.0 ({\rm stat})
\ee
at the physical point. Our obtained value is about 20\% smaller that 
a recent estimation $g_{\pi N N}=13.32\pm 0.09$ 
($g_{\pi N N}^2/4\pi=14.11\pm 0.20$) obtained from forward 
$\pi N$ scattering data~\cite{Ericson:2000md}.

\subsection{Pseudo-scalar channel}
\subsubsection{Pseudo-scalar form factor $G_P(q^2)$}
\label{Sec5:Gp}

In this study, we also calculate the pseudo-scalar nucleon matrix element
\be
\langle N(P')|P^a(x)|N(P)\rangle={\overline u}_{_N}(P') 
{\cal O}^{P}(q)t^a
u_{_N}(P) e^{iq\cdot x},
\ee
which is associated with the axial-vector matrix element through the axial Ward-Takahashi identity. Here, $P^a(x)$ is a local pseudo-scalar 
density, $P^a(x)\equiv \bar{\psi}(x)\gamma_5t^a \psi(x)$. The pseudo-scalar 
matrix element can be described only by a single form factor, which is called 
the pseudo-scalar form factor $G_P(q^2)$:
\be
{\cal O}^{P}(q)=\gamma_5 G_P(q^2).
\ee
To extract the form factor, therefore, we simply calculate the following trace of 
${\cal G}^P(q)$, which represents the 
spinor structure of the corresponding three-point function, with the projection
operator ${\cal P}^z_5$:
\be
\frac{1}{4}{\rm Tr}\left\{
{\cal P}^z_5 {\cal G}^P(q)
\right\}=iq_z\mnuc G_P(q^2),
\ee
where the definition of ${\cal G}^P(q)$ is given by Eqs.~(\ref{Eq:three-pt}) 
and (\ref{Eq:spinor-part-of-three-pt}) with the local current 
${\cal J}^{P}_5(x)={\bar u}(x)\gamma_5u(x)-{\bar d}(x)\gamma_5d(x)$.
It is apparent that non-zero three-momentum ${\bf q}\neq {\bf 0}$ 
is required to access the pseudo-scalar form factor. In other words, $G_P(q^2)$ in
the vicinity of $q^2=0$ cannot be evaluated without $q^2$ extrapolation.

\begin{figure}[htbp]
\vspace{2mm}
\begin{center}
\includegraphics[width=0.52\linewidth,clip]{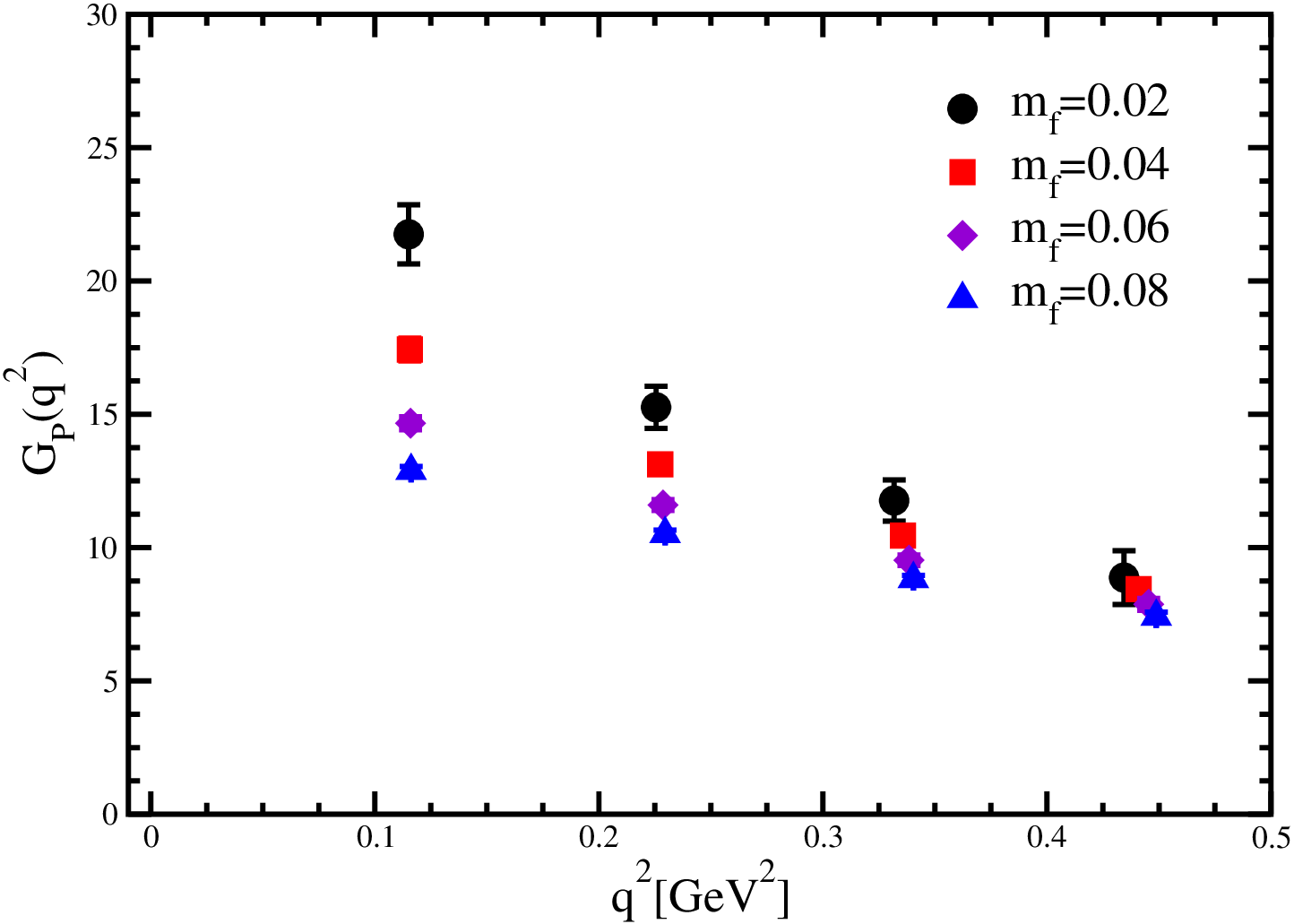}
\end{center}
\caption{The bare pseudo-scalar form factor $G_P(q^2)$ as a function of 
four-momentum squared $q^2$.}
\label{FIG:Gp_qsqr}
\end{figure}

In Fig.~\ref{FIG:Gp_qsqr}, we show the $m_f$-dependence of the pseudo-scalar
form factor $G_P(q^2)$. Significant $m_f$-dependence is observed in the
lower $q^2$ region, similar to the induced pseudo-scalar form factor $F_P(q^2)$.
We will discuss the $q^2$-dependence on $G_P(q^2)$ from the viewpoint of pion-pole
dominance later. 

\subsubsection{Test for the axial Ward-Takahashi identity}
First, we address the question whether our domain wall fermion (DWF) calculations of
nucleon form factors satisfy the axial Ward-Takahashi identity.  
A similar study has been recently done with the Wilson fermions
in both quenched and unquenched simulations~\cite{Alexandrou:2007xj}.

In the limit where the fifth-dimensional extent $L_s$ is taken to infinity,
domain wall fermions preserve the axial Ward-Takahashi identity at nonzero lattice 
spacing~\cite{Furman:1995ky}. The axial Ward-Takahashi identity for 
the DWF with $SU(2)$ iso-spin symmetry is
\be
\del_{\mu} {\cal A}^{a}_{\mu}(x)=2 m_f J_5^{a}(x) + 2 J_{5q}^{a}(x),
\label{Eq:DWF-AWT}
\ee
where ${\cal A}^{a}_{\mu}$ is the partially-conserved axial-vector current, which is
point split and requires sums over the extra fifth dimension of the DWF,
$J_5^a$ is a usual bilinear pseudo-scalar density corresponding to
$P^a$, and $J_{5q}^a$
is a similar pseudo-scalar density defined at the midpoint of the fifth
dimension. The ``midpoint" term $J_{5q}^{a}$ is responsible for the explicit 
chiral symmetry breaking due to the finiteness of the fifth-dimension~\cite{Furman:1995ky}. With moderate $L_s$, this effect
can be described by the so-called residual mass term $m_{\rm res}$. 
Then, Eq.(\ref{Eq:DWF-AWT}) can be approximately represented by
\be
\del_{\alpha} {\cal A}^{a}_{\alpha}(x)\approx 2 (m_f + m_{\rm res}) P^{a}(x).
\label{Eq:EFF-AWT}
\ee
This residual mass $m_{\rm res}$ is determined 
by $\langle 0|J^a_{5q}|\pi\rangle/\langle 0|J^a_5|\pi\rangle$.
The value of $m_{\rm res}$ is known to be small in this calculation. 
(See Table~\ref{tab:old_simulation_results}.)

For a practical reason, we did not use the conserved axial-vector current
for evaluating the nucleon axial matrix element in this study.  Instead, we use
the local axial-vector current $A_{\mu}^a=\bar{\psi}\gamma_\mu\gamma_5 t^{a}\psi$,
which may be related to the conserved axial-vector current as ${\cal A}_{\mu}
=Z_A A_{\mu}+{\cal O}(a^2, m_f a^2)$. It is worth mentioning that
$Z_V=Z_A$ is satisfied up to small discretization errors of ${\cal O}(a^2)$ in the chiral limit~\cite{Sasaki:2003jh}. 
We also did not measure the nucleon matrix element of $J_{5q}^{a}$, therefore
we cannot fully check the axial Ward-Takahashi identity (AWT) in terms of the 
nucleon matrix element in this paper. Instead, we test the following ratio, which 
may have no apparent $q^2$-dependence.
\be
\alpha_{_{\rm AWT}}=\frac{2 \mnuc F^{\rm ren}_A(q^2) - q^2 F^{\rm ren}_{P}(q^2)}{2m_f G_{P}(q^2)},
\label{Eq:R_AWT}
\ee
This ratio (\ref{Eq:R_AWT}) is associated with the following identity~\footnote{
Strictly speaking, the relativistic dispersion relation (\ref{Eq:RelDisp}), which ensures
that $(i\Pslash+\mnuc)u_{_N}(p)=0$, should be well satisfied in simulations. 
}:
\be
Z_A\langle N|\del_{\alpha} A^{a}_{\alpha}(x)|N\rangle
= 2 m_{_{\rm AWT}}\langle N|P^{a}(x)|N\rangle,
\label{Eq:mod_AWT}
\ee
where $m_{_{\rm AWT}}\equiv\alpha_{_{\rm AWT}}m_f$, which is expected to be
comparable to $m_f+m_{\rm res}$ up to terms of order ${\cal O}(a^2, m_f a^2)$.

\begin{figure}[htbp]
\vspace{2mm}
\label{FIG:AWT}
\begin{center}
\includegraphics[width=0.52\linewidth,clip]{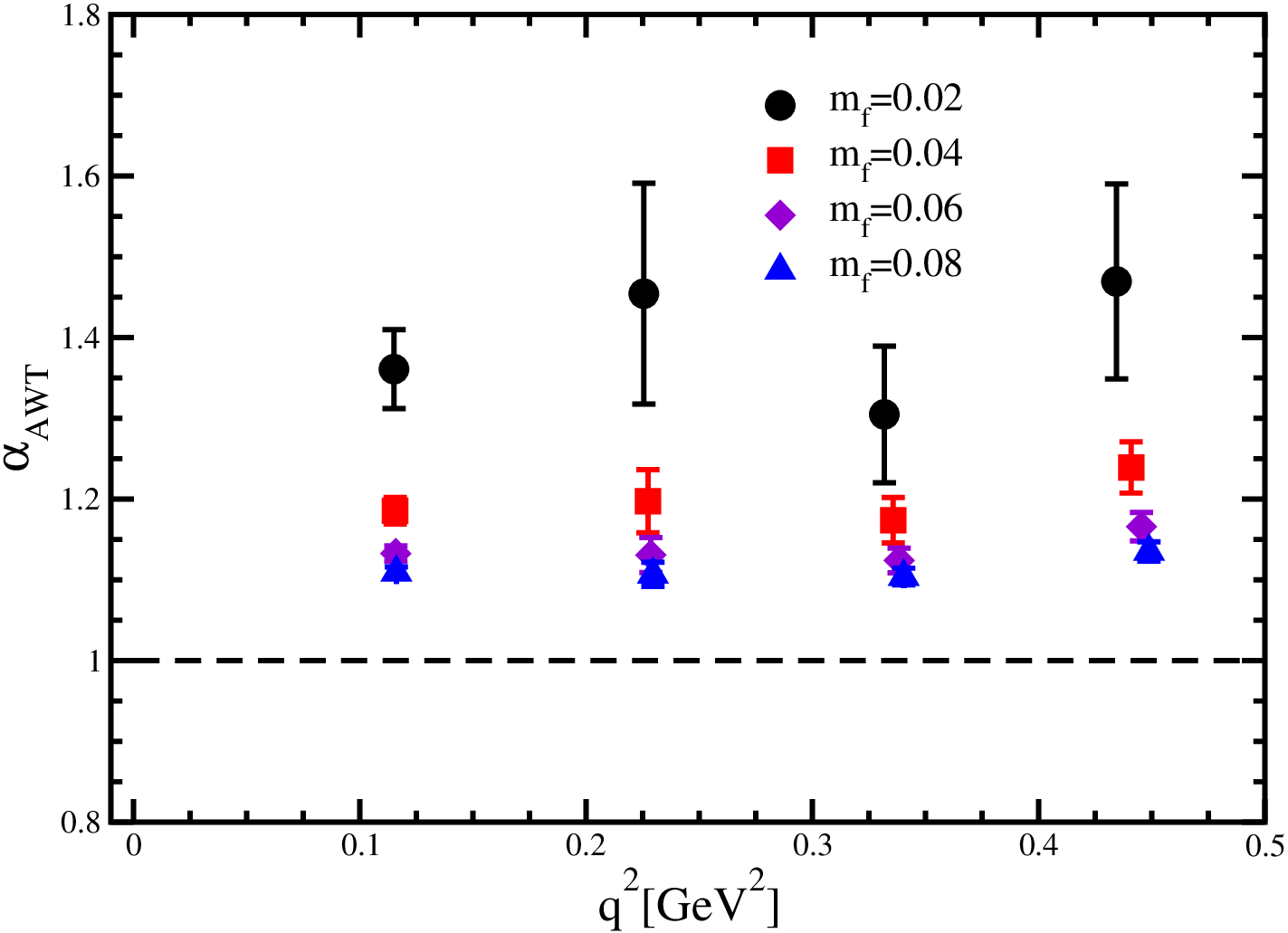}
\end{center}
\caption{
The ratio $\alpha_{_{\rm AWT}}$ defined in Eq.(\ref{Eq:R_AWT}) as a function
of four-momentum squared $q^2$.
}
\label{FIG:R_AWT}
\end{figure}

\begin{figure}[htbp]
\vspace{2mm}
\begin{center}
\includegraphics[width=0.52\linewidth,clip]{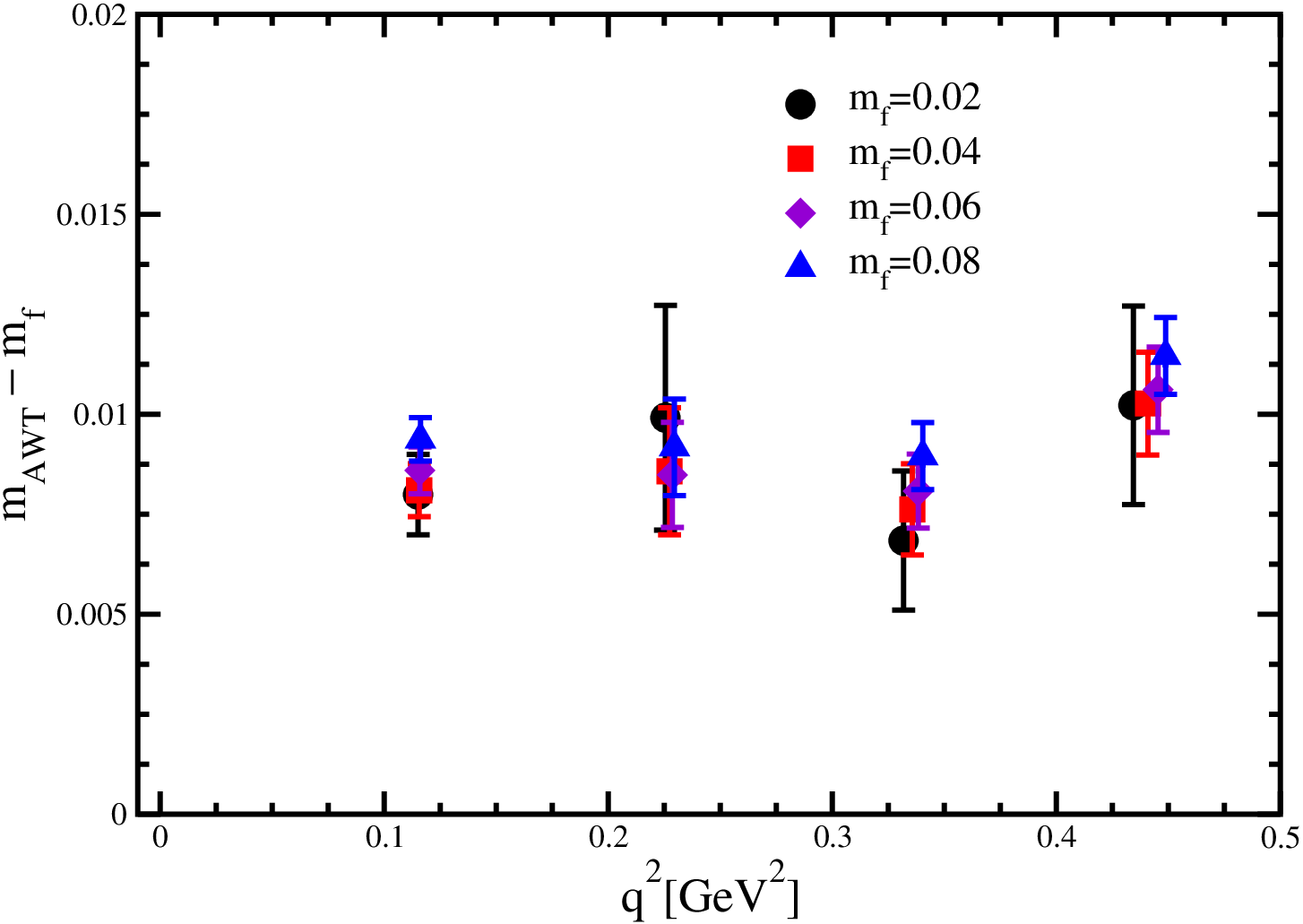}
\end{center}
\caption{
The modified ratio $m_f(\alpha_{_{\rm AWT}}-1)$, which may be expressed as
$m_{\rm AWT}-m_f$, as a function of four-momentum squared $q^2$.
}
\label{FIG:D_AWT}
\end{figure}

\begin{figure}[htbp]
\vspace{2mm}
\begin{center}
\includegraphics[width=0.52\linewidth,clip]{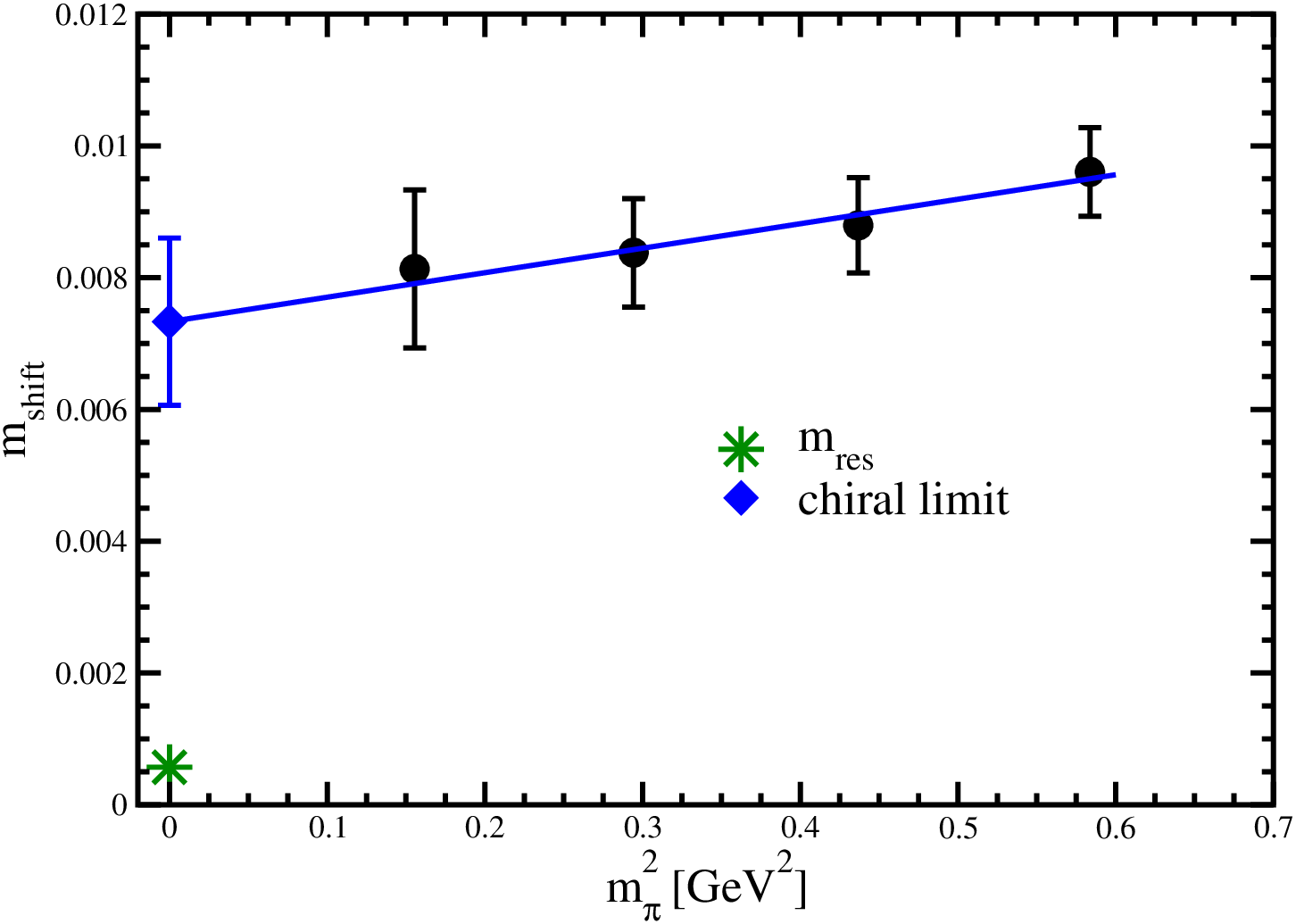}
\end{center}
\caption{
The mass shift $m_{\rm shift}=m_{_{\rm AWT}}-m_f$ as a function 
of pion mass squared. For a comparison, the value of $m_{\rm res}$ is included 
as a asterisk symbol.
}
\label{FIG:R_AWT_chi}
\end{figure}

As shown in Fig.\ref{FIG:R_AWT}, indeed, there is no appreciable $q^2$-dependence in 
the ratio $\alpha_{_{\rm AWT}}$ for each $m_f$. Four different $q^2$ points of $\alpha_{_{\rm AWT}}$ reveal a $q^2$-independent plateau within the statistical errors. 
We evaluate the weighted average of $\alpha_{_{\rm AWT}}$ by using all four measured $q^2$ points. The obtained values of $\alpha_{_{\rm AWT}}$ are tabulated in 
Table~\ref{Tab:AWT_summary}. 
Deviation from unity is getting large as $m_f$ decreases~\footnote{
In Ref.~\cite{Alexandrou:2007xj}, the similar deviation from unity 
as $\alpha_{_{\rm AWT}}>1$ can be read off from their Fig. 19 where an 
inverse of $\alpha_{_{\rm AWT}}$ is given in the limit of $q^2\rightarrow 0$.}. 
This indicates that $\alpha_{_{\rm AWT}}$ may possess a $1/m_f$ term, which
is induced by the presence of the additive mass shift in the axial Ward-Takahashi identity
such as $m_{\rm res}$. To see this point clearly, we plot the modified ratio as $m_f(\alpha_{_{\rm AWT}}-1)$, which can be interpreted as the difference between
$m_{_{\rm AWT}}$ and $m_f$. Fig.~\ref{FIG:D_AWT} shows modified ratios
as functions of four-momentum squared $q^2$ for all four values of $m_f$. There is again
no visible $q^2$ dependence. Moreover, all $m_f$ results are consistent
with each other within statistical errors. 
$m_{\rm shift}\equiv m_{_{\rm AWT}}-m_f$, which is given by taking the weighted average 
of four $q^2$ points in Fig.~\ref{FIG:D_AWT}, corresponds to the relative amplitude 
of $\langle N| J_{5q}^a |N\rangle$ to the usual pseudo-scalar matrix element 
$\langle N| J_{5}^a |N\rangle$. More precisely, $m_{\rm shift}$ is expressed by
\be
m_{\rm shift}=
\frac{\langle N| J_{5q}^a |N\rangle}
{\langle N| J_{5}^a |N\rangle}+{\cal O}(a^2, m_f a^2).
\ee
Therefore, supposing that $\langle N| J_{5q}^a |N\rangle/\langle N| J_{5}^a |N\rangle
\approx {\langle 0| J_{5q}^a |\pi\rangle}/ {\langle 0| J_{5}^a |\pi\rangle}$, 
we expect $m_{\rm shift}\approx m_{\rm res}$ 
besides ${\cal O}(a^2, m_f a^2)$ corrections.
We plot $m_{\rm shift}$ against the pion mass squared in Fig.~\ref{FIG:R_AWT_chi}. 
The mild $m_f$-dependence allows us to take a linear extrapolation  
for $m_{\rm shift}$ to the chiral limit. 
At $m_f=0$, we obtain $m_{\rm shift}=0.0073(12)$, which is about one order of magnitude
larger than $m_{\rm res}=5.69(26)\times 10^{-4}$~\cite{Aoki:2002vt} 
contrary to our naive expectation.  A few \% level ${\cal O}(a^2)$ correction, 
which is observed in the difference between $Z_V$ and $Z_A$ cannot account for this 
discrepancy. To resolve it, it is necessary to calculate the relative amplitude of 
$\langle N| J_{5q}^a |N\rangle$ to the usual pseudo-scalar matrix element $\langle N| J_{5}^a |N\rangle$ directly. We plan to study $\langle N| J_{5q}^a |N\rangle$ as well as
$\langle N| {\cal A}^a_{\alpha} |N\rangle$ with the conserved axial-vector current ${\cal A}_{\alpha}$ in our extended work~\cite{ST}.

\subsubsection{Test for the pion-pole dominance on $G_P(q^2)$}

\begin{figure}[htbp]
\vspace{2mm}
\begin{center}
\includegraphics[width=0.52\linewidth,clip]{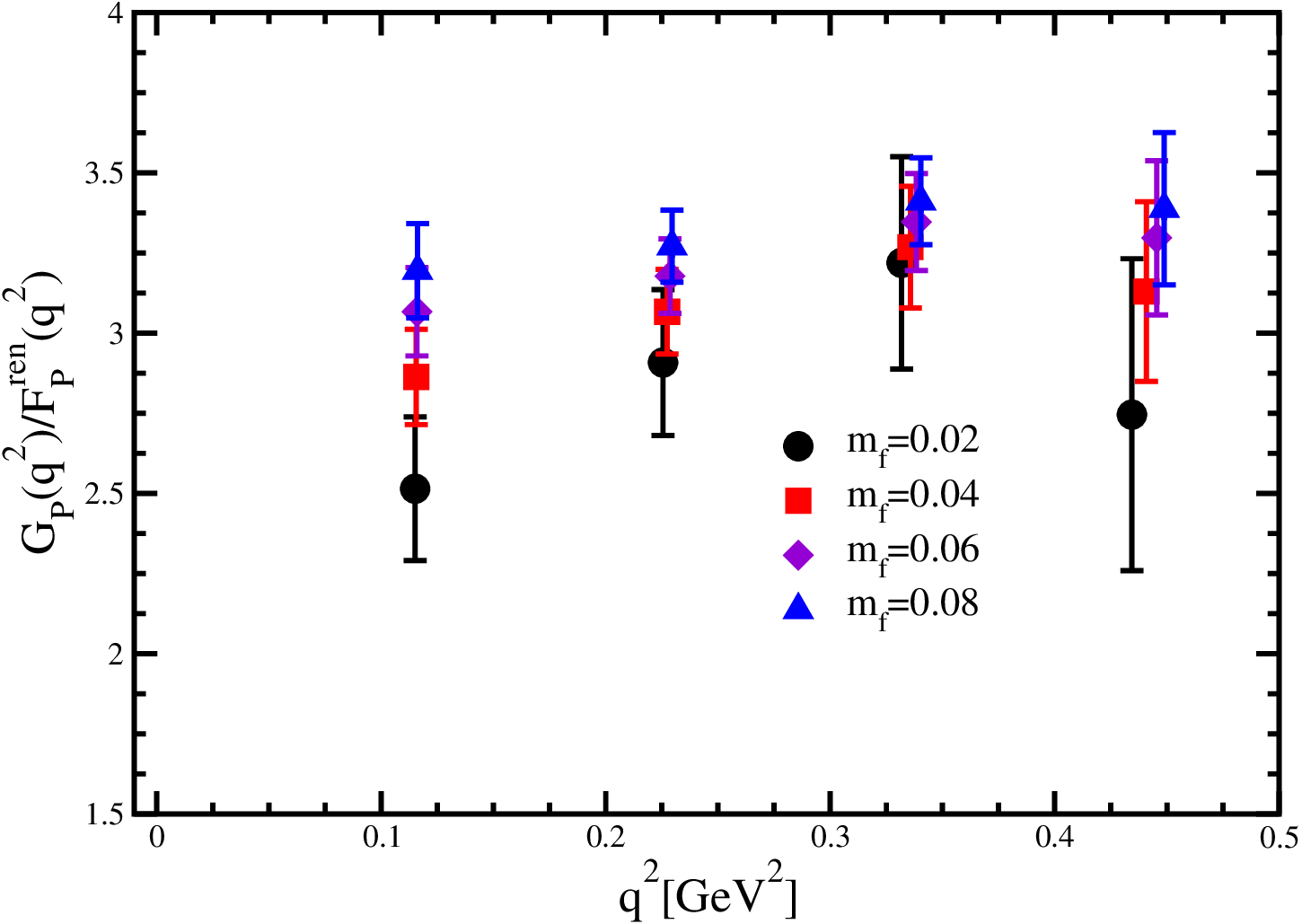}
\end{center}
\caption{
The ratio of $G_P(q^2)/F_P^{\rm ren}(q^2)$ as a function of four-momentum
squared $q^2$. A slight $q^2$-dependence remains against the naive expectation
from the pion-pole dominance hypothesis. 
}
\label{Fig:Ratio_Gp_Fp}
\end{figure}

According to the pion-pole dominance of $F_P(q^2)$, we may expect
that the pion-pole dominance holds even in $G_P(q^2)$. As described in Appendix B, 
a naive pion-pole dominance hypothesis predicts the ratio of
the pseudo-scalar form factor and induced pseudo-scalar form factor,
will not depend on $q^2$ at low $q^2$ but will exhibit a 
constant value, related to the low energy constant $B_0$.
Indeed, this is not the case. In Fig.~\ref{Fig:Ratio_Gp_Fp}, we show the ratio 
of our measured $G_P(q^2)$ and $F_P(q^2)$ as a function of momentum squared $q^2$. 
There is a linear-like $q^2$ dependence, which strengthens as the quark mass decreases.
However, as we will describe below, we confirm that the pion-pole dominance on $G_P(q^2)$ 
still remains valid in our calculation.

\begin{figure}[htbp]
\vspace{2mm}
\begin{center}
\includegraphics[width=0.52\linewidth,clip]{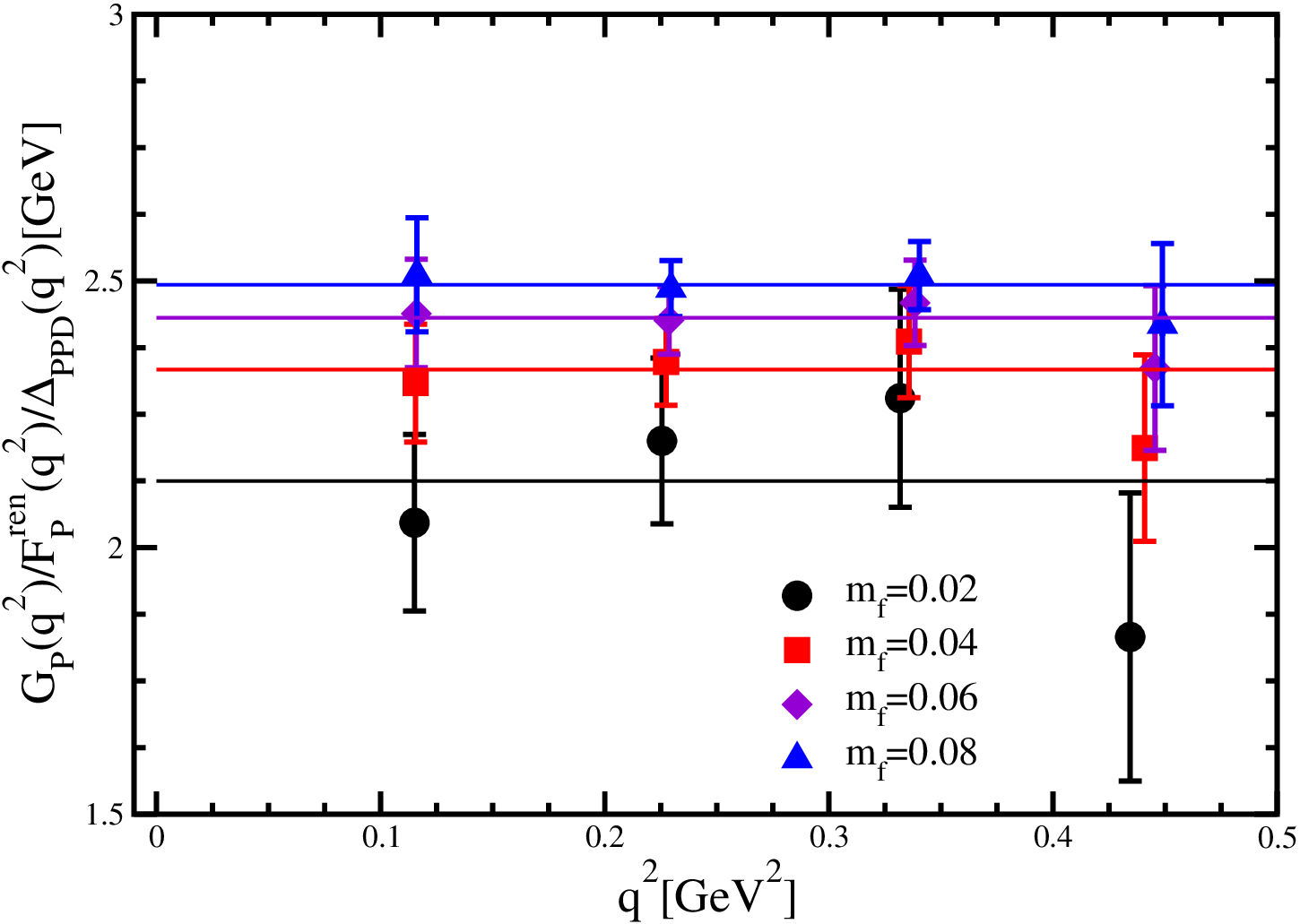}
\end{center}
\caption{
The ratio of $G_P(q^2)/F_P^{\rm ren}(q^2)/\Delta_{\rm PPD}(q^2)$ 
as a function of four-momentum squared $q^2$. 
The residual $q^2$-dependence in the ratio of $G_P(q^2)/F_P^{\rm ren}(q^2)$ 
disappears by a multiplication of $1/\Delta_{\rm PPD}(q^2)$, which is responsible for 
the fact of $\alpha_{\rm PPD}\neq 1$. 
}
\label{Fig:Ratio_Gp_Fp_sub}
\end{figure}

What we have observed in the previous subsection can be interpreted 
as a consequence of the axial Ward-Takahashi identity 
among three nucleon form factors:
\be
2 \mnuc F^{\rm ren}_A(q^2) - q^2 F^{\rm ren}_{P}(q^2)
\approx 2m_{_{\rm AWT}} G_{P}(q^2).
\label{Eq:mod_AWT_FF}
\ee
Combined with this relation and 
the important observation of pion-pole dominance 
in $F_P(q^2)$ (Eq.(\ref{Eq:Fp_PPD_form})), 
one may expect that the $q^2$-dependence of $G_P(q^2)$ is mostly described by 
the pion-pole dominance
form with a slight modification, which corresponds to an extra $q^2$-dependence 
caused by the fact that $\alpha_{_{\rm PPD}}\neq1$:
\be
G_P(q^2)\approx \frac{1+(1-\alpha_{_{\rm PPD}})\frac{q^2}{m_{\pi}^2}}
{\alpha_{_{\rm AWT}}}
\times G_P^{\rm PPD}(q^2),
\label{Eq:Gp_PPD_form}
\ee
where the naive pion-pole dominance form~\cite{Adler:1965ga} is defined as 
\be
2m_f G^{\rm PPD}_P(q^2)=2\mnuc F^{\rm ren}_A(q^2)\frac{m_{\pi}^2}{q^2+m_{\pi}^2}.
\ee
This residual $q^2$-dependence due to $\alpha_{_{\rm PPD}}\neq 1$
is supposed to be responsible for the $q^2$-dependence in 
the ratio of our measured $G_P(q^2)$ and $F^{\rm ren}_P(q^2)$
\be
\frac{G_P(q^2)}{F^{\rm ren}_P(q^2)}\approx
\frac{1+(1-\alpha_{_{\rm PPD}})\frac{q^2}{m_{\pi}^2}}
{\alpha_{_{\rm PPD}}\alpha_{_{\rm AWT}}}
\frac{G_P^{\rm PPD}(q^2)}{F_P^{\rm PPD}(q^2)}
=
\Delta_{\rm PPD}(q^2)
\frac{m_{\pi}^2}
{2m_{_{\rm AWT}}},
\ee
where $\Delta_{\rm PPD}(q^2)\equiv (1+(1-\alpha_{_{\rm PPD}})\frac{q^2}{m_{\pi}^2})/\alpha_{_{\rm PPD}}$.
It is clear that the residual $q^2$-dependence becomes large as $m_\pi^2$ goes to zero.
This feature is in agreement with what we observe in Fig.\ref{Fig:Ratio_Gp_Fp}.
If we multiply the ratio by the corresponding factor $\Delta_{\rm PPD}(q^2)$, which is 
evaluated with measured $\alpha_{_{\rm PPD}}$ and $m_{\pi}$, 
the linear-like $q^2$-dependence indeed disappears 
as indicated in Fig.\ref{Fig:Ratio_Gp_Fp_sub}. Four different $q^2$ points of this ratio 
reveal $q^2$ independent plateau within their statistical errors. We then can evaluate
the weighted average by using all four measured $q^2$ points to obtain the value of
$G_P(q^2)/F_P^{\rm ren}(q^2)$ in the limit of $q^2 \rightarrow 0$, which corresponds to $m_{\pi}^2/(2m_{_{\rm AWT}}\alpha_{_{\rm PPD}})$.

\begin{figure}[htbp]
\vspace{2mm}
\begin{center}
\includegraphics[width=0.52\linewidth,clip]{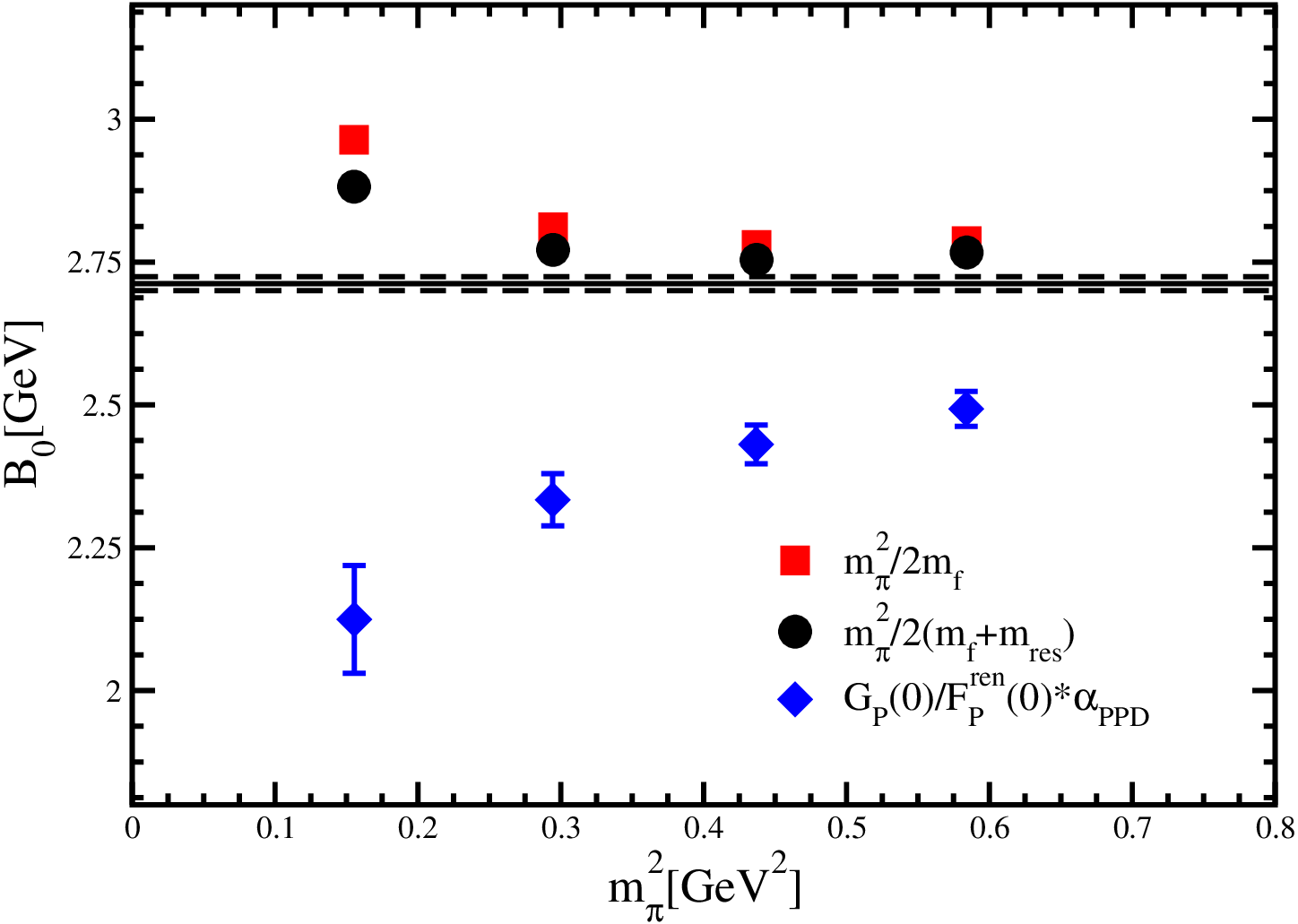}
\end{center}
\caption{
The ratio of $G_P(0)/F_P^{\rm ren}(0)/\Delta_{\rm PPD}(0)$,
$m_{\pi}^2/2m_f$, and $m_{\pi}^2/2(m_f + m_{\rm res})$ as functions
of $m_\pi^2$.
}
\label{Fig:B0_vs_others}
\end{figure}

Although $G_P(0)/F_P^{\rm ren}(0)$ is associated with the bare value of 
the low-energy constant $B_0$
in the pion-pole dominance model as discussed in Appendix B, this prediction
is slightly modified as
\be
\lim_{q^2 \rightarrow 0}\frac{G_P(q^2)}{F^{\rm ren}_P(q^2)}
=\frac{1}{\alpha_{_{\rm PPD}}}
\frac{m_f + m_{\rm res}}{m_f + m_{\rm shift}}B_0,
\ee
where the low-energy constant $B_0$ is defined by the relation 
$m_{\pi}^2=2(m_f + m_{\rm res})B_0$. Because of the fact that
$m_{\rm res}\neq m_{\rm shift}$, 
the values of $G_P(0)/F_P^{\rm ren}(0)\times \alpha_{_{\rm PPD}}$ deviate from $B_0$
as shown in Fig.~\ref{Fig:B0_vs_others}. The horizontal solid and dashed lines
represent the reference value of $B_0$ and its standard deviation, which
are evaluated from fitting squared pion masses to the linear function 
$c_0 + c_1 \cdot m_f$. The fit yields the low-energy constant 
$B_0=c_1/2 = 2.0705(93)$ in lattice units. 
The ratio of fitting parameters ${c_0}$ and ${c_1}$ gives rise to 
the value of $1.75(15)\times 10^{-3}$, which slightly overestimates
$m_{\rm res}$ quoted in Ref.~\cite{Aoki:2002vt}. This is simply because
our fitting is performed in the rather heavier quark mass region 
($0.02 \le m_f \le 0.08$). For comparison, values of $m_{\pi}^2/2m_f$
and $m_{\pi}^2/2(m_f + m_{\rm res})$ are also plotted as 
square and circle symbols in Fig.~\ref{Fig:B0_vs_others}.
This observation may raise a question whether the single universal ``residual mass"
parameter exists.
However, what we observe here is very sensitive to the correct chiral behavior
of the nucleon matrix elements. The quenched approximation may provide
unknown quenching effects in nucleon matrix elements near the chiral limit. 
In this sense, dynamical simulations are much preferable to investigate it
further.

\section{Finite volume effect on nucleon form factors}
\label{Sec:FVE}

As we discussed in Sec.~\ref{Sec4:coupling}, we have found a significant 
finite volume effect on the axial-vector coupling $g_A$, while there is
no appreciable effect on the vector-coupling $g_V$. In this section, we test
for finite volume effects on all of the five nucleon form factors computed in this study. 
Unlike those couplings $g_V$ and $g_A$, which are defined at $q^2=0$, 
it is hard to compare values of the form factor at non-zero $q^2$ 
among different spatial sizes $L$. This is simply because non-zero $q^2$ values 
are discrete in units of $(2\pi/L)^2$. In this study, however, our largest spatial
size ($L=24$) is twice as large as the smallest one ($L=12$). 
The $q^2$ value for ${\bf q}=\frac{2\pi}{L}(2,0,0)$ at $L=24$ coincides with
the one for ${\bf q}=\frac{2\pi}{L}(1,0,0)$ at $L=12$. 
Therefore, at least, a single value of non-zero 
$q^2$ is common between two different lattice sizes. 

\begin{figure}[htbp]
\vspace{2mm}
\begin{center}
\includegraphics[width=0.49\linewidth,clip]{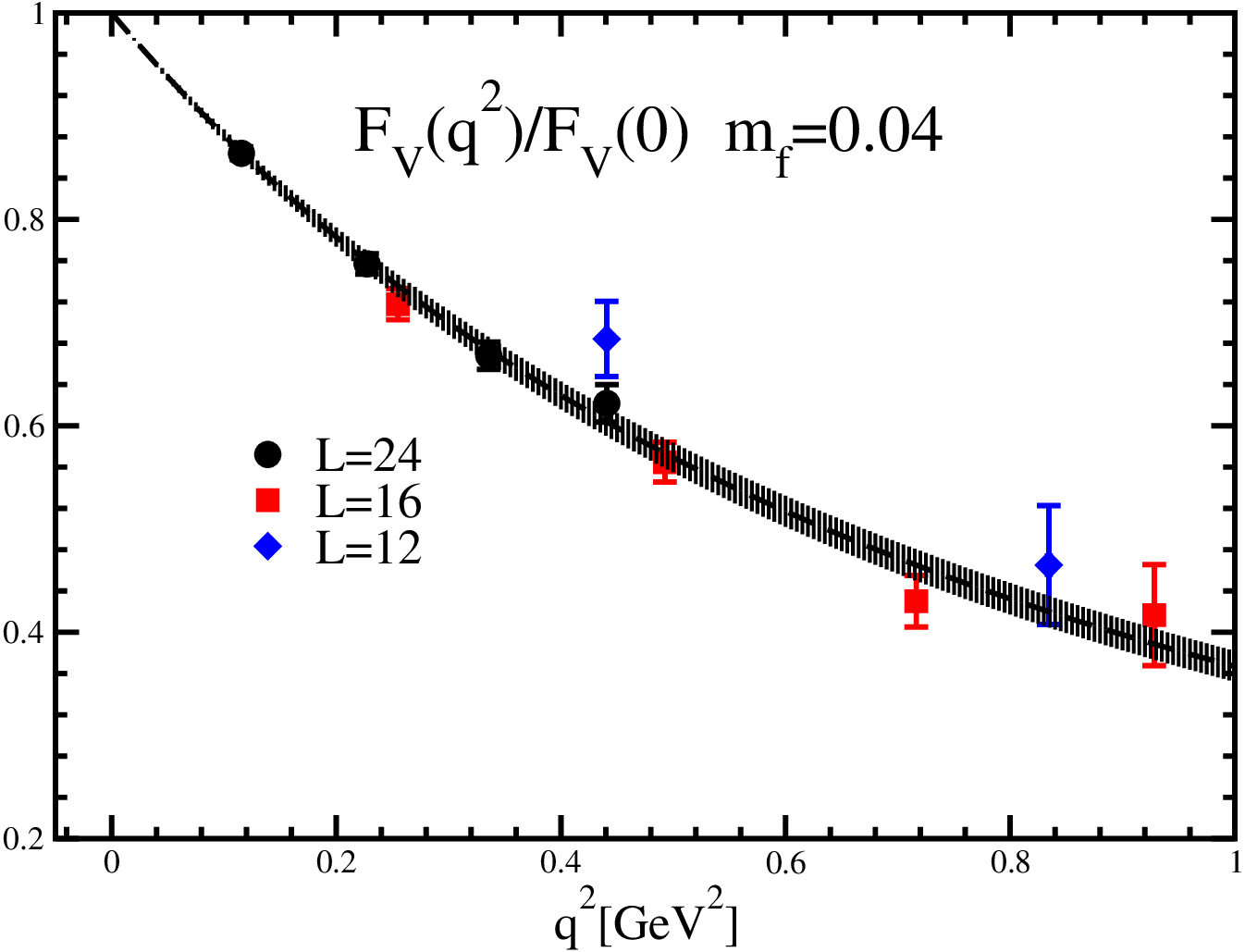}
\includegraphics[width=0.49\linewidth,clip]{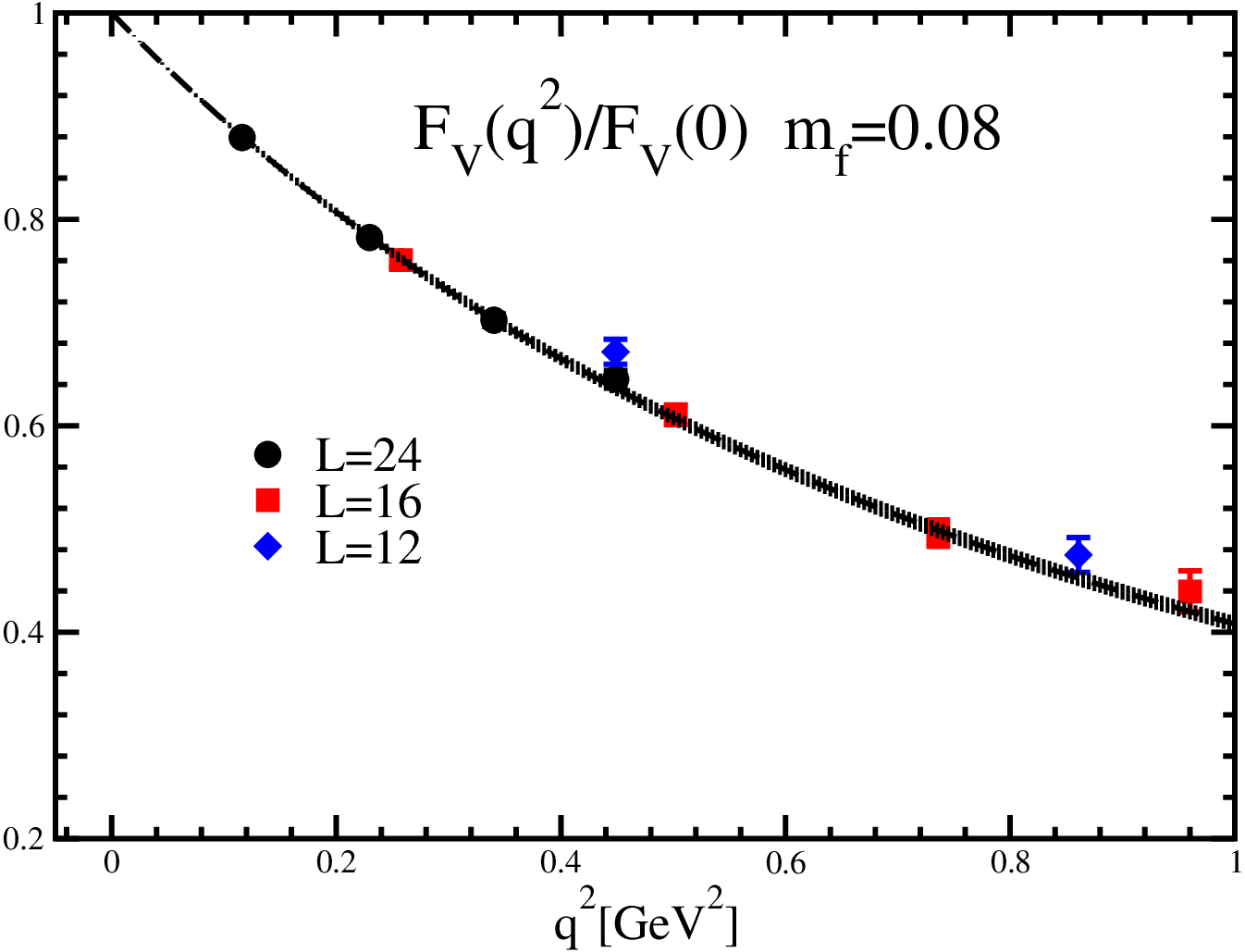}
\end{center}
\caption{The normalized vector form factor $F_V(q^2)/F_V(0)$ obtained from
simulations on lattices with three different spatial sizes. 
The left (right) panel is for $m_f=0.04$ ($m_f=0.08$).
Filled circles, squares and diamonds are obtained from 
$L=24$, $L=16$ and $L=12$. The curves represent the dipole form fits on results
of the largest volume ($L=24$).}
\label{Fig:FV_Fv}
\end{figure}

In Fig.~\ref{Fig:FV_Fv}, we show results for the vector form factor $F_V(q^2)$ 
for three spatial sizes. The left (right) panel is for $m_f=0.04$ ($m_f=0.08$). 
Different symbols represent the values obtained from simulations
with different spatial lattice sizes. 
Solid curves are dipole form fits results on the largest volume ($L=24$) as
we described in Sec.\ref{Sec5:Fv}. These curves should be capable of exposing
finite volume effects on the form factor. Data points from smaller 
lattice sizes of either $L=16$ or $L=12$ at heavier $m_f$ 
agree well with the dipole fits, while results obtained from the 
smallest lattice ($L=12$) at lighter $m_f$ seem to be slightly away from them.
However, the values obtained from $L=24$ and $L=12$ at $q^2\approx  0.44$ GeV
are not significantly different in the statistical sense. 

\begin{figure}[htbp]
\vspace{2mm}
\begin{center}
\includegraphics[width=0.49\linewidth,clip]{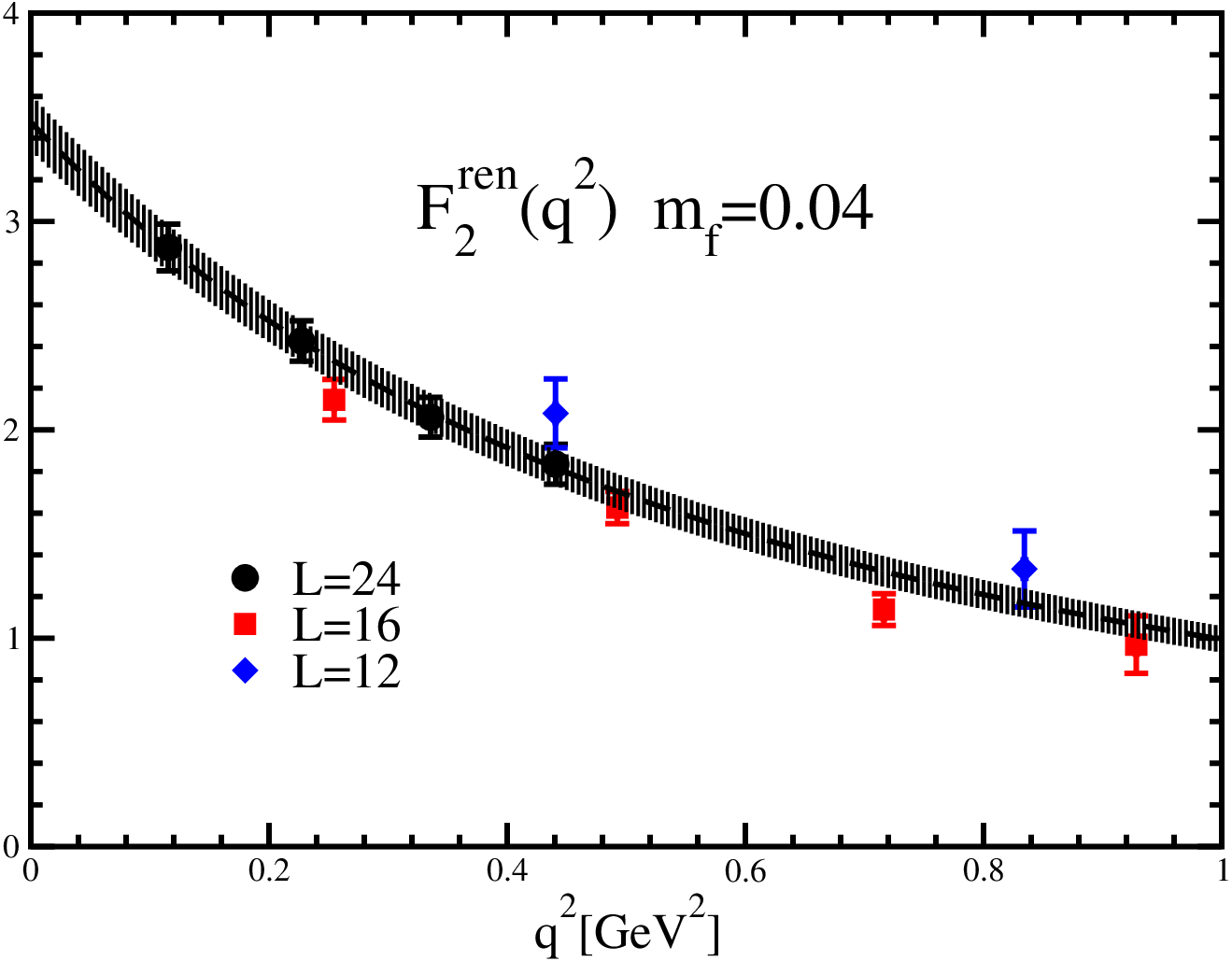}
\includegraphics[width=0.49\linewidth,clip]{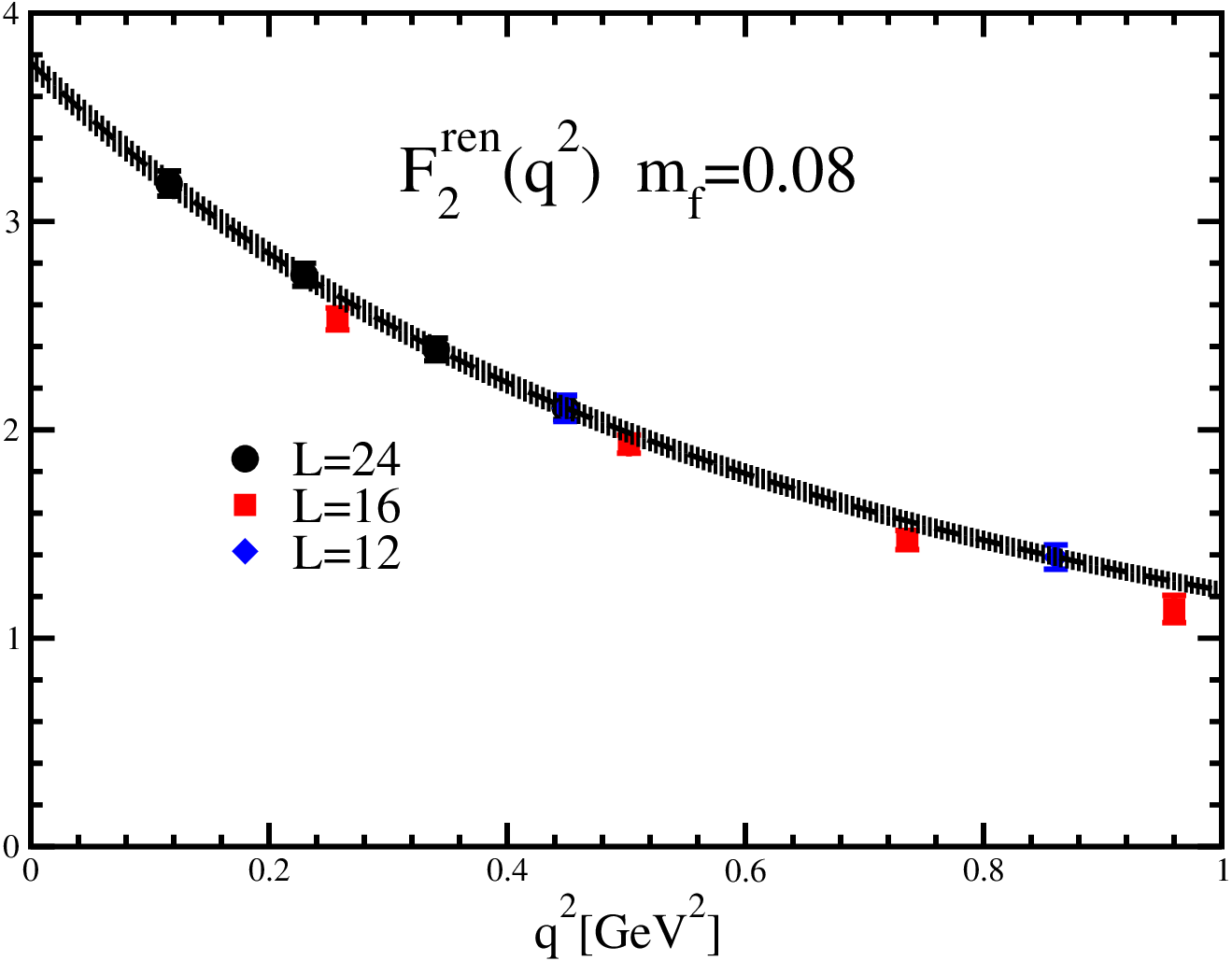}
\end{center}
\caption{The renormalized and dimensionless induced-tensor form factor $F^{\rm ren}_2(q^2)=2\mnuc F_T(q^2)/F_V(0)$ obtained from simulations on lattices with three different spatial sizes. 
The left (right) panel is for $m_f=0.04$ ($m_f=0.08$).
Symbols and solid curves are defined as in Figs.\ref{Fig:FV_Fv}.
}
\label{Fig:FV_Ft}
\end{figure}

We next show the same type of figures for the induced tensor form 
factor $F_T(q^2)$ in Fig.~\ref{Fig:FV_Ft}.
Qualitative features are quite similar to the case of the vector form factor. 
In the case of the heavier
quark mass ($m_f=0.08$), all data points follow the solid curve fairly well, 
which is the dipole form fitted to the data of the largest volume ($L=24$). 
On the other hand, at lighter quark mass ($m_f=0.04$), the data points obtained
from the smallest lattice volume ($L=12$) slightly overestimate the solid curve.
Again, the value for $L=12$ at $q^2\approx  0.44$ ${\rm GeV}^2$  is not significantly
away from the value for $L=24$ in the statistical sense. 

\begin{figure}[htbp]
\vspace{2mm}
\begin{center}
\includegraphics[width=0.49\linewidth,clip]{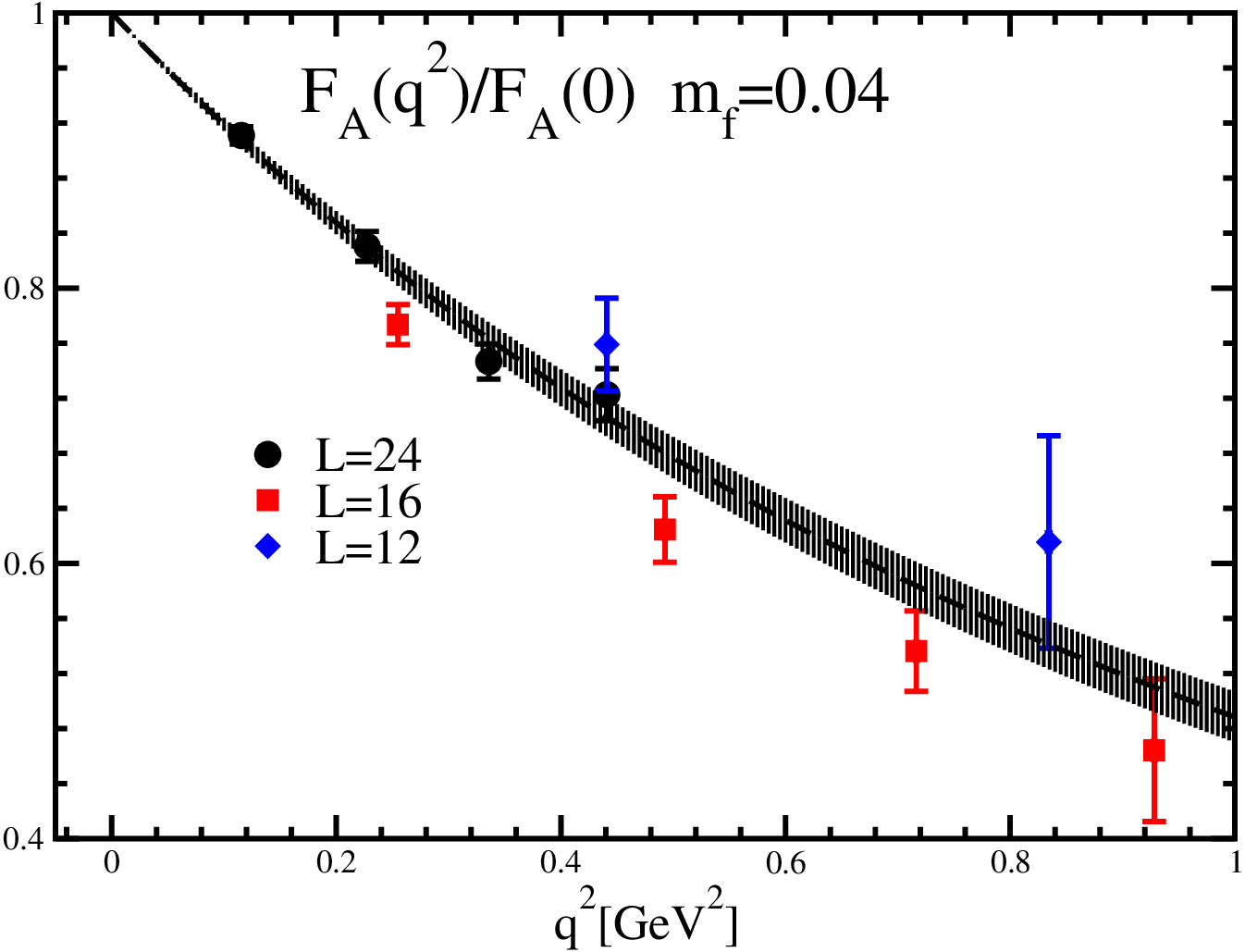}
\includegraphics[width=0.49\linewidth,clip]{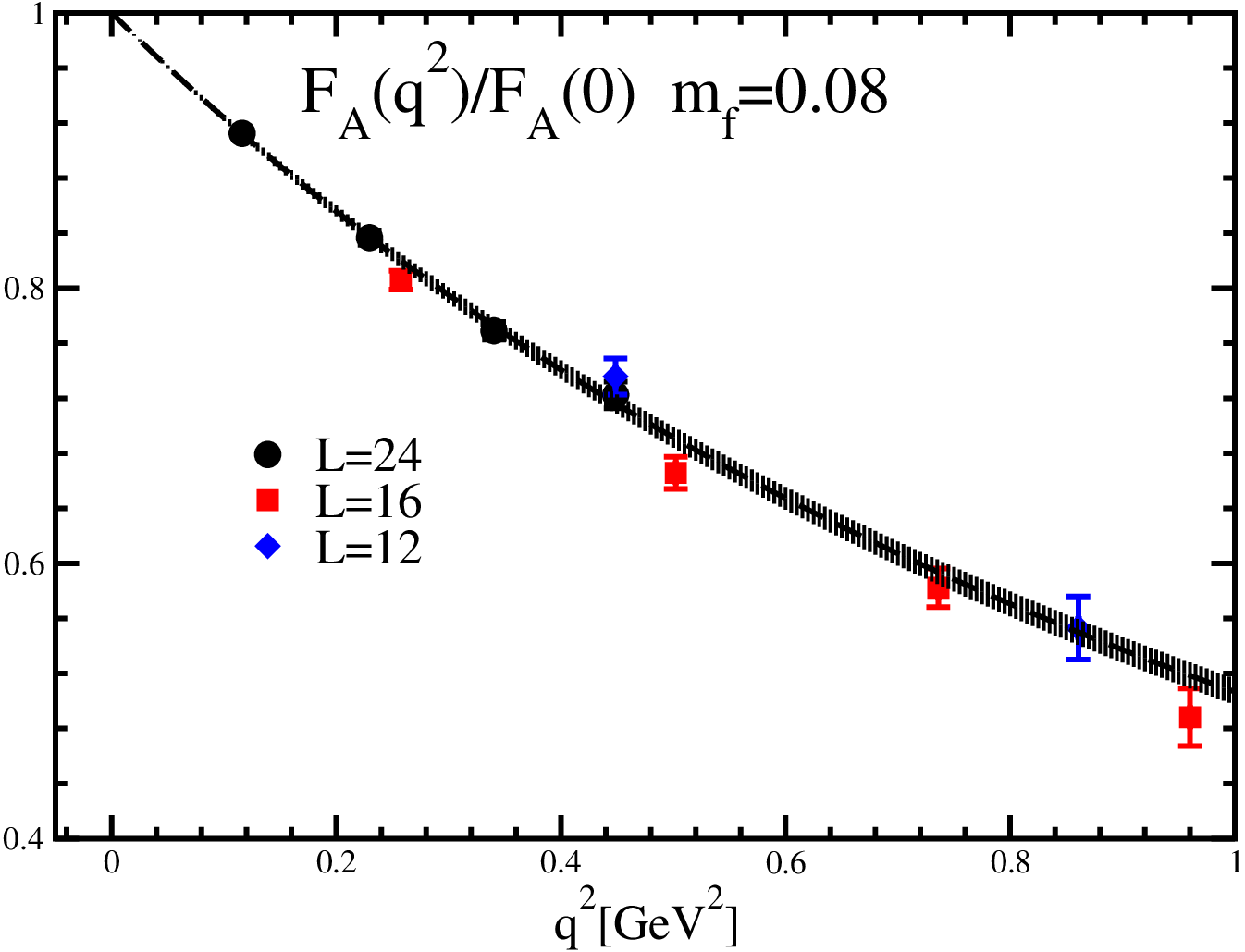}
\end{center}
\caption{The normalized axial-vector form factor $F_A(q^2)/F_A(0)$ 
obtained from simulations on lattices with three different spatial sizes. 
The left (right) panel is for $m_f=0.04$ ($m_f=0.08$).
Symbols and solid curves are defined as in Figs.\ref{Fig:FV_Fv}.
}
\label{Fig:FV_Fa}
\end{figure}

\begin{figure}[htbp]
\vspace{2mm}
\begin{center}
\includegraphics[width=0.49\linewidth,clip]{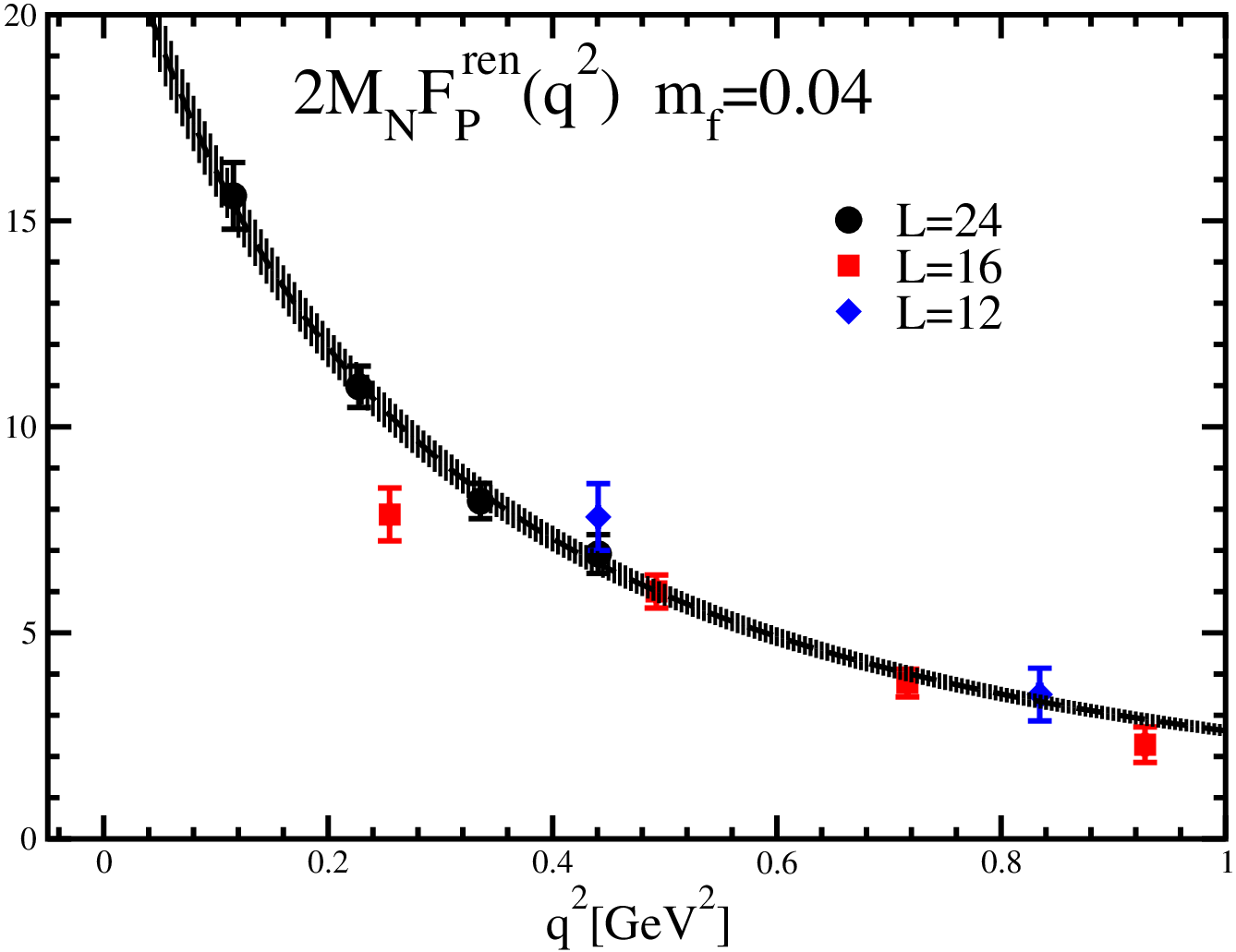}
\includegraphics[width=0.49\linewidth,clip]{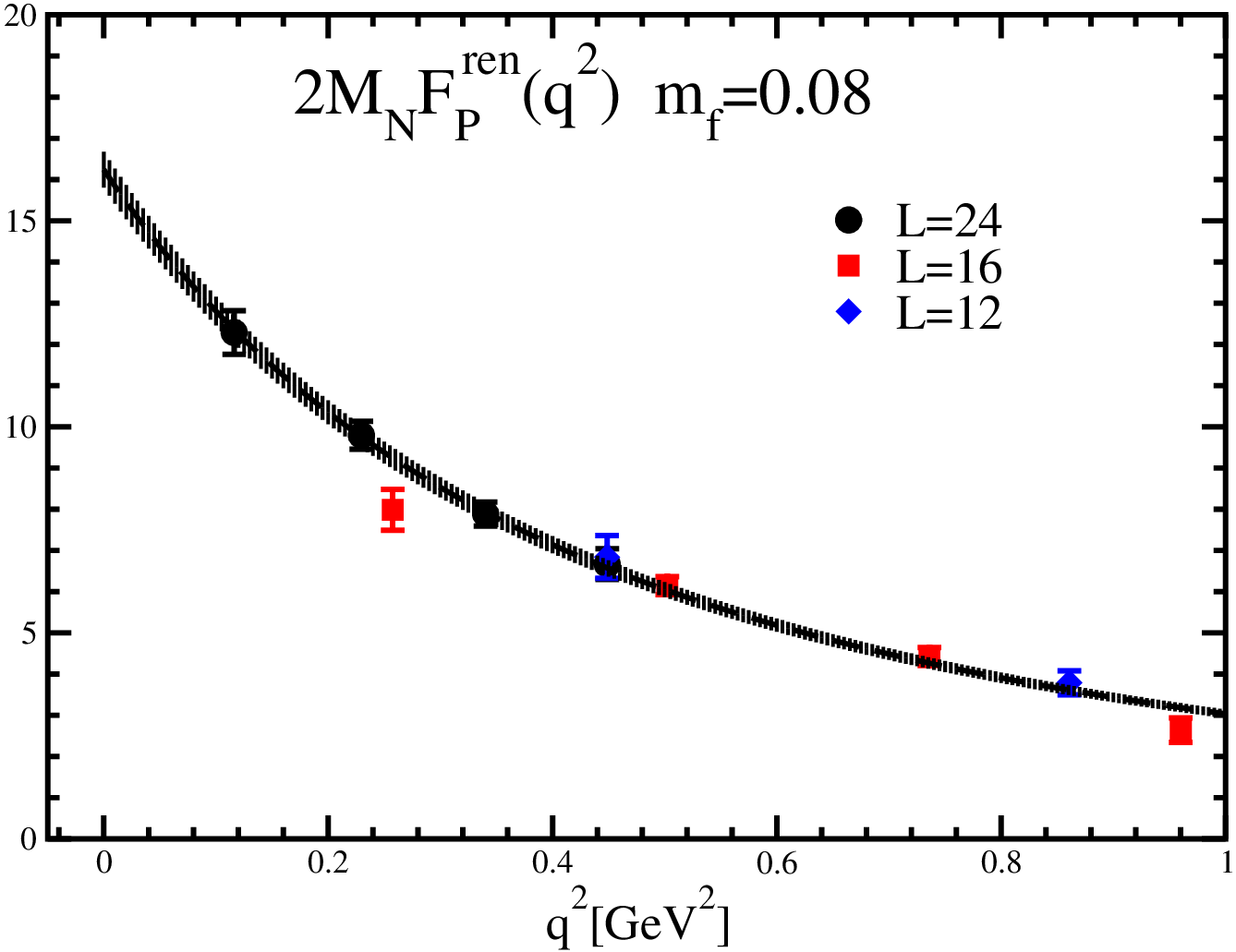}
\end{center}
\caption{The renormalized and dimensionless 
induced-pseudo-scalar form factor, $2\mnuc F^{\rm ren}_P(q^2)$ 
obtained from simulations on lattices with three different spatial sizes. 
The left (right) panel is for $m_f=0.04$ ($m_f=0.08$).
Symbols are defined as in Figs.\ref{Fig:FV_Fv}. Solid curves are resulting
fits with the form (\ref{Eq:Fp_PPD_form}) on results of the largest volume ($L=24$).
}
\label{Fig:FV_Fp}
\end{figure}

\begin{figure}[htbp]
\vspace{2mm}
\begin{center}
\includegraphics[width=0.49\linewidth,clip]{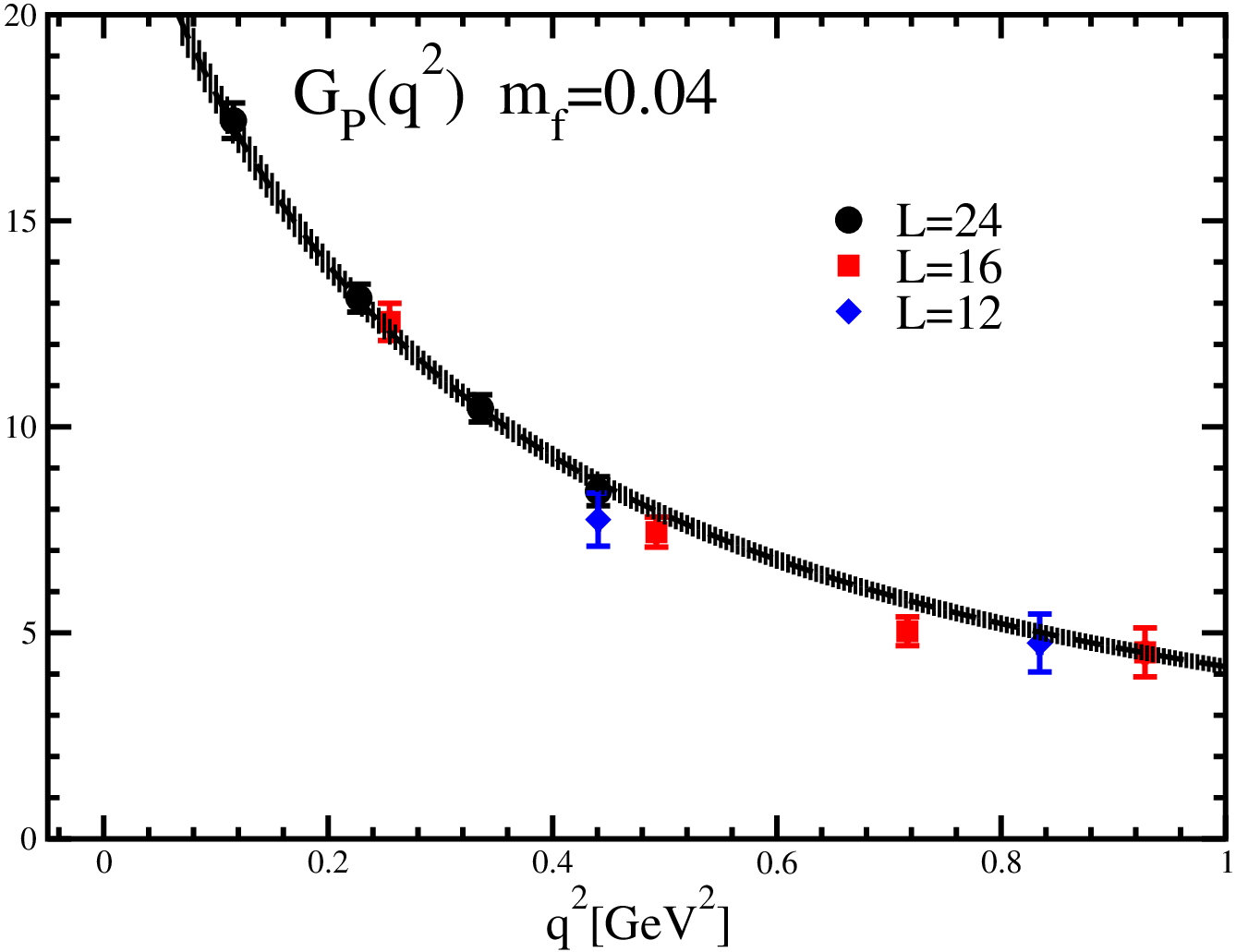}
\includegraphics[width=0.49\linewidth,clip]{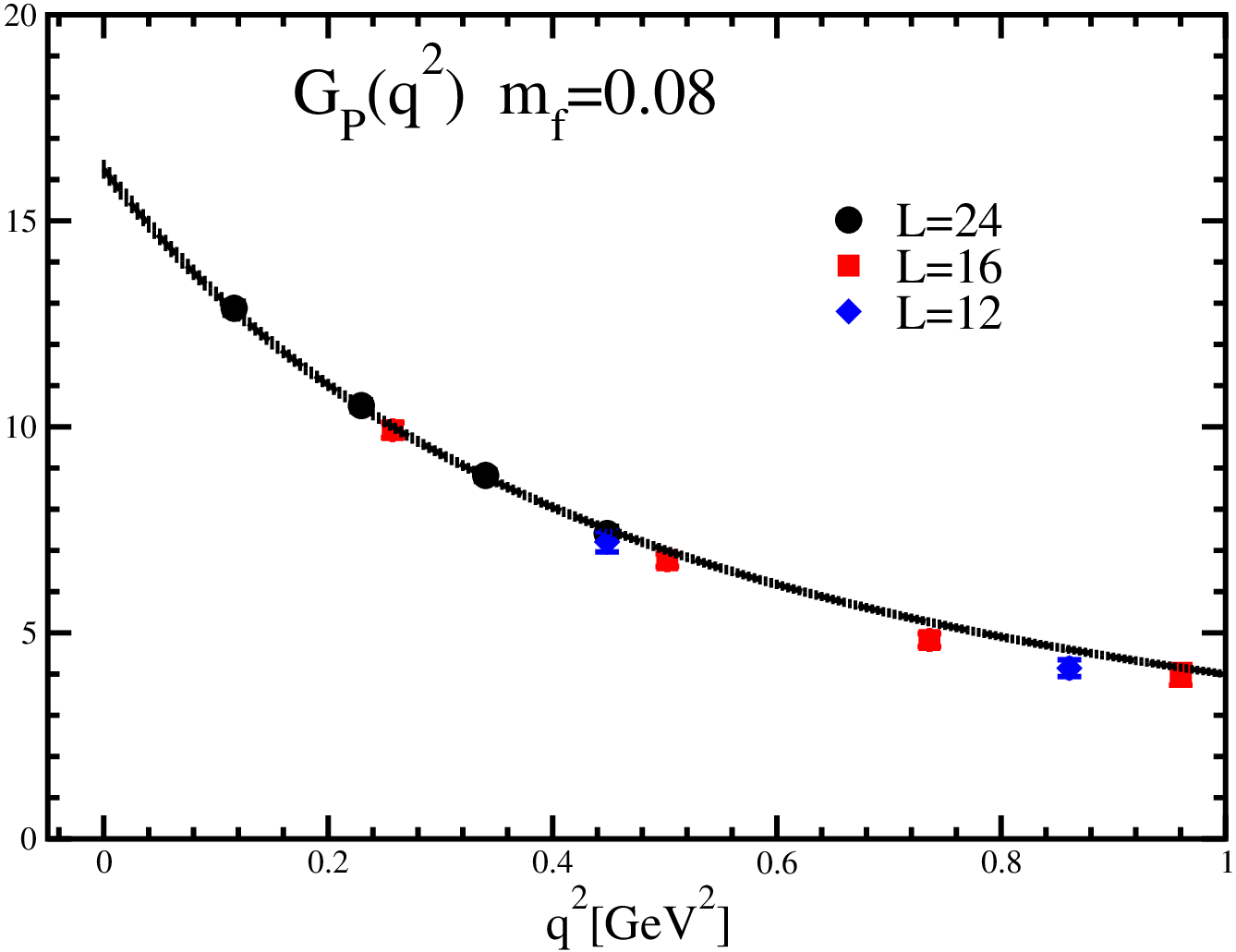}
\end{center}
\caption{The bare pseudo-scalar form factor $G_P(q^2)$ 
obtained from simulations on lattices with three different spatial sizes. 
The left (right) panel is for $m_f=0.04$ ($m_f=0.08$).
Symbols are defined as in Figs.\ref{Fig:FV_Fv}. Solid curves are resulting
fits with the form (\ref{Eq:Gp_PPD_form}) on results of the largest volume ($L=24$).
}
\label{Fig:FV_Gp}
\end{figure}

Next, let us examine two form factors in the axial-vector channel,
where the axial-vector coupling $g_A$ significantly suffers from the finite
volume effect. In Fig.~\ref{Fig:FV_Fa}, the axial-vector form factor $F_A(q^2)/F_A(0)$ for
three spatial sizes is shown. Gross features are mostly similar to both 
the vector form factor and the induced tensor form factor. At the heavier
quark mass, it is observed that all data points follow the solid curve fitted to data of the largest volume with the dipole form (\ref{fig:dipole_fit_a}). Although two values at $q^2\approx 0.44$ ${\rm GeV}^2$ obtained from both largest and smallest volumes agree each other
within statistical errors, the finite volume effect seems to be non-negligible in the lighter quark mass region. On the other hand, the finite volume effects do not show up in the induced pseudo-scalar form factor except for the lowest $q^2$ value for $L=16$, as indicated in Fig.~\ref{Fig:FV_Fp}.
The solid curves are obtained by fitting data of the largest volume with the PPD-like form (\ref{Eq:Fp_PPD_form}), where the pion-pole structure is essential at low $q^2$.
At this moment, we do not have any explanation why the lowest $q^2$ data for $L=16$ 
deviates from the solid curve.
We also show the form factor in the pseudo-scalar channel, namely $G_P(q^2)$, for three lattice sizes in Fig.~\ref{Fig:FV_Gp}. At either quark mass $m_f=0.04$ or 0.08, 
all data points agree well with the solid curves fitted to data of the largest volume by the 
modified PPD form (\ref{Eq:Gp_PPD_form}).
All data that appear in Fig.~\ref{Fig:FV_Fv}-\ref{Fig:FV_Gp} 
are tabulated in Tables~\ref{tab:FF_l24}-\ref{tab:FF_l12}.

Although we may observe the common tendency 
that the three form factors $F_V(q^2)$, $F_T(q^2)$ and $F_A(q^2)$ 
become flatter as a function of $q^2$ in the smallest volume ($L=12$), 
we do not make a definite conclusion through this study. 
Rather we can say that the finite volume effects on the
values of any form factor at finite $q^2$ are less appreciable
than our expectation raised by the fact that the axial-vector coupling significantly
suffers from the finite volume effect. Indeed, we could attempt the direct comparison 
between results for two lattice sizes only at $q^2\approx 0.44$ ${\rm GeV}^2$, which
is a relatively high value. Therefore, we deduce that the finite volume effects are not so 
serious in the high $q^2$ region.  
Finally, it is worth mentioning that we confirm that our observed forms for the five form factors well describe the $q^2$-dependence of those form factors even in the relatively high $q^2$ region, up to at least $q^2\approx 1.0$ ${\rm GeV}^2$, apart from consideration of 
finite volume effects.

\section{Comparison with previous results}
\label{Sec7:Comp}

As we discussed in Sec.~\ref{sec:numeric}, the larger spatial volume enables us
to perform the shorter $q^2$-extrapolation to extract
fundamental information on the nucleon structure, e.g. the magnetic 
moment, the charge radius and the induced pseudo-scalar coupling from 
respective form factors
without large systematic uncertainties. However, some of previous studies are 
performed on relatively small volumes, where the longer $q^2$-extrapolation 
is inevitable. In this context, our lattice setup is superior to previous studies.
(See Table~\ref{Table:PvsLatt} for a summary of previous 
calculations~\cite{{Gockeler:2003ay},{Alexandrou:2006ru},{Hagler:2007xi},{Alexandrou:2007xj}}.)

\begin{figure}[htbp]
\vspace{2mm}
\begin{center}
\includegraphics[width=0.49\linewidth,clip]{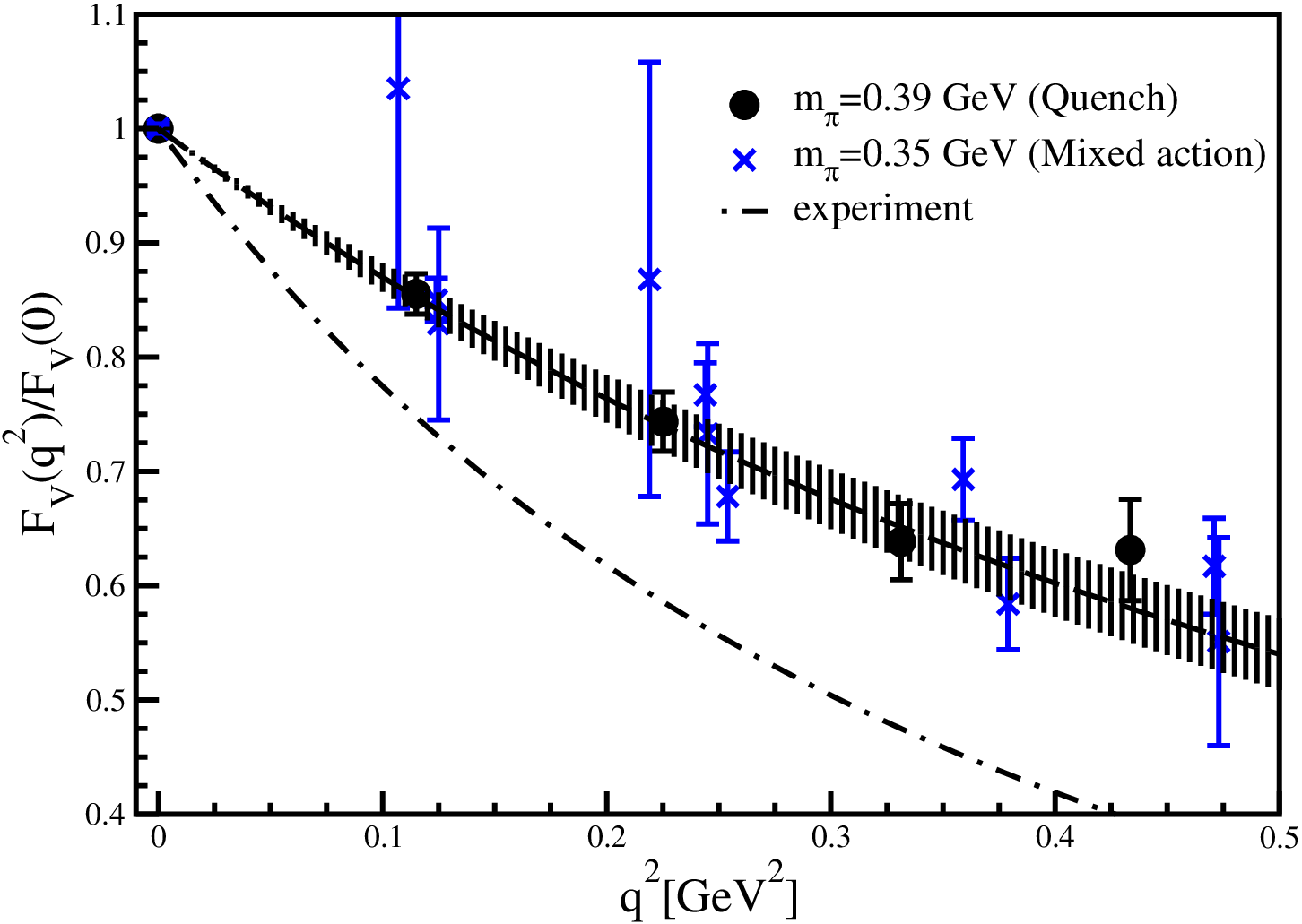}
\includegraphics[width=0.49\linewidth,clip]{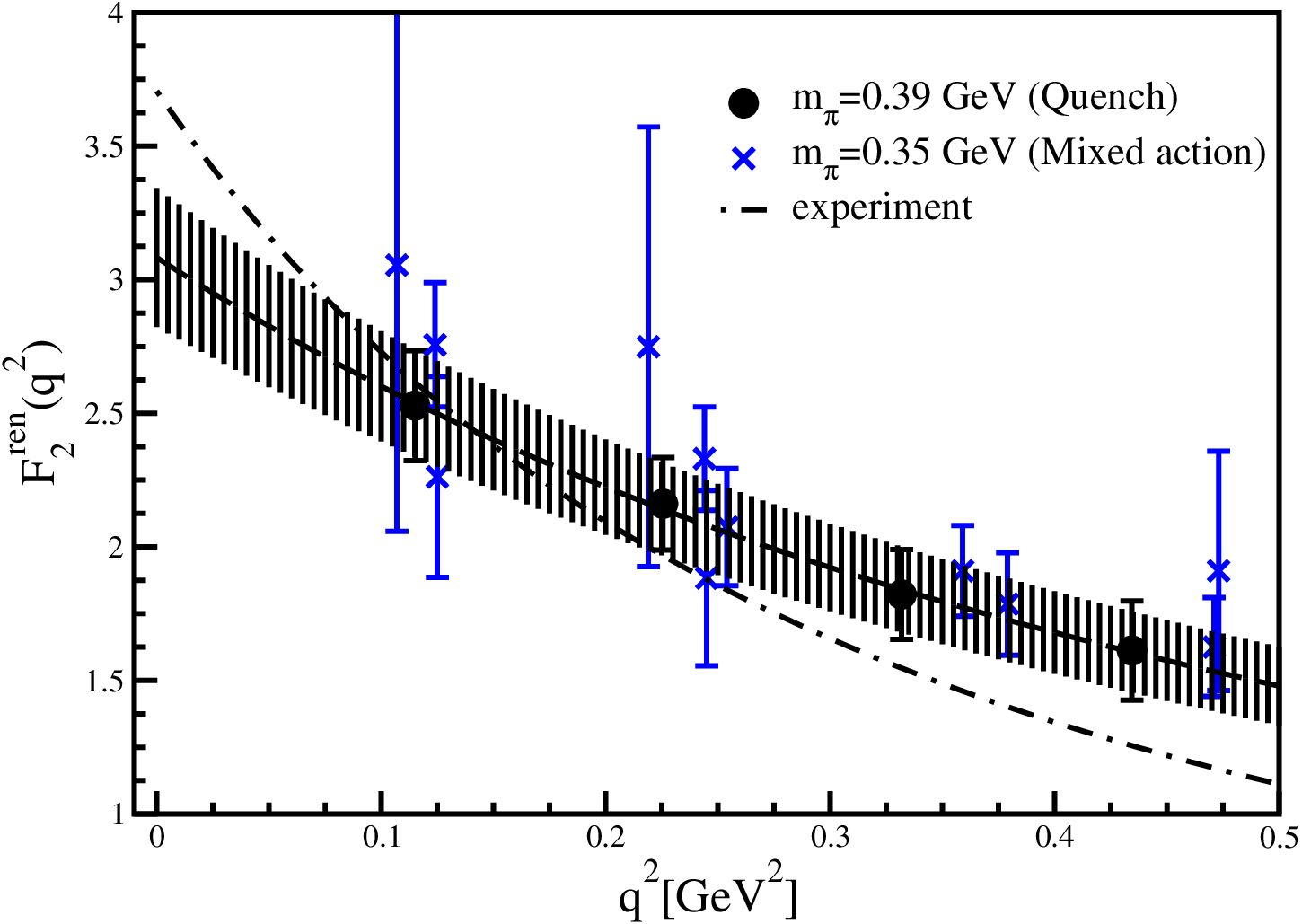}
\includegraphics[width=0.49\linewidth,clip]{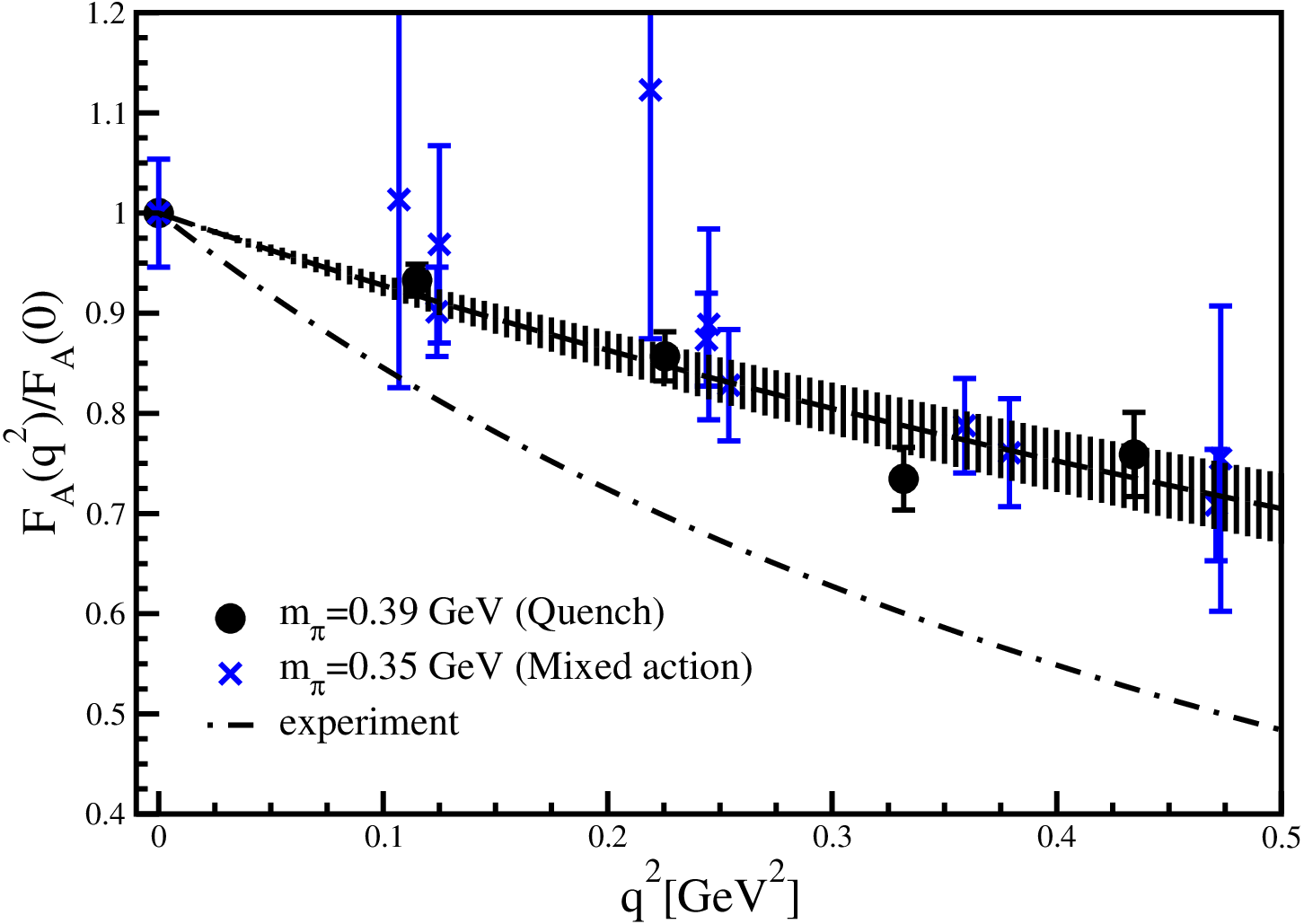}
\includegraphics[width=0.49\linewidth,clip]{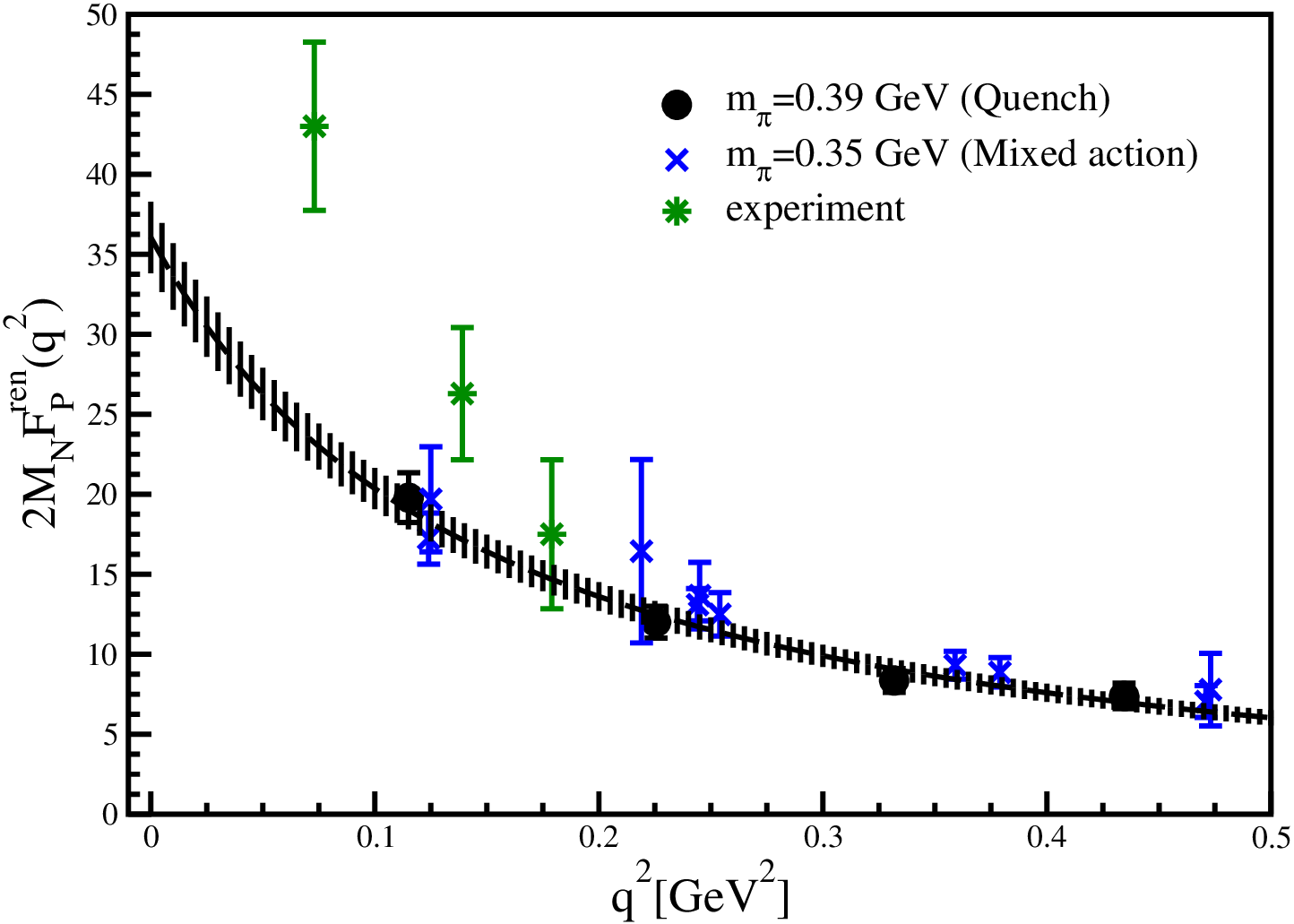}
\end{center}
\caption{Comparisons to results obtained from the LHPC 
mixed action results~\cite{Hagler:2007xi}.
Results for all four form factors at low $q^2$ are consistent with each other.
This suggest that unquenching effects on these form factors are still
small for $m_{\pi}\simge 0.35$ GeV.
}
\label{Fig:COMP_mix}
\end{figure}

A large volume simulation, which is comparable to our lattice volume, $(3.6\;{\rm fm})^3$,
has been done by the LHPC Collaboration with the mixed action simulation using 
DWF valence quarks on the asqtad-improved gauge configurations with 
fourth-rooted staggered sea quarks. The quenching effect in our simulations 
could be observed through a comparison with the mixed action results.
However, there is no detailed analysis of the usual form factors in Ref~\cite{Hagler:2007xi}. 
Instead, we can access their raw data of four nucleon form factors from their tables. 
We simply compare our measured form factors at the lightest quark mass ($m_\pi$=0.39 GeV) 
with their results with the lightest pion mass of 0.35 GeV and the largest volume of $(3.5\;{\rm fm})^3$ in Fig.~\ref{Fig:COMP_mix}.
Surprisingly, all figures for four form factors show good consistency between our 
quenched results and the LHPC mixed action results within statistical errors, 
at least, in the low $q^2$ region. 
This indicates that (un)quenching effects on these form factors are still small 
for $m_{\pi}\simge 0.35$ GeV. However, this conclusion is rather premature
since the mixed action simulation is {\it not} a fully dynamical simulation, rather a 
partially quenched simulation~\cite{Yamazaki:2008py}. 
We must wait for fully dynamical DWF 
simulation to make a firm conclusion. The RBC and UKQCD Collaborations
have begun 2+1 flavor DWF calculations with large physical volume~\cite{{Boyle:2007fn},{Yamazaki:2007mk}}. We will do such a comparison in a future publication.

Finally, it is worth comparing our DWF results with the results obtained
using Wilson fermions.  In Table~\ref{Table:CompPvs}, our results of the axial-vector 
coupling $g_A$ and  the rms radii of the iso-vector Dirac and Pauli form factors, 
the iso-vector nucleon magnetic moment $\mu_p- \mu_n$, the axial dipole mass,
the induced pseudo-scalar coupling $g_P$ and the pion-nucleon coupling $g_{\pi NN}$ 
in the chiral limit are compared with previous 
quenched Wilson results from Refs.~\cite{Alexandrou:2006ru} and 
\cite{Alexandrou:2007xj}. Their lightest pion mass of 0.41 GeV 
and physical volume of $(2.9\;{\rm fm})^3$ are relatively similar 
to our lattice set up. One finds that the rms radii and the axial dipole mass are
quite consistent with each other, while the better agreement with the experiment 
for the axial-vector coupling and the iso-vector nucleon magnetic moment appear 
in our quenched DWF results. Note that although the induced pseudo-scalar form 
factor $F_P(q^2)$ was calculated in Ref.~\cite{Alexandrou:2007xj}, the value of 
the induced pseudo-scalar coupling $g_P$ was not evaluated there. 

Let us estimate $g_P$ from their fit parameters given in 
Table III of Ref.~\cite{Alexandrou:2007xj} as follows. 
First we read off $\alpha_{_{\rm PPD}}$ from their parameters obtained by 
the monopole fit $c_0/(1+q^2/\Lambda^2)$ to the ratio of $2M_N F_P(q^2)/F_A(q^2)$. 
The authors reported that the monopole mass $\Lambda$ is larger than their measured pion mass, while the value of $c_0$ is smaller than $4M_N^2/m_{\pi}^2$ evaluated by the measured pion and nucleon masses. Therefore, their $\alpha_{_{\rm PPD}}$ may have appreciable $q^2$-dependence, which is given by $(1+q^2/m_{\pi}^2)/(1+q^2/\Lambda^2)$. 
On the other hand, the value of $\alpha_{_{\rm PPD}}$ at $q^2=0$ can be given by $c_0 m_{\pi}^2/(4M_N^2)$, which is ranged from 0.68 to 0.62 when the pion mass vary from
0.56 GeV to 0.41 GeV. There is the descending tendency with the decrease of the pion mass.
A simple linear extrapolation with respect to the pion mass squared leads to a value 
in the chiral limit less than 0.6, that is significantly deviated from unity. 
Thus, one may easily deduce that their $g_P$ should be much smaller than our observed 
$g_P$ since their corresponding $\alpha_{_{\rm PPD}}$ and measured $F_A(0)$, both of 
which are main ingredients in Eq.~(\ref{Eq:ps-coupling}), are about 40\% and 20\% smaller 
than our DWF values respectively.
Indeed, our quoted $g_P$ of the quenched Wilson results in Table~\ref{Table:CompPvs}, 
which is determined with the value of $\alpha_{_{\rm PPD}}$ evaluated at $q^2=0.88m_{\mu}^2$, is almost a half of our quenched DWF value.
This discrepancy is attributed to the fact that the Wilson fermions do not yield the correct pion-pole structure of $F_P(q^2)$~\cite{Alexandrou:2007xj}, while the PPD form provides a good description of the $q^2$-dependence of $F_P(q^2)$ at low $q^2$ in our DWF calculation.
\section{Conclusion}

In this paper, we have studied the weak nucleon form factors at low $q^2$
in quenched lattice QCD. We have used domain wall fermions in a very large 
physical volume $(3.6\;{\rm fm})^3$. There are two reasons for 
requiring  such large volume. 
As shown in the early calculation of the axial-vector 
coupling $g_A$~\cite{Sasaki:2003jh}, the nucleon matrix element 
may suffer significantly large finite volume effects. However, we really 
did not know whether the spatial volume $(2.4\;{\rm fm})^3$, that was utilized in 
Ref.~\cite{Sasaki:2003jh}, was large enough for the nucleon. 
Secondly, the large spatial extent provides the capability to access 
lower non-zero momentum transfer.
For the spatial volume $(3.6\;{\rm fm})^3$, the smallest value of 
non-zero $q^2$ is about  0.1 ${\rm GeV}^2$. 

We first demonstrated that the finite volume effect on the axial-vector coupling 
$g_A$ is well described by the power-law behavior, while the vector coupling
$g_V$ has no appreciable finite volume effect. 
However, it is found that a serious finite volume effect on the axial-vector coupling 
$g_A$ is not seen in the range of the spatial lattice size from 2.4 fm 
to 3.6 fm. Finally, we obtain the ratio of the axial-vector to the
vector coupling $(g_A/g_V)^{\rm ren} = 1.219 \pm 0.038({\rm stat})\pm0.024({\rm norm})\pm0.002({\rm vol})$
at the physical point from the largest volume $(3.6\;{\rm fm})^3$, which
agrees with our early estimate from the volume $(2.4\;{\rm fm})^3$~\cite{Sasaki:2003jh} 
and underestimates the experimental value of 1.2695(29) by less than 5\%.

Using the largest volume $(3.6\;{\rm fm})^3$, we studied four of the weak nucleon 
form factors and also the pseudo-scalar form factor.
The $q^2$-dependences of all measured form factors at low $q^2$ are 
discussed with great interest~\footnote{Detailed knowledge about the $q^2$-dependence
of the weak form factors in neutron beta decay is required for our lattice 
study of flavor $SU(3)$ breaking effects in hyperon beta decays~\cite{Sasaki:2006jp}.}.
It is observed that the vector ($F_V$), induced tensor ($F_T$) and the axial-vector ($F_A$) form factors are well described by
the dipole form as $F_i(q^2)=F_i(0)/(1+q^2/M_i^2)$ $(i=V,T, A)$ 
at low $q^2$ ($q^2 < 0.44$ ${\rm GeV}^2$). 
Each measured dipole mass overestimates the corresponding experimental value 
by about 20 \%. This fact indicates that corresponding rms radii are somewhat
smaller than the actual nucleon size. However, interestingly, the ratios of dipole 
masses, $M_A/M_V=1.285 \pm 0.073$ and $M_T/M_V=0.869 \pm 0.057$ are 
fairly consistent with respective experimental ones. We also calculated
the difference of the proton and neutron magnetic moments,  $\mu_p -\mu_n =4.13 \pm 0.23$,
from the value of $F_T(q^2)$ at $q^2=0$. Our obtained value is about 14\%
smaller than experimental value of 4.70589.

We have presented a detailed study of the induced pseudo-scalar form factor 
$F_P(q^2)$, which is less well-known experimentally.  It is observed that the $q^2$-dependence of $F_P(q^2)$ exhibits the strong quark-mass dependence in the low $q^2$ region. This is 
associated with the pion-pole contribution. Indeed, we confirm
that the measured value of $F_P(q^2)$ is well described by the pion-pole dominance (PPD) form
$F^{\rm PPD}_P(q^2)=2\mnuc F_A(q^2)/(q^2+m_\pi^2)$ with
a multiplicative factor $\alpha_{\rm PPD}(< 1)$. With the help of such the PPD-like form, we
can evaluate the induced pseudo-scalar coupling as $(g_P)^{\rm ren}
=8.15\pm0.54({\rm stat})\pm0.16({\rm norm})$.
This value is to be compared with the most recent experimental value of $7.3\pm 1.1$ from the MuCap experiment. Furthermore, we evaluated the pion-nucleon coupling $g_{\pi NN}$ from
the residue of the pion pole in the induced pseudo-scalar form factor and found $g_{\pi NN}=10.4 \pm 1.0({\rm stat})$.

We have also studied the axial Ward-Takahashi identity
in terms of the nucleon matrix elements, which may be referred to as the 
generalized Goldberger-Treiman relation. For this purpose, we also calculated 
the pseudo-scalar matrix element, which is described by the single form factor 
called as the pseudo-scalar form factor $G_P(q^2)$. We have found that the
measured $q^2$-dependence of $G_P(q^2)$ is quite consistent with an expected behavior 
associated with $F_A(q^2)$ and $F_P(q^2)$ in consequence of the axial Ward-Takahashi identity, or the generalized Goldberger-Treiman relation. 
This fact ensures that the PPD form provides a good description 
of the $q^2$-dependence of $G_P(q^2)$ as well. 

In the case of large but finite fifth dimension, the axial Ward-Takahashi identity 
for domain wall fermions can be slightly modified by introducing an additive shift of 
the quark mass due to the presence of the midpoint contribution to the divergence 
of the axial-vector current. Such an additive constant shift known as the residual quark mass is 
usually measured through mesonic two-point correlation functions. In an earlier calculation, 
the residual quark mass is observed much smaller than the lightest quark mass 
utilized here~\cite{Aoki:2002vt}. However, our observed quark mass shift 
$m_{\rm shift}$, which is required for satisfaction of the generalized 
Goldberger-Treiman relation, is close to 50\% of our lightest quark mass 
and an order of magnitude larger than the residual mass quoted 
in Ref.~\cite{Aoki:2002vt}. This issue may be connected with
the correct chiral behavior of the nucleon matrix elements. 
In the quenched approximation, there may be unknown 
quenching effects in nucleon matrix elements in the vicinity of the chiral limit. 
In this context, the above issue is beyond the scope of this quenched study.
Rather, dynamical simulations are much preferable to investigate it further.
The RBC and UKQCD Collaborations have begun 
$N_f=2+1$ flavor domain wall fermion calculations with large physical volume
$V\approx (2.7\;{\rm fm})^3$ and the lightest $ud$ quark mass 
down to 1/7 the strange quark mass ($m_{\pi}\approx 330$ MeV)~\cite{Boyle:2007fn}.
We plan to develop the present calculation for a precise determination of nucleon form factors
and also to address all of unsolved issues described in this paper. 
Such planning is now underway~\cite{Yamazaki:2007mk}.

{\bf Note added: } After the completion of this work, 
we became aware of a paper~\cite{Alexandrou:2007xj} where
the nucleon axial-vector form factor $F_A(q^2)$ and
induced pseudo-scalar form factor $F_P(q^2)$ are calculated
in quenched and unquenched lattice QCD using Wilson fermions.

\begin{acknowledgments}
It is a pleasure to acknowledge S. Choi for his private communication 
providing actual values of the form factor $F_P(q^2)$ in his experiment.
We would like to thank our colleagues in the RBC collaboration and especially  
T. Blum for helpful suggestions and his careful reading of the manuscript, 
and H.-W. Lin and S. Ohta for fruitful discussions. 
We also thank RIKEN, Brookhaven National Laboratory and the U.S. DOE
for providing the facilities essential for the completion of this work. The results
of calculations were performed by using of QCDOC at RIKEN BNL Research Center.
S.S. is supported by the JSPS for a Grant-in-Aid for Scientific Research (C)
(No. 19540265). T.Y. is supported by US DOE grant DE-FG02-92ER40716 and the
University of Connecticut.
\end{acknowledgments}

\section*{Appendix A: Various rms radii in the vector channel}

In Table~\ref{Table:ExpValues}, the electric charge and magnetization radii 
for the proton and neutron are summarized. Using these experimental values,
the iso-vector electric charge and iso-vector magnetization radii can be
evaluated by the following relations~\cite{Thomas:2001kw, Alexandrou:2006ru}
\bea
\langle (r_E^v)^2\rangle &\equiv&  -6 \left.\frac{1}{G^{v}_E(q^2)}\frac{d G^{v}_E(q^2)}{dq^2}\right|_{q^2=0}
=\langle (r_E^p)^2\rangle - \langle (r_E^n)^2\rangle,  \\
\langle (r_M^v)^2\rangle &\equiv& -6 \left.\frac{1}{G^{v}_M(q^2)}\frac{d G^{v}_M(q^2)}{dq^2}\right|_{q^2=0}
=\frac{\mu_p}{\mu_v}\langle (r_M^p)^2\rangle - \frac{\mu_n}{\mu_v}\langle (r_M^n)^2\rangle,
\eea
where $G_{E(M)}^v(q^2)=G_{E(M)}^p(q^2)-G_{E(M)}^n(q^2)$ and $\mu_{v}=\mu_p-\mu_n$. Then one obtains 
$\sqrt{\langle (r_E^v)^2\rangle}= 0.939(5)$ fm and 
$\sqrt{\langle (r_M^v)^2\rangle}=0.862(14)$ fm.  
Similarly, the rms radii for the iso-vector Dirac form factor $F_1^v(q^2)=F_1^p(q^2)-F_1^n(q^2)$ and the iso-vector Pauli form factor $F_2^v(q^2)=F_2^p(q^2)-F_2^n(q^2)$
can be given through the following relations~\cite{Thomas:2001kw, Alexandrou:2006ru}:
%
%
\bea
\langle (r_1^v)^2 \rangle &=& \langle (r_E^v)^2 \rangle - \frac{3}{2}\frac{F_2^v(0)}{\mnuc^2},
\\
\langle (r_2^v)^2 \rangle &=& \frac{1}{\mu_v-1}\left( \mu_v\langle (r_M^v)^2 \rangle - 
 \langle (r_1^v)^2\rangle \right),
\eea
which yield $\sqrt{\langle (r_1^v)^2\rangle}= 0.797(4)$ fm and 
$\sqrt{\langle (r_2^v)^2\rangle}= 0.879(18)$ fm. 

\section*{Appendix B: Generalized Goldberger-Treiman relation and pion-pole dominance}

The generalized Goldberger-Treiman relation is derived from the nucleon matrix elements 
of the currents on both sides of the axial Ward-Takahashi identity~\cite{Weisberger:1966ip};
$\del_{\alpha}A^{a}_{\alpha}(x) = 2 {\hat m} P^{a}(x)$ where 
the exact iso-spin symmetry is considered as ${\hat m}=m_{u}=m_{d}$.
The nucleon matrix element of the divergence of the axial-vector current 
is represented in the following form:
%
%
\bea
\langle N(p^{\prime})| \del_{\alpha}A^a_{\alpha}(0) |N(p) \rangle
&=&
\bar{u}_{_N}(p^{\prime})[i(\Pslash-{\Pslash}{}^{\prime})F_{A}(q^2)
-q^2 F_{P}(q^2)]\gamma_5 t^a u_{_N}(p) \nonumber\\
&=&
[2\mnuc F_{A}(q^2)
-q^2 F_{P}(q^2)]\bar{u}_{_N}(p^{\prime})\gamma_{5}t^au_{_N}(p).
\eea
Here, it is worth mentioning that we have used the Dirac equation for the nucleon, $\bar{u}_{_N}(p)(i\Pslash+\mnuc)=(i\Pslash+\mnuc)u_{_N}(p)=0$ 
to get from the first line to the second line. 
Then one easily finds that the $q^2$-dependences 
of three form factors are constrained by the following relation
\be
2\mnuc F_{A}(q^2)=q^2F_{P}(q^2) + 2{\hat m}G_{P}(q^2),
\label{Eq:GAWT}
\ee
which is a consequence of the axial Ward-Takahashi identity. 
This expression may be referred to as the generalized Goldberger-Treiman relation~\cite{Weisberger:1966ip}.

Here we discuss the case where 
the limits ${\hat m}\rightarrow 0$ and $q^2 \rightarrow 0$ are taken on 
Eq.(\ref{Eq:GAWT}).
Of course, the left-hand side (l.h.s.) of Eq.(\ref{Eq:GAWT}) yields a non-zero value in the double limit.
First, we consider the case where the chiral limit is first taken 
before the limit of $q^2\rightarrow 0$.
\be
\lim_{q^2\rightarrow 0}\left(
\lim_{{\hat m} \rightarrow 0}
2\mnuc F_A(q^2)
\right)
=\lim_{q^2\rightarrow 0}\left(q^2 \lim_{{\hat m}\rightarrow 0}F_P(q^2)\right),
\label{Eq:q0_af_m0}
\ee
which requires the massless pion pole in $F_P(q^2)$ 
in the chiral limit~\cite{Nambu:1960xd} as $\lim_{{\hat m}\rightarrow 0}F_P(q^2) \propto \frac{1}{q^2}$
%
%
%
%
for non-vanishing of the l.h.s. of Eq. (\ref{Eq:q0_af_m0}). 
Secondly, the chiral limit is taken after the limit of $q^2\rightarrow 0$:
\be
\lim_{{\hat m}\rightarrow 0}\left(
\lim_{q^2 \rightarrow 0}
2\mnuc F_A(q^2)
\right)
=\lim_{{\hat m}\rightarrow 0}\left(2{\hat m} \lim_{q^2\rightarrow 0}G_P(q^2)\right),
\label{Eq:q0_bf_m0}
\ee
which requires the $1/{\hat m}$ singularity in $G_P(q^2)$ at $q^2=0$ as
$\lim_{q^2\rightarrow 0}G_P(q^2) \propto \frac{1}{{\hat m}}\sim \frac{1}{m_\pi^2}$
%
%
%
%
for non-vanishing of the l.h.s. of Eq. (\ref{Eq:q0_bf_m0}). 
As a result, $F_P(q^2)$ and  $G_P(q^2)$ must have
the pion-pole structure which should become dominant at low $q^2$~
\cite{Nambu:1960xd}.
Therefore, one can deduce that $F_P(q^2)$ and $G_P(q^2)$ are described by the following
forms, at least, in the vicinity of the pole position 
$q^2=-m_\pi^2$~\cite{Nambu:1960xd,Adler:1965ga}.
\bea
F^{\rm PPD}_P(q^2) &=& \frac{2 \mnuc F_A(q^2) }{q^2 + m_\pi^2},  \\
2 {\hat m} G^{\rm PPD}_P(q^2) &=& 2 \mnuc F_A(q^2) \frac{m_\pi^2}{q^2 + m_\pi^2},
\eea
which we call the pion-pole dominance (PPD) forms.
Consequently, we realize that the ratio of $G^{\rm PPD}_{P}(q^2)$ and 
$F^{\rm PPD}_{P}(q^2)$ gives the low-energy constant $B_0$ as
\be
\frac{G^{\rm PPD}_{P}(q^2)}{F^{\rm PPD}_{P}(q^2)}=B_0,
\ee 
where $m_\pi^2=2{\hat m}B_0$.

\clearpage

\clearpage

%
%
\begin{table}[htbp]
\caption{
Experimental values of magnetic moments, electric charge and magnetization radii of the proton and neutron. }
\begin{ruledtabular}
\begin{tabular}{|lcc|}
\hline
Observable & Experimental value & Reference \\
\hline
$\mu_p$ & $+2.792847351 (28)$ & \cite{Yao:2006px}\\
$\mu_n$ & $-1.91304273 (45)$ & \cite{Yao:2006px}\\
${\langle (r^{p}_{E})^{2}\rangle}^{1/2}$& $0.8750 (68)\; {\rm fm}$ 
& \cite{Yao:2006px} \\
$\langle (r^{n}_{E})^2 \rangle$               & $-0.1161 (22)\; {\rm {fm}^2}$ 
& \cite{Yao:2006px} \\
${\langle (r^{p}_{M})^{2}\rangle}^{1/2}$& $0.855 (35)\;{\rm fm}$
&\cite{Hyde-Wright:2004gh} \\
${\langle (r^{n}_{M})^{2}\rangle}^{1/2}$& $0.873 (11)\;{\rm fm}$& 
\cite{Kubon:2001rj} \\
\hline
\end{tabular}
\end{ruledtabular}
\label{Table:ExpValues}
\end{table}
%

%
%
\begin{table}[htbp]
\caption{Summary of available experimental data for 
the induce pseudo-scalar form factor $F_P(q^2)$.
The smallest $q^2$ point is given by the MuCap experiment, 
while other three $q^2$ points are obtained from 
a single experiment of pion electroproduction at threshold.}
\begin{ruledtabular}
\begin{tabular}{|ccl|}
\hline
$q^2$ $({\rm GeV}^2)$
& $F_P(q^2)$ $({\rm MeV}^{-1})$ & Experiment (reference)\\
\hline
0.0098
& 0.069  $\pm$ 0.010 & ordinary muon capture \cite{Andreev:2007wg}\\
0.073  
& 0.0229 $\pm$ 0.0028 & pion electroproduction 
\cite{{Choi:1993vt},{Choi:private}} \\
0.139 
& 0.0140 $\pm$ 0.0022 & pion electroproduction
\cite{{Choi:1993vt},{Choi:private}} \\
0.179  
& 0.00932 $\pm$ 0.00248 & pion electroproduction
\cite{{Choi:1993vt},{Choi:private}} \\
\hline
\end{tabular}
\end{ruledtabular}
\label{Table:ExpPsF}
\end{table}


%
%
\begin{table}[htbp]
\caption{
Simulation parameters for each volume studied in this work.}
\begin{ruledtabular}
\begin{tabular}{|lcccrccc|}
\hline
Gauge action ($\beta$) &  $L^3 \times T$ & $L_s$ & $M_5$ & Quark mass values ($m_f$) & spatial size $L$ [fm] & Statistics & \# of sources\\
\hline
DBW2 (0.87) & $24^3 \times 32$ & 16 & 1.8 & 0.02, 0.04, 0.06, 0.08 
	&3.6 & 70 & 3\\ 
& $16^3 \times 32$ & 16 & 1.8 & 0.04, 0.06, 0.08 
	&2.4 & 377 & 1\\ 
& $12^3 \times 32$ & 16 & 1.8 & 0.04, 0.06, 0.08 
    	&1.8 & 800 & 1\\ 
\hline
\end{tabular}
\end{ruledtabular}
\label{tab:simulation_param}
\end{table}

%
%
\begin{table}[htbp]
\caption{
The residual mass $m_{\rm res}$, inverse lattice spacing ($a_{\rho}^{-1}$,
set by the $\rho$ meson mass), the renormalization factor of the axial-vector
current ($Z_A$), and the pion decay constant ($F_{\pi}$). 
Those values are taken from Ref.~\cite{Aoki:2002vt}, where
simulations are performed on a $16^3\times 32$ volume.}
\begin{ruledtabular}
\begin{tabular}{|lcccccc|}
\hline
Gauge action ($\beta$) & $M_5$ & $L_s$ & $m_{\rm res}$ & $a_{\rho}^{-1}$ [GeV]& $Z_A(m_f=-m_{\rm res})$
& $F_{\pi}$ [MeV]  \\
\hline
DBW2 (0.87) & 1.8 & 16 & 5.69(26)$\times 10^{-4}$ & 1.31(4) & 0.77759(45) & 91.2(5.2) \\
\hline
\end{tabular}
\end{ruledtabular}
\label{tab:old_simulation_results}
\end{table}

%
%
\begin{table}[htbp]
\caption{
Fitted masses of the pseudo-scalar meson state and 
fitted energies of the nucleon state with the five
lowest momenta for each volume.
All tabulated values are given in lattice units.
Results for the nucleon energies with nonzero momenta are averaged over
all possible permutations of the lattice momentum ${\bf p}=(n_x, n_x, n_z)$
in units of $2\pi/L$, including both positive and negative directions. }
\begin{center}
\begin{ruledtabular}
\begin{tabular}{|cc| l | l l l l l |}
\hline
&&&$E_N({\bf p})$ &&&&\\
size $L$ &
$m_f$ & $m_{\pi}$ & $(0,0,0)$  & (1,0,0)
& (1,1,0) & (1,1,1) & (2,0,0)\\
\hline
24
&0.02&0.3003 (10)   & 0.8651 (70)& 0.9038 (79)& 0.9432 (94)& 0.9824 (117)& 1.0157 (130) \\
&0.04&0.4143 (11)   & 0.9801 (52)& 1.0152 (58)& 1.0496 (66)& 1.0829 (77)& 1.1161 (83)\\
&0.06&0.5040 (11)   & 1.0783 (47)& 1.1098 (50)& 1.1408 (55) & 1.1710 (60) & 1.2006 (66) \\
&0.08&0.5829 (11)   & 1.1690 (43)& 1.1976 (45)& 1.2257 (49)& 1.2531 (53) & 1.2793 (57) \\
\hline
16
&0.04&0.4148 (9)   & 0.9869 (50)& 1.0595 (58) & 1.1251 (74) & 1.1822 (106) & 1.2436 (150)\\
&0.06&0.5050 (8)   & 1.0821 (42)& 1.1483 (48) & 1.2095 (61) & 1.2632 (84)  & 1.3189 (111) \\
&0.08&0.5837 (8)   & 1.1703 (37)& 1.2313 (43) & 1.2886 (54) & 1.3391 (71) & 1.3890 (92)\\
\hline
12
&0.04&0.4150 (10)&  0.9795 (75)& 1.097 (10)& 1.190 (17)  & 1.332 (68) & 1.236 (79)  \\
&0.06&0.5046 (9)  &  1.0729 (55)& 1.1844 (70) & 1.280 (12) & 1.419 (41) & 1.466 (22)\\
&0.08&0.5832 (8)  &  1.1703 (37)& 1.2313 (43) & 1.289 (54) & 1.339 (71)& 1.389 (92) \\
\hline
\end{tabular}       
\end{ruledtabular}
\end{center}
\label{tab:mass_spect}
\end{table}

%
%
\begin{table}[htbp]
\caption{Results for the vector coupling $g_V^{\rm lattice}$, the axial-vector coupling
$g_A^{\rm lattice}$ and their ratio $g_A^{\rm ren}=(g_V/g_A)^{\rm lattice}$.
Gauss-smeared-to-gauss-smeared quark propagators are used in the present study, while
box-to-local quark propagators were used in the previous calculation~\cite{Sasaki:2003jh}. 
$L=16$ results in the present calculations agree well with the previous $L=16$ 
results.}
\begin{ruledtabular}
\begin{tabular}{|lcc|lll|}
\hline
Smearing type (Ref.) & $L^3 \times N_{t}$ &  $m_f$ & $(g_V)_{\rm lattice}$ & $(g_A)_{\rm lattice}$ & $(g_A)_{\rm ren}$ \\
\hline\hline
Gauss-Gauss (this work) & 
$24^3 \times 32$ 
    &0.02 & 1.2435(60) & 1.509(52) & 1.212(42)\\
 &&0.04 & 1.2386(18) & 1.537(27) & 1.240(22)\\
 &&0.06 & 1.2306(12) & 1.533(18) & 1.245(15)\\
 &&0.08 & 1.2206(10) & 1.529(14) & 1.252(11)\\
\hline
&
$16^3 \times 32$
    &0.04 & 1.2429(24) & 1.494(29) &  1.202(24)\\
 &&0.06 & 1.2332(12) & 1.497(17) &  1.214(14)\\
 &&0.08 & 1.2224(9)   &  1.499(13) &  1.226(11)\\
\hline
&
$12^3 \times 32$
    &0.04 & 1.2465(34) & 1.441(51) & 1.156(41) \\
 &&0.06 & 1.2328(16) & 1.462(27) & 1.185(22) \\
 &&0.08 & 1.2213(11) & 1.474(17) & 1.206(14) \\
\hline\hline 
 Box-Local \cite{Sasaki:2003jh} & 
$16^3 \times 32$ 
    &0.02 & 1.2440(30) & 1.531(60) & 1.229(49) \\
 &&0.04 & 1.2323(14) & 1.523(24) & 1.230(20) \\
 &&0.06 & 1.2220(10) & 1.510(15) & 1.230(12) \\
 &&0.08 & 1.2106(8)    & 1.505(11) & 1.236(9) \\
\hline
&
$8^3 \times 24$ 
    &0.04 & 1.223(10) & 1.303(146)& 1.059(120) \\
 &&0.06 & 1.214(5)   & 1.342(74)   & 1.099(62) \\
 &&0.08 & 1.203(4)   & 1.373(46)   & 1.136(39) \\
\end{tabular}
\end{ruledtabular}
\label{tab:ga_gv_all}
\end{table}

%
%
\begin{table}[htbp]
\caption{
Fitted results of $F_1(q^2)$ and  $F_2(q^2)$ with the dipole form [Eqs.(\ref{fig:dipole_fit})
and (\ref{fig:dipole_fit_t})] and their extrapolated values to the chiral limit and the physical point.
}
\begin{ruledtabular}
\begin{tabular}{|c|cc|ccc|}
\hline
& $F_1(q^2)=F_V(q^2)$ &  & $F_2(q^2)=2\mnuc F_T(q^2)$ & &\\
$m_f$ & $M_V$ (GeV) & $\langle r_V^2\rangle^{\frac{1}{2}}$ (fm) 
& $F^{\rm ren}_2(0)=F_2(0)/F_V(0)$ & $M_T$ (GeV) & $\langle r_T^2\rangle^{\frac{1}{2}}$ (fm)\\
\hline
0.08 & 1.330(17)  & 0.514(7)   & 3.76(7)   &  1.155(18) &  0.592(9)   \\
0.06 & 1.286(21)  & 0.532(9)   & 3.65(9)   &  1.112(22) &  0.615(12)  \\
0.04 & 1.239(32)  & 0.552(14)  & 3.48(13)  &  1.072(32) &  0.637(19)  \\
0.02 & 1.177(65)  & 0.581(32)  & 3.08(26)  &  1.062(78) &  0.644(47)  \\
\hline
phys. point  & 1.148(55)  & 0.589(25)  & 3.13(23)  & 1.002(58)  & 0.676(34) \\
chiral limit & 1.142(57)  & 0.592(25)  & 3.11(24)  & 0.997(60)  & 0.679(35) \\
\hline
Empirical values  & 0.857(8)  &  0.797(4) & 3.70589  & 0.778(23)  & 0.879(18)    \\
\hline
\end{tabular}
\end{ruledtabular}
\label{tab:summary_vector}
\end{table}
%

%
%
\begin{table}[htbp]
\caption{
Fitted results of $F_A(q^2)$ with the dipole form [Eq.(\ref{fig:dipole_fit})]
and  $F_P(q^2)$ with the PPD-like form [Eq.(\ref{Eq:Fp_PPD_form})], 
and their extrapolated values to the chiral limit and the physical point.
}
\begin{ruledtabular}
\begin{tabular}{|c|ccc|cc|}
\hline
& $F_A(q^2)$ &  &  & $F_P(q^2)$ & \\
$m_f$ & $(g_A)^{\rm ren}=F_A(0)/F_V(0)$ & $M_A$ (GeV) & $\langle r_A^2\rangle^{\frac{1}{2}}$ (fm) 
&  $(g_P)^{\rm ren}$     
&$\alpha_{_{\rm PPD}}$ \\
\hline
0.08 & 1.252(11) & 1.572(26)  & 0.435(7)  & 
11.08(28) & 0.815(18)  \\
0.06 & 1.245(15) & 1.541(32)  & 0.444(9)  & 
10.37(28)& 0.831(18)  \\
0.04 & 1.240(22) & 1.523(47)  & 0.449(14) & 
9.66(34)& 0.853(20)  \\
0.02 & 1.212(42) & 1.618(126) & 0.422(33) & 
8.73(51)& 0.884(31)  \\
\hline
phys. point   & 1.219(38) & 1.502(83) & 0.453(24) & 
8.15(54) & 0.897(32) \\
chiral limit  & 1.218(40) & 1.500(85) & 0.454(25) & 
8.04(55)& 0.900(33) \\
\hline
Experimental values  &  1.2695(29) & 1.026(21) & 0.666(14) & 7.3 (1.1) &  1 (Theor.) \\
\hline
\end{tabular}
\end{ruledtabular}
\label{Tab:FaFp_summary}
\end{table}
%

%
%
\begin{table}[htbp]
\caption{
Ratio of the dipole masses and their extrapolated values to the chiral limit and the physical point.
}
\begin{ruledtabular}
\begin{tabular}{|ccc|}
\hline
$m_f$ & $M_T/M_V$ & $M_A/M_V$
\\
\hline
0.08 & 0.869(14) &  1.182(17)  \\
0.06 & 0.865(19) &  1.198(24)  \\
0.04 & 0.865(32) &  1.229(42)  \\
0.02 & 0.902(84) &  1.374(131) \\
\hline
phys. point   & 0.869(57)  &  1.285(73) \\
chiral limit  & 0.869(58)  &  1.289(75) \\
\hline
Empirical values & 0.908(28)& 1.197(27)\\
\hline
\end{tabular}
\end{ruledtabular}
\label{tab:summary_ratio_mdp}
\end{table}
%

%
%
\begin{table}[htbp]
\caption{Results of the ratio $\alpha_{_{\rm AWT}}$, 
the quark mass ($m_{_{\rm AWT}}$) defined in the generalized 
Goldberger-Treiman relation [Eq.(\ref{Eq:mod_AWT_FF})]
and the mass shift $m_{\rm shift}$.
}
\begin{ruledtabular}
\begin{tabular}{|c|ccc|}
\hline
$m_f$ & $\alpha_{_{\rm AWT}}$ & $m_{_{\rm AWT}}$ & $m_{\rm shift}=m_{_{\rm AWT}}-m_f$\\
\hline
0.08 &1.12(1)& 0.0896(7)&0.0096(7)\\
0.06 &1.15(1)& 0.0688(7)&0.0088(7)\\
0.04 &1.21(2)& 0.0484(8)&0.0084(8)\\
0.02 &1.41(6)& 0.0281(12)&0.0081(12) \\
\hline
0  & N/A & 0.0073(12) &0.0073(12)\\
\hline
\end{tabular}
\end{ruledtabular}
\label{Tab:AWT_summary}
\end{table}
%

%
%
\begin{table}[htbp]
\caption{
Results of five (dimensionless) form factors computed on a $24^3 \times 32$ volume.
All form factors are renormalized except for the pseudo-scalar form factor $G_P(q^2)$. 
For notational simplicity, we use the momentum ${\bf q}$ in units of $2\pi/L$.}
\begin{ruledtabular}
\begin{tabular}{|c c c  l l l l l |}
\hline
$m_f$ & ${\bf q}$ & $q^2$ (GeV$^2$) & $F^{\rm ren}_V(q^2)$ & $2\mnuc F^{\rm ren}_T(q^2)$ 
& $F^{\rm ren}_A(q^2)$ & $2\mnuc F^{\rm ren}_P(q^2)$ & $G_P^{\rm bare}(q^2)$ \\
\hline\hline
 0.02 & (0,0,0) & 0.000 & 1.0000(48) & N/A      & 1.212(42) & N/A         & N/A        \\
      & (1,0,0) & 0.113 & 0.852(15)  & 2.53(21) & 1.132(41) & 19.79(1.55) & 21.75(1.11)\\
      & (1,1,0) & 0.222 & 0.739(23)  & 2.16(17) & 1.040(40) & 12.00(99)   & 15.26(79)  \\
      & (1,1,1) & 0.326 & 0.633(29)  & 1.82(17) & 0.892(40) &  8.36(74)   & 11.76(78)  \\
      & (2,0,0) & 0.427 & 0.617(42)  & 1.61(19) & 0.921(62) &  7.40(79)   &  8.88(1.01)\\
\hline
 0.04 & (0,0,0) & 0.000 &1.0000(15) & N/A      & 1.240(22) & N/A       & N/A      \\
      & (1,0,0) & 0.114 & 0.864(6)   & 2.88(11) & 1.131(21) & 15.61(81) & 17.43(43)\\
      & (1,1,0) & 0.224 & 0.757(10)  & 2.42(10) & 1.031(21) & 10.97(50) & 13.12(34)\\
      & (1,1,1) & 0.331 & 0.668(13)  & 2.06(9)  & 0.927(19) &  8.20(43) & 10.45(33)\\
      & (2,0,0) & 0.434 & 0.622(18)  & 1.83(10) & 0.897(28) &  6.91(47) &  8.44(35)\\
\hline
 0.06 & (0,0,0) & 0.000 & 1.0000(10) & N/A      & 1.245(15) & N/A       & N/A      \\
      & (1,0,0) & 0.114 & 0.872(4)   & 3.05(8)  & 1.133(14) & 13.46(59) & 14.66(25)\\
      & (1,1,0) & 0.225 & 0.771(6)   & 2.60(7)  & 1.036(14) & 10.27(39) & 11.60(20)\\
      & (1,1,1) & 0.333 & 0.687(8)   & 2.24(7)  & 0.946(13) &  8.01(33) &  9.53(20)\\
      & (2,0,0) & 0.439 & 0.633(11)  & 1.98(7)  & 0.895(18) &  6.72(39) &  7.87(22)\\
\hline
 0.08 & (0,0,0) & 0.000 &1.0000(8)  & N/A      & 1.252(11) & N/A       & N/A      \\
      & (1,0,0) & 0.114 &0.879(3)   & 3.18(6)  & 1.143(10) & 12.29(53) & 12.88(17)\\
      & (1,1,0) & 0.226 &0.782(4)   & 2.74(6)  & 1.048(10  &  9.80(34) & 10.51(14)\\
      & (1,1,1) & 0.335 &0.702(6)   & 2.39(5)  & 0.963(10) &  7.88(29) &  8.82(14)\\
      & (2,0,0) & 0.442 &0.645(8)   & 2.11(6)  & 0.905(14) &  6.67(37) &  7.41(17)\\
\hline
\end{tabular}          
\end{ruledtabular}
\label{tab:FF_l24}
\end{table}
%

%
%
\begin{table}[htbp]
\caption{
The same as Table~\ref{tab:FF_l24}, but for a $16^3 \times 32$ volume.}
\begin{ruledtabular}
\begin{tabular}{|c c c l l l l l |}
\hline
$m_f$ & ${\bf q}$ & $q^2$ (GeV$^2$) & $F^{\rm ren}_V(q^2)$ & $2\mnuc F^{\rm ren}_T(q^2)$ 
& $F^{\rm ren}_A(q^2)$ & $2\mnuc F^{\rm ren}_P(q^2)$ & $G_P^{\rm bare}(q^2)$ \\
\hline\hline
\hline
 0.04 & (0,0,0) & 0.000 &1.0000(19) & N/A      & 1.202(24) & N/A       & N/A      \\
      & (1,0,0) & 0.251 & 0.718(15)  & 2.14(10) & 0.939(23) &  7.87(64) & 12.54(45)\\
      & (1,1,0) & 0.485 & 0.565(19)  & 1.53(8)  & 0.751(30) &  6.00(40) &  7.45(36)\\
      & (1,1,1) & 0.706 & 0.430(25)  & 1.14(8)  & 0.645(35) &  3.78(34) &  5.04(35)\\
      & (2,0,0) & 0.915 & 0.417(49)  & 0.97(14) & 0.558(64) &  2.29(43) &  4.53(59)\\
\hline
 0.06 & (0,0,0) & 0.000& 1.0000(10) & N/A      & 1.214(14) & N/A       & N/A      \\
      & (1,0,0) & 0.253 & 0.744(9)   & 2.37(7)  & 0.963(15) &  8.00(52) & 10.93(25)\\
      & (1,1,0) & 0.491 & 0.593(13)  & 1.80(5)  & 0.788(20) &  6.04(27) &  7.09(21)\\
      & (1,1,1) & 0.717 & 0.471(17)  & 1.33(6)  & 0.688(23) &  4.17(25) &  4.93(22)\\
      & (2,0,0) & 0.933 & 0.430(29)  & 1.06(9)  & 0.580(37) &  2.48(32) &  4.15(34)\\
\hline
 0.08 & (0,0,0) & 0.000 & 1.0000(7)  & N/A      & 1.226(11) & N/A       & N/A      \\
      & (1,0,0) & 0.254 & 0.760(7)   & 2.53(5)  & 0.988(11) &  7.99(50) &  9.92(18)\\
      & (1,1,0) & 0.495 & 0.611(10)  & 1.93(4)  & 0.816(15) &  6.15(22) &  6.76(15)\\
      & (1,1,1) & 0.725 & 0.496(13)  & 1.47(5)  & 0.714(17) &  4.42(22) &  4.83(16)\\
      & (2,0,0) & 0.946 & 0.440(20)  & 1.14(7)  & 0.599(27) &  2.64(30) &  3.98(25)\\
\hline
\end{tabular}          
\end{ruledtabular}
\label{tab:FF_l16}
\end{table}
%

%
%
\begin{table}[htbp]
\caption{
The same as Table~\ref{tab:FF_l24}, but for a $12^3 \times 32$ volume.}
\begin{ruledtabular}
\begin{tabular}{|c c c l l l l l |}
\hline
$m_f$ & ${\bf q}$ & $q^2$ (GeV$^2$) &$F^{\rm ren}_V(q^2)$ & $2\mnuc F^{\rm ren}_T(q^2)$ 
& $F^{\rm ren}_A(q^2)$ & $2\mnuc F^{\rm ren}_P(q^2)$ & $G_P^{\rm bare}(q^2)$ \\
\hline\hline
\hline
 0.04 & (0,0,0) & 0.000 & 1.0000(27) & N/A      & 1.156(41) & N/A       & N/A      \\
      & (1,0,0) & 0.434 & 0.681(34)  & 2.02(14) & 0.883(43) &  7.47(74) &  7.81(61)\\
      & (1,1,0) & 0.822 & 0.466(53)  & 1.32(16) & 0.693(81) &  3.44(56) &  4.79(66)\\
\hline
 0.06 & (0,0,0) & 0.000 & 1.0000(13) & N/A      & 1.185(22) & N/A       & N/A      \\
      & (1,0,0) & 0.439 & 0.672(17)  & 2.03(8)  & 0.885(23) &  7.00(54) &  7.60(33)\\
      & (1,1,0) & 0.837 & 0.468(25)  & 1.33(8)  & 0.666(38) &  3.59(34) &  4.41(30)\\
\hline
 0.08 & (0,0,0) & 0.000 & 1.0000(9)  & N/A      & 1.206(14) & N/A       & N/A      \\
      & (1,0,0) & 0.442 & 0.671(11)  & 2.09(6)  & 0.891(16) &  6.74(48) &  7.23(23)\\
      & (1,1,0) & 0.848 & 0.477(15)  & 1.38(5)  & 0.669(25) &  3.74(25) &  4.20(19)\\
\hline
\end{tabular}          
\end{ruledtabular}
\label{tab:FF_l12}
\end{table}
%

%
%
\begin{table}[htbp]
\caption{Previous lattice calculations for nucleon form factors.
Note that our lowest non-zero $q^2$ ($q_{\rm min}^2$) is
smaller than previous calculations. This is an essential point in  
determinations of the induced pseudo-scalar coupling ($g_P$) 
from $F_P(q^2)$ and the nucleon magnetic moments ($\mu_N$) from 
$F_T(q^2)$ in order to reduce the systematic
uncertainties stemming from long $q^2$-extrapolation.
}
\begin{ruledtabular}
\begin{tabular}{| llccccc |}
\hline
Group (reference) & Type & Fermion (valence)& $a$ [fm] & spatial size [fm] & $m_{\pi}$ [GeV] & $q^2_{\rm min}$ [${\rm GeV}^2$]\\
\hline\hline
QCDSF~\cite{Gockeler:2003ay}~\footnote{Only electric-magnetic form factors
($F_V$, $F_T$) are studied.}
 & Quench & Clover & 0.11 & 1.8 & 0.54, 0.64, 0.74, 0.91, 0.98 & 0.47 \\
&&&0.08 & 1.9 & 0.61, 0.76, 0.89, 1.03 & 0.40 \\
&&&0.06 & 1.9 & 0.63, 0.79, 0.79, 0.93, 1.05 & 0.39\\
\hline
Cyprus-MIT~\cite{Alexandrou:2006ru,Alexandrou:2007xj} 
& Quench & Wilson & 0.09 & 2.9 &  0.41, 0.49, 0.56  & 0.17\\
& Full ($N_f=2$) & Wilson & 0.08 &1.9 & 0.38, 0.51,0.69  & 0.42\\
\hline
LHPC~\cite{Hagler:2007xi}~\footnote{Raw data of $F_V$, $F_T$, $F_A$ and $F_P$
are available in tables, while detail analysis is not found.}
& Mixed ($N_f=2+1$)~\footnote{DWF valence quarks on the asqtad-improved
gauge configurations with fourth-rooted staggered sea quarks.}
 & DWF & 0.124  &3.5& 0.35 &  0.11\\
 &&& 0.124  & 2.5 & 0.36, 0.50, 0.60, 0.68, 0.76  & 0.18$-$0.20\\
\hline
This work & Quench & DWF & 0.15 & 3.6& 0.39, 0.54, 0.66,  0.76  & 0.11\\
&&&0.15&2.4 & 0.54, 0.66, 0.76 & 0.25 \\
&&&0.15&1.8 & 0.54, 0.66, 0.76 & 0.43 \\
\hline
\end{tabular}
\end{ruledtabular}
\label{Table:PvsLatt}
\end{table} 
%

%
%
\begin{table}[htbp]
\caption{Comparisons with previous quenched Wilson results obtained 
from Refs.~\cite{Alexandrou:2006ru} and \cite{Alexandrou:2007xj}.
The extrapolated values in the chiral limit are evaluated by a simple linear quark mass dependence.}
\begin{ruledtabular}
\begin{tabular}{| llllllll |}
\hline
Reference
& $\mu_{p}-\mu_{n}$ 
&$\langle r_V^2\rangle^{\frac{1}{2}}$ [fm] &
$\langle r_T^2\rangle^{\frac{1}{2}}$ [fm] & $(g_A)^{\rm ren}$
 &$M_A$ [GeV] &
$(g_P)^{\rm ren}$ & $g_{\pi NN}$\footnote{$g_{\pi NN}$ was evaluated
at $q^2=0$ in Ref.\cite{Alexandrou:2007xj}, while our $g_{\pi N N}$ 
is defined at the pion pole, $q^2=-m_{\pi}^2$ with the physical pion mass.}
\\
\hline\hline
Refs.~\cite{Alexandrou:2006ru} and \cite{Alexandrou:2007xj}
& 3.73(13) 
& 0.585(13) & 0.72(2) & 1.065(24)~\footnote[2]{
The chiral extrapolated values are not available in Ref~\cite{Alexandrou:2007xj}.
Therefore, we performed correlated  fits with their measured data to evaluate them.}
& 1.500(59) ${}^b$ & 3.54(61)~\footnote[3]{
Although $F_P(q^2)$ was calculated in Ref.~\cite{Alexandrou:2007xj}, the value
of the induced pseudo-scalar coupling was not evaluated. The quoted value
is estimated by us as described in the text.}
& 11.8(0.3)\\
\hline
This work & 4.11(24) & 0.592(25) & 0.679(35)  & 1.218(40) & 1.500(85) & 8.04(55)
&10.4(1.0)\\
\hline
\end{tabular}
\end{ruledtabular}
\label{Table:CompPvs}
\end{table} 
%

\clearpage
%
%

\clearpage
%
%

\clearpage
%
%
%

\clearpage
%
%
%

%
%
%

%
%
%

%
%

\clearpage

%
%
%

\clearpage
%
%
%

%
%

\clearpage
%
%
%

%
%
%


\clearpage
%
%
%

%
%

\clearpage

\clearpage
%
%
%

%
%

\clearpage
%
%
%

%
%
%

\clearpage
%
%
%

%
%

\clearpage
%
%
%

%
%

\clearpage
%
%
%

%
%

\clearpage
%
%
%

%
%

\clearpage

%
%
%

%
%

\clearpage

%
%
%

%
%
%


\clearpage
%
%

\end{document}